\documentstyle{amsppt}

\define \ie{{\it i.e.\/}}
\define \dlim{\displaystyle\lim}
\define \FB{\partial^+}
\define \PB{\partial^-}
\define \BB{\partial_B}
\define \Lp{\Cal L_1}
\define \Pa{\text{P}}
\define \Cz{\Cal C_0}
\define \Bu{\Cal B}
\define \Dtx{\Cal D}
\define \bDtx{\overline{\Dtx}}
\define \ip{\{i^+\}}
\define \IP{\Cal I \Cal P}

\define \inv{^{-1}}
\define \mk{\Bbb L^}
\define \sph{\Bbb S^}
\define \euc{\Bbb R^}
\define \R{\euc{}} 
\define \st{\,|\,}
\define \V{V^+}

\define \M{M^*}
\redefine \L{\hat L}
\define \<{\langle}
\define \>{\rangle}
\define \({\left(}
\define \){\right)}
\define \LI{\roman{LI}}
\define \LS{\roman{LS}}
\define \rhm{\rho_{\text{min}}}

\define \sch{\text{\bf{Sch}}}
\define \sds{\text{\bf{SdS}}}
\define \sads{\text{\bf{SAdS}}}
\define \rn{\text{\bf{RN}}}
\define \ext{{\text{ext}}}
\define \intr{{\text{int}}}
\define \med{{\text{med}}}
\define \alt{{\text{alt}}}
\define \altint{{\text{alt-int}}}
\define \wth{{\text{whh}}}
\define \und{{\text{und}}}
\define \crt{{\text{crt}}}
\define \ovr{{\text{ovr}}}

\input amstex

\topmatter

\title 
Topology of the Causal Boundary for \\
Standard Static Spacetimes \endtitle
\rightheadtext {Topology of Boundary for Standard Static}

\author 
Jos\'e L. Flores and
Steven G. Harris \endauthor

\address
Department of Mathematics, Saint Louis University,
St. Louis, MO 63103, USA \endaddress
\curraddr
Departamento de \'Algebra, Geometr\'ia, y Topolog\'ia, Facultad de
Ciencias, Universidad de M\'alaga, 29071 M\'alaga,
Spain \endcurraddr
\email 
floresj\@agt.cie.uma.es \endemail

\address
Department of Mathematics, Saint Louis University,
St. Louis, MO 63103, USA and Department of Mathematics,
University of Missouri--Columbia, Columbia, MO 65221, USA
\endaddress
\curraddr Department of Mathematics, Saint Louis University,
St. Louis, MO 63103, USA \endcurraddr
\email 
harrissg\@slu.edu \endemail

\thanks 
J. L. Flores was supported in part by
MCyT-FEDER Grant BFM2001-2871-C04-01, MECyD Grant
EX-2002-0612, and MEC Grant RyC-2004-382 and gratefully acknowledges
the hospitality of the Department of Mathematics of Saint Louis
University.  S. G. Harris gratefully acknowledges
the hospitality of the Department of Mathematics of the
University of Missouri--Columbia.  Both authors thank the
Isaac Newton Institute for Mathematical Sciences, Cambridge,
UK, for support during the Programme on Global Problems in
Mathematical Relativity.\endthanks


\endtopmatter

\document

\head 1. Introduction and Chief Results \endhead

The causal boundary of a strongly causal spacetime,
introduced by Geroch, Kronheimer, and Penrose in
\cite{GKP}, is a tool for examining the causal nature of a
spacetime ``at infinity".  As it is conformally
invariant, it is a property of the conformal structure of
the spacetime.  The addition of the causal boundary to a
spacetime results in an object whose structure might be
most naturally characterized as a ``chronological set",
\ie, a set with a chronology relation, a relation
extending that on the spacetime.  (In a spacetime, $p$ is
said to chronologically precede $q$, or $p \ll q$, if
there is a future-directed timelike curve from $p$ to
$q$.  There is also the causality relation:  $p$ causally
precedes $q$, or $p \prec q$, if there is a
future-directed causal curve from $p$ to $q$.)  What is
desired is a topology on this object so that the causal
boundary puts appropriate endpoints on timelike curves
which are endless in the spacetime; but there is a deal of
controversy on how this might best be done.  The topology
suggested in \cite{GKP} has some severe problems, most notable of
which is that in the simplest of all cases, Minkowski $n$-space
$\mk n$, it fails to give what most consider the obvious
topology for the spacetime-cum-boundary, that obtained
from the conformal embedding of $\mk n$ into the Einstein
static space, $\mk 1 \times \sph{n-1}$ (see \cite{HE}). 
Some attempts have been made to ameliorate the general
problems with this topology (such as \cite{BS} and
\cite{S}), but they have no effect on the issue with
Minkowski space.  A new approach can be found in
\cite{MR}, suggesting a substantial modification of the
GKP method in a somewhat complex but also elegant manner; a
related approach that also combines methods used here is to be
found in \cite{F}.

None the less, the basic ideas in the causal boundary
have much to recommend them; in particular, the separate
elements of the future and past causal boundaries---that 
portion of the GKP construction designed to place future
endpoints on future-endless timelike curves or,
separately, past endpoints on past-endless timelike
curves---seem both simple and natural.   A project to
regularize this naturality of just the future causal
boundary was initiated in \cite{H1}.  It was shown there
that a complete categorical treatment is possible for the
future causal boundary (called there the future chronological 
boundary):  In a category of chronological sets, the addition of
the future-chronological (or causal) boundary resulted in a
``future-complete" object, and future-completion is both
functorial and categorically universal, in the category of
chronological sets.  In
\cite{H2} it was shown how to define a topology on any
chronological set, what might be called the future-chronological 
topology; for a future-completed spacetime, this topology has a
number of desirable properties, including giving the right result
for the future-completion of  $\mk n$.

In \cite{H3} it was shown how to construct the future-completion for 
a standard static spacetime, \ie, a spacetime $V$ conformal to a
product spacetime $\mk 1
\times M$, where $M$ is any Riemannian manifold and the
conformal factor is independent of $\mk1$.  The future-completion 
$\V$ of $V$ is most naturally expressed in terms of real-valued
functions on $M$, leading to the possibility of imbuing $\V$ with a
function-space topology.  This results in a very simple structure
for $\V$ as $\euc 1\times \M$ plus one additional point
$\ip$, where $\M$ is a kind of geometric completion of
$M$, adding points at ``geometric infinity" (plus the
Cauchy completion of $M$, should it not be complete); this
geometric completion of $M$ is closely related to the
boundary sphere construction for Hadamard manifolds (see
\cite{BGS}).  The function-space topology on $\V$ yields a
corresponding topology on $\M$, which may likewise be
called the function-space topology there.

However, the function-space topology on $\V = (\mk1 \times
M)^+$ is not the future-chronological topology; the latter
may have more convergence of sequences than the former.  The
future-chronological topology on $V^+$ yields a topology on $\M$;
this may be called the chronological topology on $M^*$, as the
same topology is imparted to $\M$ from using the
past-chronological topology on the past-completion $V^-$ of $V$. 
Contrary to what was said in [H3], the result of using these
topologies may very well be that $V^+$ does not have the simple
product structure over $\M$ that the function-space topology
yields; and although
$\M$ and $\V$ are always Hausdorff in the function-space
topology, this may not be true in the future-chronological
topology for $\V$ and the chronological topology for
$\M$.  This is a benefit, as the non-Hausdorffness on
$\V$ is actually better representative of the physical
conditions in the spacetime; furthermore, the
chronological topology on
$\M$ is always compact, which is not so for the
function-space topology.

It is the purpose of this paper to examine in detail the
circumstances that lead the future-chronological topology
to differ from the function-space topology on $\V$.  In
particular, we show (Theorem 5.15) that if the
future-chronological topology on $\V$ is not the same as the
function-space topology, then the former must be non-Hausdorff.

Section 2 provides a recapitulation of the basic framework and
definitions, followed by a detailed analysis of the
prime example space.  This 2+1 example space is not likely of much
physical interest (it could perhaps represent a static
spacetime based around an infinite band of dielectric material in
an otherwise empty plane), but it presents a sharp focus on the
manner in which the geometry of the Riemannian factor can lead to
topology in the boundary which may not be naively expected: lack
of Hausdorffness and no product structure in the boundary.  The
paper thereafter follows two nearly independent tracks: 

Sections 3 and 4 present a few general results on convergence in
the future-chronological topology.  Section 3 lays out technical
lemmas, and section 4 shows how convergence is related to
rays (minimizing semi-infinite geodesics).   

Sections 5 and 6 present the most important results, many of which
are applicable very generally.  Notable are Corollary 5.12
(detailing when a sequence of points in any strongly causal
spacetime possesses a limit in the future-chronological
topology), Theorem 5.14 (showing that adding the boundary to a
standard static spacetime compactifies the Riemannian factor in
the chronological topology), Theorem 5.15 (showing Hausdorffness
is the sole key to the question of whether or not the naive
topology---a product topology---is the proper one), Theorem 6.2
(a condition on the Riemannian factor guaranteeing the naive
topology, a condition occurring among classical spacetimes), and
Corollary 6.6 (an application to such spacetimes as external
Schwarzschild, external or internal Reissner-Nordstr\"om, or parts
of Schwarzschild--de Sitte.).

\head 2. Preliminaries and Example \endhead

The following can be found in \cite{GKP}, \cite{HE}, or
\cite{H1}:  Let $V$ be a strongly causal, time-oriented
spacetime.  For any point $p \in V$, the past of $p$ is
$I^-(p) = \{q \in V \st q \ll p\}$; for a subset $A \subset
V$, the past of $A$ is $I^-[A] = \bigcup_{p \in
A}I^-(p)$.  A subset $P \subset V$ is a past set if $I^-[P]
= P$; it is an indecomposable past set, or IP, if it is
not the union of two proper subsets, both being past
sets.  Any IP can be expressed as $I^-[\gamma]$ for
$\gamma$ a timelike curve; if $P$ is not the past of any
point, then it is the past of a future-endless timelike
curve.  The future causal boundary of $V$ is
$\FB(V) = \{ P \subset V \st P \text{ is an IP and is not
} I^-(p) \text{ for any } p \in V\}$.  Let $\V = V \cup
\FB(V)$; then we can define an extension to $\V$ of the
chronology relation $\ll$ on $V$ as follows: For $p \in V$
and $P,	Q \in \FB(V)$,
\roster
\item"" $p \ll P$ iff $p \in P$;
\item"" $P \ll p$ iff for some $z \in V$ with $z \ll p$, $P
\subset I^-(z)$; and
\item"" $P \ll Q$ iff for some $z \in V$ with $z \in Q$, $P
\subset I^-(z)$.
\endroster
(In a complete treatment of the future causal boundary,
additional chronology relations would be defined within
$V$---the process called past determination in
\cite{H1}---but in the spacetimes of interest for this
paper, that is unnecessary, as they are already
past-determined.  The usage of $\V$ here is that of
$\widehat V$ in \cite{H1}; in \cite{H1} $\V$ denotes the
further application of past-determination.  We will
ignore that distinction for this paper, as $\widehat V = \V$ for
the spacetimes considered here.)  

As is shown in \cite{H1}, this same process applies equally
well to any ``chronological set": a set $X$, with a
transitive and anti-reflexive relation $\ll$, which has a
countable subset $D$ such that for any $\alpha \ll \beta$
in $X$, there is some $\delta \in D$ with $\alpha \ll
\delta \ll \beta$, and for which no point is unrelated to all other
points.  The only difference is that ``timelike curve" must be
replaced by ``future chain", which means a sequence of points
$\alpha_1 \ll \cdots \ll \alpha_n \ll  \alpha_{n+1} \ll \cdots$.
(Also:  In a spacetime the past of any point is an IP, but there
are chronological sets where the past of a point may be
decomposable; but no such chronological sets appear in this
paper.)  A point $\alpha
\in X$ is a future limit of the future chain $c =
\{\alpha_n\}$ if $I^-(\alpha) = I^-[c]$; if
$X$ is a strongly causal, time-oriented spacetime, this is
equivalent to $\alpha$ being the future endpoint of a timelike
curve through the points $\{\alpha_n\}$.  

The identical process of defining the future causal
boundary, $\FB$, works for a chronological set, and we define
$X^+ = X \cup \FB(X)$, called the future-completion of $X$, with
the same extensions of
$\ll$ as above.  Call a chronological set future-complete if
every future chain has a future limit.  Then for any
chronological set
$X$, 
\roster
\item $X^+$ is again a chronological set, 
\item $X^+$ is future-complete, 
\item future-completion is a functor in an appropriate
category of chronological sets, and 
\item future-completion is universal, hence, categorically
unique for providing a process of future-completing any
chronological set (in the language of [M]: future-completion is 
left-adjoint to the forgetful functor).  
\endroster
These are the reasons for believing that the
future causal boundary is a natural construction.

Although the causality relation is not crucial for
study of the effects of the chronological relation, it can
add some additional insights to boundary considerations. 
The causality relation in $V$ can also be extended to $\V$:
\roster
\item"" $p \prec P$ iff $I^-(p) \subset P$;
\item"" $P \prec p$ iff $P \subset I^-(p)$; and
\item"" $P \prec Q$ iff $P \subset Q$.
\endroster

A static spacetime is one with a timelike Killing field
whose perpendicular-space is integrable.  A standard
static spacetime is a warped product of the form $\mk1
\times M$ with metric $-\Omega \,dt\,^2 + \bar h$, where
$\bar h$ is a Riemannian metric on $M$ and $\Omega : M \to
\R$ is a positive function (this is always strongly
causal and time-orientable).  Thus, any standard static
spacetime is conformal to a metric product,  $\mk1
\times M$ with metric $-\,dt\,^2 + h$, where $h =
(1/\Omega)\bar h$.  As the causal boundary construction is
conformally invariant (relying solely upon the chronology
relation), we will be able to encompass results for all
standard static spacetimes even if we restrict our study
to metric products.  Accordingly, from this point on, the
spacetime $V$ will always mean a metric product $\mk1
\times M$, metric $-\,dt\,^2 + h$.

For any function $f : M \to \R$, let $\Pa(f)$ denote the
past of the graph of $f$, \ie, $ \{(t,x)
\in V \st t < f(x)\}$.  It is shown in \cite{H3} that the
past sets of $V$ are precisely the set $V$ itself and
all subsets of the form $\Pa(f)$, where $f$ is any
Lipschitz-1 function on $M$, \ie, one satisfying $|f(x) - f(y)|
\le d(x,y)$, where $d$ is the distance function on $M$
coming from its Riemannian metric $h$; we can even think of
$V$ as $\Pa(\infty)$.   Let $\Lp(M)$ denote the
set of Lipschitz-1 functions on $M$.

The IPs of $V$ have the form $\Pa(f)$ for special
elements of $\Lp(M)$:  For any $t \in \R$ and $x \in M$, let
$d^t_x : M \to \R$ be the map $d^t_x(y) = t - d(x,y)$; then
for $p = (t,x) \in V$, $I^-(p) = \Pa(d^t_x)$.  For the
elements of $\FB(V)$ we must look at IPs of the form
$I^-[\gamma]$ for $\gamma$ a future-endless timelike
curve; actually, we can use null curves instead of
timelike with no loss of generality.  Any null curve
$\gamma$ in $V$ can be parametrized so that it is in the
form $\gamma(t) = (t,c(t))$ for $c$ a unit-speed curve in
$M$; $\gamma : [\alpha, \omega) \to V$ is future-endless
precisely if $c$ has no endpoint at $\omega$ ($\omega
= \infty$ is allowed).  For any endless unit-speed curve
$c: [\alpha,\omega) \to M$, the {\it Busemann function \/}
for $c$ is the function $b_c: M \to \R$ given by $b_c =
\dlim_{t \to \omega} d^t_{c(t)}$, \ie,  
$$b_c(x) = \lim_{t \to \omega} \; t-d(c(t),x).$$
For any $x \in  M$, the function $t \mapsto t-d(c(t),x)$ is
monotonic increasing, so the limit above always exists, if
we allow $\infty$ as a possible limit; but either $b_c(x) =
\infty$ for all $x$, or $b_c(x)$ is finite for all $x$, in
which case $b_c$ is Lipschitz-1.  In either case,
$I^-[\gamma] = \Pa(b_c)$.   Let $\Cz(M)$ be the set of all
unit-speed endless curves in $M$ with finite Busemann
function; then $\FB(V)$ consists of $\{\Pa(b_c) \st c \in
\Cz(M)\}$ together with $\Pa(\infty)$ (\ie, $V$
itself as a single point of the boundary).  
In parallel with the nomenclature for $\mk n$
and other spacetimes, we may call $\Pa(\infty)$ ``timelike
infinity" and label it $i^+$.

Any ray---a semi-infinite geodesic which is minimizing on
all intervals---is always in $\Cz(M)$; accordingly, we may
call the curves in $\Cz(M)$ {\it asymptotically
ray-like\/}.  If $M$ is a Hadamard manifold---complete,
simply connected, and with non-positive curvature---then
the rays yield the totality of all finite Busemann functions,
and they are used to construct the boundary sphere for
$M$; see \cite{BGS}.  But positive curvature allows for the
existence of asymptotically ray-like curves $c$ such that
$b_c$ is not the Busemann function of any geodesic (see the
example in section 4).

Let $\Bu(M)$ denote the finite Busemann functions on $M$. 
Note that for any $c:[\alpha,\omega) \to M$ in $\Cz(M)$,
for any $a \in \R$, the curve $c^a : [\alpha+a,\omega+a)
\to M$ given by $c^a(t) = c(t-a)$ is also in $\Cz(M)$ and
has $b_{c^a} = b_c + a$.  Thus, there is a natural
$\R$-action on $\Bu(M)$ with $a \cdot b_c = b_c + a$. 
Thus we have an $\R$-action on $\V$ with $a \cdot (t,x) =
(t+a,x)$, $a \cdot \Pa(b_c) = \Pa(a \cdot b_c)$, and $a
\cdot i^+ = i^+$.

If the metric on $M$ is complete, then all endless curves
have infinite length, hence, have $\omega = \infty$.  In
that case, the chronology relation on $\V$ has $p \ll
i^+$ for any $p \in V$ and $p \ll \Pa(b_c)$ for all  $p \in
I^-[\gamma_c]$ (where $\gamma_c(t) = (t,c(t))$\,); there are no
other chronology relations involving $\FB(V)$.  (Within
$\FB(V)$ there are causal relations, such as $\Pa(b_c)
\prec i^+$ and $\Pa(b_c) \prec a \cdot \Pa(b_c)$ for $a \ge
0$; there may also be other causal relations.)  Simple
example:  $M = \euc n$.  For any $x \in \euc n$ and any
unit vector $u$ in $\sph{n-1}$, we have the ray
$c^x_u(t) = x + tu$; then $b_{c^x_u}(y) = \<y-x,u\>$, where
$\<-,-\>$ denotes the Euclidean inner product; thus, this
function is just a shift of $\<-,u\>$.  As with any
Hadamard manifold, the rays in $\euc n$ account for all
finite Busemann functions.  Thus, aside from the
$\R$-action, $\Bu(\euc n)$ is parametrized by
$\sph{n-1}$.  This corresponds to the future causal
boundary of $\mk{n+1}$ being a null cone on $\sph{n-1}$, as is
also seen in the conformal embedding of $\mk{n+1}$ into the
Einstein static spacetime $\mk1 \times \sph{n}$ (see
\cite{HE} or \cite{H2}).

If $M$ is incomplete, then there are asymptotically
ray-like curves with $\omega$ finite.  In that case,
besides the same relations that occur when $M$ is
complete, for any $c \in \Cz(M)$ with $\omega < \infty$,
$\Pa(b_c) \ll i^+$ obtains, and some instances of
$\Pa(b_c) \ll p$ for $p \in V$ and of $\Pa(b_c) \ll
\Pa(b_{c'})$ will occur.  Simple example:  With $\bar M$ a
complete Riemannian manifold and $*$ any point in $\bar M$,
let $M = \bar M - \{*\}$.  Let $c: [\alpha, \omega] \to
\bar M$ be any finite-length curve with $c(\omega) = *$. 
Then $c$, restricted to $[\alpha, \omega)$, is in $\Cz(M)$
and corresponds to the point $(\omega, *)$ in $\bar V =
\mk1 \times \bar M$.  Thus, the future causal boundary of
$V$ is that of $\bar V$ together with the timelike line
$\mk1 \times \{*\}$.  

What about topology for the boundary?  The function space
$\Lp(M)$ has the compact-open topology; convergence in
that topology is the same as pointwise convergence.  We
can add $\infty$ (as a function) to this topological space
by saying that a sequence of functions converges to
$\infty$ if and only if the sequence converges pointwise to
infinity; this is equivalent to convergence to infinity on
any one point.  Aside from $i^+$, we can identify
$\FB(V)$ with $\Bu(M)$, a subset of $\Lp(M)$; $i^+$ we
can identify with $\infty$; and we can identify $V$ with
the functions $\Dtx(M) = \{d^t_x \st t \in \R, x \in M\}$, also
lying in $\Lp(M)$.  Then the function-space topology on $\V$
is that derived from these identifications, \ie, as $\Dtx(M) \cup
\Bu(M) \cup \ip$; the induced topology on $V$ is just the manifold
topology.  In this topology, the $\R$-action on $\V$ is
continuous; this action is degenerate on $i^+$, but the action is
free on $\V_0 = \V - \ip$ (let $\FB_0(V) = \FB(V)
-\ip$, so $\V_0 = V \cup \FB_0(V)$, corresponding to $\Dtx(M)
\cup \Bu(M)$).  

Let $\M = \V_0/\R$, with the quotient topology
(function-space topology on $\V_0$); then
$\pi : \V_0 \to \M$ is a line bundle.  Indeed, it is a
trivial line bundle:   If we choose any point $x \in
M$ and define the map $e: \Lp(M) \to \R$ as evaluation
at $x$, $e(f) = f(x)$, then $e$ is continuous in the
function-space topology.  Thus, thinking of $\V_0$ in its
guise as a subset of $\Lp(M)$, we get a continuous
cross-section $\zeta : \M \to \V_0$ of $\pi$ via
$\zeta([f]) = f - e(f)$ (where $[f]$ denotes the equivalance
class of $f$ under the real action) and a continuous
bijection $\phi : \R \times \M \to \V_0$ via $\phi(a,[f])
= a \cdot \zeta([f])$; as $\phi\inv(f) = (e(f),[f])$,
we see $\phi^{\inv}$ is evidently continuous, so this is a
homeomorphism.  If we restrict $\pi$ to $V$, we just have the
obvious projection $\pi : \mk1 \times M \to M$.  The upshot is that
$\M$ is $M$ with a sort of boundary attached, $\FB_0(V)/\R$, \ie,
$\Bu(M)/\R$; we can call this the {\it Busemann
boundary\/} of $M$, denoted $\BB(M)$.  (If $M$ is a
Hadamard manifold, then this is precisely the boundary
sphere of $M$, as explicated, for instance, in
\cite{BGS}.)  We thus have that, in the
function-space topology, $\FB_0(V) = \R \times \BB(M)$; if
$M$ is complete then each line $\R \times \{[b_c]\}$ is a null
line, and adding in $\ip$, we have that $\FB(V)$ is a null cone
on $\BB(M)$. 

But the function-space topology on $\V$ is not the one
that is important for understanding the chronological
structure of $V$ in the future; and for all that the
function-space topology leads to a simple
product-structure for $\FB_0(V)$, it may not be entirely
reflective of the physics of the spacetime.  In \cite{H2}
it was shown how to formulate a topology---what we might
call the future-chronological topology---on any
chronological set $X$.  It has a number of desirable
properties:  If $X$ is a strongly causal spacetime, this
results in the manifold topology; using this topology for the
future-completion $X^+$ of the spacetime, 
$\FB(X)$ is closed in $X^+$, $X$ is dense in $X^+$, and
the subspace topology on $X$ induced from that on $X^+$ is, again,
the manifold topology.  Future limits of future
chains are topological limits in this topology.  If $X =
(\mk n)^+$, this topology is the same as that coming from
the conformal embedding of
$\mk n$ into $\mk1 \times \sph{n-1}$.  In the topological
category of chronological sets with spacelike future
boundaries, using the future-chronological topology,
future-completion retains the functorial, natural, and
universal qualities it has in the category of
chronological sets.  These results are contained in
\cite{H2}, section 2 (section 5.1 for $\mk n$).

The future-chronological topology is defined by means of
a limit-operator on sequences:  For $X$ a chronological
set, for any sequence $\sigma$ of points
$\{\alpha_n\}$ in $X$, $\L(\sigma)$ is to be thought of as
first-order limits of $\sigma$.  More concretely, define a 
set $A$ to be closed if $\L(\sigma) \subset A$ for every
sequence $\sigma$ contained in $A$. So long as a
limit-operator $L$ obeys the property that for any
subsequence $\tau \subset \sigma$, $L(\tau) \supset
L(\sigma)$, this method defines a topology in which $L(\sigma)$
consists of limits of $\sigma$ (technically, these are the
first-order limits).  Let
$X$ be a chronological set in which the past of every point is
indecomposable (this includes both spacetimes and
future-completions of spacetimes); then the
future-chronological limit-operator in $X$ is defined
thus:  $\alpha \in \L(\{\alpha_n\})$ if and only if
\roster
\item for all $\beta \in I^-(\alpha)$, $\beta \ll \alpha_n$ for
$n$ sufficiently large, and
\item for any IP $P \supsetneq I^-(\alpha)$, $P$ contains
an element $\beta$ such that $\beta \not\ll \alpha_n$ for
$n$ sufficiently large.
\endroster
(If $X$ contains some $\alpha$ with $I^-(\alpha)$ 
decomposable, then a more complicated definition is
needed, but the idea is largely the same.)  

So long as $X$ is past-distinguishing
(\ie, $x \neq y$ implies $\Pa(x) \neq \Pa(y)$; this includes any
strongly causal spacetime or its future-completion), points are
closed in $X$; but
$X$ need not be Hausdorff.  Indeed, it is possible that
$\L(\sigma)$ may contain more than one point---and that represents
an important piece of physical information in $X$.  Conceivably,
$\L(\sigma)$ might contain an entire sequence $\tau$ of points, in
which case $\L(\tau)$ would be second-order limits of $\sigma$;
but no examples of such a chronological set have come to
light, and that possibility plays little role in things
(it just means that a very few of the proofs of how $\L$
works must invoke transfinite induction).

In this paper we are concerned only with the application
of the future-chrono\-logical topology to $\V$ for $V =
\mk1 \times M$.  As above, let us identify $\V_0$ with the 
appropriate subset of $\Lp(M)$, i.e, $\Dtx(M) \cup \Bu(M)$.  We can
also include $i^+$ in the discussion by treating that as the
function $\infty$; let us call this class of functions $\IP(M)$
(as they represent the IPs of $\mk1 \times M$), with $\IP_0(M) =
\Dtx(M) \cup \Bu(M)$ denoting the finite-valued ones.  Then we have
the following characterization of the future-chronological topology
in $\V$ from \cite{H3} (a derivation of this is indicated in
section 5):  

\definition{Future-Chronological Topology in $(\mk1 \times
M)^+$} Let
$\sigma =
\{f_n\}$ be a sequence of functions in $\IP(M)$.  Then
$\L(\sigma)$ consists of all $f \in \IP(M)$ satisfying
\roster
\item $f \le \displaystyle \liminf_{n\to\infty} f_n$\;\; and
\item for all $g \in \IP(M)$ with $f \le g \le \displaystyle
\limsup_{n\to\infty} f_n$,\;\; $g = f$.
\endroster
\enddefinition 

With this topology, we again have that the $\R$-action on
$\V$ is continuous, so we can still consider the
projection $\pi : \V_0 \to \M$; only now $\M = \V_0/\R$
must be imbued with the quotient topology from the
future-chronological topology on $\V_0$.  Let us call this
the chronological topology on $\M$.  (We could just as
easily have time-reversed everything, applying the
past-chronological topology---with an appropriate $\check
L$ operator---to $V^-$, the past-completion of $V$.  The
Busemann functions obtained thereby are just the negatives
of the ones for $\V$, and essentially the same structure is
obtained for the quotient $\M$.)

The future-chronological topology on $\V$ differs from the
function-space topology in that more sequences may be
convergent:  If $f \in \IP(M)$ is the pointwise limit of a
sequence  $\{f_n\}$ in $\IP(M)$, then $f \in \L(\{f_n\})$;
but the converse does not necessarily hold.  The
function-space topology is always Hausdorff, but the
future-chronological topology may not be; and this
difference passes through to the quotient $\M$ as well. 
When there is some $f \in \L(\{f_n\})$ which is not the
pointwise limit of the sequence, there must be some
$x \in M$ with $f(x) \neq \lim f_n(x)$, which means that
the $x$-based evaluation map $e: \V_0 \to \R$ is not
continuous; thus, $\V_0$, in the future-chronological
topology, may not be a product over $\M$ in the
chronological topology (contrary to what was said in
\cite{H3}).

These important features of the chronological topology
will be illustrated in our prime example space (Example 4
in \cite{H3}):

\example{The Unwrapped Grapefruit on a Stick} 

Let $G$ be a grapefruit on a stick: a sphere 
transfixed symmetrically by an infinite cylinder of
smaller radius.  For definiteness, take the sphere to be
the unit sphere at the origin of $\euc3$ and the cylinder
to be $\{x^2 + y^2 = r^2\}$ for some fixed positive $r <
1$, best imagined as small; the surface $G$ is that
portion of the sphere outside the cylinder combined with
that portion of the cylinder outside the sphere.  This is
not differentiable where the cylinder meets the sphere, at
$z = \pm r'$ (where $r' = \sqrt{1-r^2}$), so we must
introduce a smoothing, say between $z = r'$ and $z =
r'-\delta$ for some positive $\delta$ much smaller than
$r'$, and also between $z = -r'$ and $z = -(r'-\delta)$.

Let $M = \tilde G$, the universal cover of $G$.  We can
think of $M$ as parametrized by rectangular coordinate $z$
and polar-angle coordinate $\theta$; it is a plane which is
flat for $|z| > r'$ but with uniform positive curvature in
the region $|z| < r'$ (actually, the uniform positive
curvature is in $|z| < r' - \delta$, with some negative
curvature appearing where $r' - \delta < |z| < r'$; but
this has essentially no practical effect on the asymptotic
geometry of the space).  Specifically, the metric in the
flat part is $r^2\,d\theta\,^2 + dz\,^2$, while in the
round part it is $(1-z^2)\,d\theta\,^2 +
(1/(1-z^2))\,dz\,^2$ (with smoothing between those two in
the negative-curvature part).

Geodesics in $M$ are most easily seen by looking in $G$. 
A typical geodesic in the cylinder is, of course, a
helix; where it encounters the sphere, it turns into a
great circle and thus travels from one intersection of
sphere with cylinder to the other, upon which encounter
it again becomes a helix.  In $M$, this is a straight
line in the flat part, turning into a curve through the
round part that traverses no more than $\pi$ in the
$\theta$ coordinate before it again becomes a line in
the other flat part.  Clearly, such a geodesic will be
minimizing.  Other geodesics will not be minimizing, save
for short arcs: geodesics which are endlessly repeated
great circles in the sphere of $G$, corresponding to
periodically oscillating curves in the round part of $M$. 
So what is a minimizing geodesic between two points in the round
part of $M$, separated by a large amount of $\theta$ (say, much
greater than $\pi$)?  Ask the question in $G$:  How may
one travel efficiently between two points on the
grapefruit, while requiring that one make several
revolutions before stopping?  The clear answer is to
travel quickly off the grapefruit and wrap around the
stick the requisite number of times (thus accomplishing
the necessary increase in $\theta$) before returning
to the grapefruit to reach the endpoint.  Translated to
$M$, this is (approximately) a curve which goes from the
round part to the flat part, then holds at $z = \pm r'$
for some distance, then goes back into the round part.  In
actuality, the curve never reaches the flat part, but
stays for a long distance in the negative-curvature part;
but the flat and negative-curvature metrics hardly
differ at all in measuring the length of such a curve. 
(There are also geodesics which are the limits of those
minimizers between distant points in the round part, as one
endpoint is fixed and the other is pushed out to
infinity:  They start in the round part, enter the
negative-curvature region, and remain there, asymptotic to
the flat part.)  Consideration of these minimizing
geodesics allows calculation (at least in asymptotic form,
for large differences in $\theta$) of the distance
function in $M$.

It will be useful to examine the functions
$d^t_x$ for $x$ sitting in the round part; in
particular, we will want to consider the sequence
$\sigma$ in $V$ consisting of $\sigma(n) = 
(t_n, x_n)$ for $t_n = rn + (1-r)\pi/2$ and
$x_n = (n,z_0)$ in $\theta$-$z$ coordinates for fixed $z_0$
with $|z_0| < r'$ (actually, $|z_0| < r' - \delta$, but we
will henceforth ignore that minor correction).  It will be
shown that $\sigma$ converges to two points in $\FB_0(V)$,
using the future-chronological topology, but has no limit
in $\V$ using the function-space topology.  This
corresponds to the sequence $\{x_n\}$ in $M$ having two limits in
$\BB(M)$ using the chronological topology but having no limit in
$\M$ using the function-space topology.  The manner of
convergence will show that in the future-chronological topology,
$\V_0$ does not have a simple product structure over $\M$.

Rather than give the minutiae of the asymptotic
formulas for $d^{t_n}_{x_n}$, we will just display the
limit for $n$ going to infinity, which is all that we are
are interested in.  Define the function
$f(z) = \sin\inv(z/r') -
r\sin\inv\left(rz/(r'\sqrt{1-z^2})\right)$; this is a
strictly increasing function taking the interval
$[-r',r']$ onto  $[-(1-r)\pi/2,\,(1-r)\pi/2]$.  Let $\bar
f$ be the extension of $f$ to all of $\R$ defined by $\bar
f(z) = (1-r)\pi/2$ for $z > r'$ and $\bar f(z) =
-(1-r)\pi/2$ for $z < -r'$.  Then $\lim d^{t_n}_{x_n} =
\delta_{z_0}$ is given by
$$\delta_{z_0}(\theta,z) = r\theta + |\bar f(z) + f(z_0)|
- (1-r)\pi/2\,.$$   Our goal is to see how this compares
with elements of $\FB( V)$. 

The geodesics in $M$ come mostly in three classes: those
eventually in $z > r'$, those eventually in $z < -r'$, and
the oscillating ones remaining within $|z| < r'$.  (There
are also those geodesics asymptotic to the first two
classes, remaining eventually within the
negative-curvature region.)  The first two classes are all
in $\Cz(M)$ and are essentially straight lines in the flat
portions of $M$.  But those in the third class, being
non-minimizing over any interval longer than
$\pi$, do not go efficiently to infinity, and they all have
infinite Busemann functions; in other words, the null
geodesics in $V$ sitting over these oscillatory geodesics
in $M$ each have the entire spacetime $V$ for their pasts,
as they represent travellers who, while going out to
infinity, do so in a dilatory manner, so that they
eventually see all events.  (The geodesics asymptotic to
the first two classes have Busemann functions the same as
$z =$ constant geodesics in the respective classes.)  

The straight-line geodesics in $M$ have basically the
same Busemann functions as lines in the Euclidean plane;
the only difference for, say, the line $c_k$ of slope $k$ given by
$z = k\theta$ (or anything parallel to that) with $k$
positive and defined for $\theta > 0$, is that
$b_{c_k}(\theta,z)$ has an additional constant added to it
(in comparison with the Busemann function for that line in
the flat plane) for those $z < -r'$---representing the
additional distance it takes to get over the hump
of the grapefruit, compared to the flat plane---and is
rather complicated for those $z$ with $|z| < r'$.  What
is truly interesting is what happens for a line of zero slope:  This
includes geodesics in both of the two $\Cz(M)$
classes, yielding similar but subtly different
Busemann functions.  Let $c^+$ be the geodesic $c^+(t) =
((1/r)t,1)$ and $c^-$ its twin in the other class, $c^-(t)
= ((1/r)t,-1)$.  Then we have
$$\align
b_{c^+}(\theta,z) & = r\theta + \bar f(z) - (1-r)\pi/2 \\
b_{c^-}(\theta,z) & = r\theta - \bar f(z) - (1-r)\pi/2 \, .
\endalign$$
The curious aspect of these Busemann functions is that
although they are not $\R$-related to one another, i.e,
$[b_{c^+}] \neq [b_{c^-}]$, they remain a bounded distance
apart (as $\bar f$ is bounded); this is not so for any of
the other Busemann functions in the space, nor for those in
Euclidean space, nor, it appears, for most geodesics in
most spaces.  One implication of this is that the null
lines generated by $b_{c+}$ and $b_{c^-}$ in $\FB(V)$ have
a most unusual leap-frog kind of relationship:  If
$\alpha$ is any element of the first line and $\beta$ any
element of the second, then there is some $s_0 \in \R$ such
that $\alpha \prec s \cdot \beta$ for all $s \ge s_0$ and
$\alpha \succ s \cdot \beta$ for all $s \le s_0 -
(1-r)\pi$.

Now compare with $\delta_{z_0}$:  Note first that
$\delta_{z_0} = \max\{b_{c^+} + f(z_0), b_{c^-} -
f(z_0)\}$.  The import of this is that $\Pa(\delta_{z_0})$
is not an IP:  It is decomposable as $\Pa(f(z_0) \cdot
b_{c^+}) \cup \Pa(-f(z_0) \cdot b_{c^-})$.  This means
that $\delta_{z_0}$ is not in $\Bu(M)$; thus, the
sequence $\sigma$ has no limit in $V^+$ with the
function-space topology.  In similar vein, the sequence 
of points $\{x_n\} = \{(n,z_0)\}$ in $M$ has no limit in
$\M$ with the function-space topology; indeed, no
subsequence has a limit, either, so $\M$ is not compact
in this topology.

On the other hand, from $\lim_{n\to\infty} d^{t_n}_{x_n} =
\max\{b_{c^+} +  f(z_0), b_{c^-} - f(z_0)\}$ and the fact that no
other element of
$\FB(V)$ comes in between $\delta_{z_0}$ and, on the one hand,
$b_{c^+} + f(z_0)$, or, on the other hand, $b_{c^-} -
f(z_0)$, it follows that both of these functions are
future-chronological limits of the sequence $\{d^{t_n}_{x_n}\}$,
\ie, $\L(\sigma) = \{b_{c^+} + f(z_0), b_{c^-} -
f(z_0)\}$.  Thus, we have additional convergence in $\V$ 
in the future-chronological topology, but at a price:  It
is not Hausdorff.  (This price must always be paid, as
will be shown in section 5.)  We also have convergence in
$\M$ with the chronological topology: $\{(n,z_0)\}$
converges to both $[b_{c^+}]$ and $[b_{c^-}]$.  

The lack of Hausdorffness in $\V$ using the future-chronological
topology should not be seen as a defect:  It accurately reflects
the physics of the universe being modeled, in that there are 
observers with pasts that are not identical and with neither
containing the other, but are none the less closely
related.  In $\M$, the lack of Hausdorffness indicates that there
are pairs of asymptotically ray-like curves which go out to
infinity in nearly the same direction, but are subtly different in
the asymptotics of their respective distance functions.

A final issue:  Can $\pi : \V_0 \to \M$ be a trivial
bundle, \ie, can $\V_0$ be realized as $\R \times \M$ in
an $\R$-covariant manner?  If this were possible, then there would
be an $\R$-covariant map $\tau: \V_0 \to \R$, \ie, one such
that $\tau(a \cdot \alpha) = \tau(\alpha) + a$, continuous
with the future-chronological topology on $\V_0$.  But if
such a map existed, then we would have 
$$\lim_{n \to \infty} \tau(\sigma(n)) = \tau(b_{c^+} + f(z_0)) =
\tau(b_{c^+}) + f(z_0)$$
and also 
$$\lim_{n \to \infty} \tau(\sigma(n)) = \tau(b_{c^-} - f(z_0)) =
\tau(b_{c^-}) - f(z_0)$$
from which we derive $\tau(b_{c^-}) - \tau(b_{c^+}) =
2f(z_0)$.  But the left side is independent of $z_0$,
while the right side is not; thus, no such $\tau$ can
exist.  Thus, $\V_0$ is not a simple product over $\M$.  (This
appears generally to be the case when the two topologies differ,
though details are somewhat murky.)

But it is worth noting that $\V_0$, in the future-chronological
topology is, none the less, a line bundle over $\M$ in the
chronological topology, though it is not a trivial line bundle. 
(It is unclear to what extent this generalizes to other standard
static spacetimes.)  Showing the details of this will help
illuminate the nature of the future-chronological topology.

First some general points.   For local trivialization,
we need open sets in $\M$.  In the quotient topology,
an open set in $\M$ is the image under $\pi$ of an
$\R$-invariant open set in $\V_0$, and likewise for a
closed set.  Note that for any point $f \in \V_0$
(thought of as $\IP_0(M)$), the $\R$-orbit $\R \cdot
f$ is a closed set:  For any sequence of reals
$\{a_n\}$, if $g \in \IP_0(M)$ satisfies $g \le \liminf
(f+a_n)$ and if there is no other $g' \in \IP_0(M)$
with $g \le g' \le \limsup (f+a_n)$, then it must be
that $\{a_n\}$ has a limit $a$ and $g = f + a$.  For
let $a^- = \liminf a_n$ and $a^+ = \limsup a_n$.  We
have $g \le f + a^-$ (so we know $a^- \neq -\infty$,
if we are assuming $\L(\{f+a_n\}) \neq \emptyset$).  If
$a^+ > a^-$, then $g' = f + b$ violates the property of
$g$ for any $b$ with $a^- < b \le a^+$; thus, we know
that $a^+ = a^-$, so there is a limit $a$.  And $a$
must be finite, since if $a = \infty$, then $g' = g+1$
violates the property for $g$.  We have $g \le f+a$;
if $g \neq f+a$, then $f+a$ violates the property of
$g$ ($g \le f+a = \limsup (f+a_n)$).  Therefore, $g =
f+a$.  It follows that for any sequence $\sigma
\subset \R \cdot f$, $\L(\sigma)$ is either empty (the
sequence heads to $-\infty$) or consists of a single
element of the same $\R$-orbit: $\R \cdot f$ is closed
in $\V_0$ with the future-chronological topology.  (In
$\V$, the closure of $\R\cdot f$ is $(\R \cdot f) \cup
\ip$.)  It follows that $\{[f]\}$ is closed in $\M$
with the chronological topology,
\ie, points are closed in this topology.

Back to the unwrapped grapefruit on a stick, $M =
\tilde G$:  There are two pairs of troublesome
non-Hausdorff-related points, $\{[b_{c^+}],[b_{c^-}]\}$
and the analogous points for going in the opposite
direction, \ie, generated by the curves $-c^+(t) =
(-(1/r)t,1)$ and $-c^-(t) = (-(1/r)t,-1)$.  Remove one
from each pair to form the trivializing open sets: Let
$U^- = \M - \{[b_{c^+}],[b_{-c^+}]\}$ and let  $U^+ =
\M - \{[b_{c^-}],[b_{-c^-}]\}$; as points are closed,
these are open sets, and they clearly cover $\M$. 
Let $W^- = \pi\inv[U^-]$ and $W^+ = \pi\inv[U^+]$. 
Pick a point $x = (\theta,z) \in M$ and let $e_x : \V_0
= \IP_0(M) \to \R$ be evaluation at $x$; is $e_x$
continuous on either $W^-$ or $W^+$?  It depends on
the $z$ chosen for $x$. For the sequence
$\{d^{t_n}_{x_n}\}$ above (with $x_n = (n,z_0)$),
$\{e_x(d^{t_n}_{x_n})\}$ converges to $e_x(b_{c^+}
+f(z_0))$ precisely when $z \ge -z_0$, and it
converges to $e_x(b_{c^-}-f(z_0))$ when $z \le -z_0$.  The problem
for continuity is that $\{d^{t_n}_{x_n}\}$ has two limits in
$\V_0$, $b_{c^+} +f(z_0)$ and $b_{c^-} -f(z_0)$. In
$W^-$, we don't have to worry about convergence to
$b_{c^+} + f(z_0)$, as that is missing in $W^-$, so we are at
liberty to choose, say, $z = -1$; then $z < -z_0$ for all possible
choices of troublesome sequences $\{d^{t_n}_{x_n}\}$, so that
$e_x : W^- \to \R$ will be continuous.  This provides 
a cross-section $\zeta^-: U^- \to W^-$ of $\pi$
defined by $\zeta^-([f]) = f - e_x(f)$.  Similarly, a
choice of $x$ with $z = 1$ yields a cross-section
$\zeta^+ : U^+ \to W^+$.

In short:  The base space in the line bundle $\pi :
\V_0 \to \M$ is non-Hausdorff, but the bundle is
non-trivial because the non-Hausdorffness is not
reflected in the total space in an $\R$-invariant
manner.  Rather, for pairs $\{x^+,x^-\}$ in the base
space which are non-Hausdorff-related, for any element
$f^+$ in the fibre over $x^+$, only some of the
elements in the fibre over $x^-$ are
non-Hausdorff-related to $f^+$.  But away from
non-Hausdorff-related elements, the bundle is trivial.

\endexample

\head 3. Technical Results on Convergence \endhead

This section presents some preliminary work of a technical nature,
needed for the more interesting results on convergence later in
the paper.  As throughout the remainder of this paper, $M$ will be
an arbitrary Riemannian manifold, $V = \mk1 \times M$, and $\omega$
will commonly be used for the limiting argument of an endless
unit-speed curve (necessarily $\omega = \infty$ if $M$ is
complete).

Suppose that $c$ is a ray in $M$ (unit-speed, as all curves
considered here will be), \ie, for all $t \ge s \ge 0$,
$d(c(s), c(t)) = t-s$.  It then eaisly follows that for all $s \ge
0$, $b_c((c(s))) = s$.  If $c$ is asymptotically ray-like (\ie,
lies in $\Cz(M)$, \ie, has a finite Busemann function), then the
same thing is asymptotically true:

\proclaim{Lemma 3.1} For any $c \in \Cz(M)$, $\dlim_{s \to \omega}
(b_c(c(s)) - s) = 0$.
\endproclaim

\demo{Proof} It suffices if we show that $d(c(s),c(t))$ is
arbitrarily close to $t-s$ for $t \ge s$ and $s$ sufficiently
large. If this is not so, then for some $\epsilon > 0$, there is
a sequence $\{s_n\}$ increasing to $\omega$ with the propety that
for all $n$,
$$d(c(s_{2n-1}),c(s_{2n})) \le s_{2n} - s_{2n-1} - \epsilon.$$
It follows that 
$$d(c(s_0),c(s_{2n})) \le s_{2n} - s_0 - n\epsilon.$$
But then $b_c(c(s_0)) = \lim(s_{2n} - d(c(s_0),c(s_{2n}))) \ge \lim
(s_0 + n\epsilon) = \infty$, violating the asymptotic ray-like
nature of $c$. \qed \enddemo

We will want to compare the asymptotic behavior of the Busemann
function of a curve $c$, possibly in other directions, such as
another curve $c' \in \Cz(M)$, to other relevant functions:
the pointwise limits of functions in $\Dtx(M)$.  Let
$\bDtx(M)$ denote the closure of $\Dtx(M)$ in $\Lp(M)$ with the
function-space topology; this includes $\Bu(M)$, but may
(depending on $M$) also include functions representing past sets
that are not IPs but are closely related to IPs (such as the
functions $\delta_{z_0}$ for $M = \tilde G$, the example in
section 2).  

\definition{Definition} For any $c$ and $c'$ in $\Cz(M)$ and any
continuous $f: M \to \R$, $c$ is {\it $c'$-asymptotically like
$f$\/} if $\dlim_{s \to \omega'} (b_c(c'(s)) - f(c'(s))) = 0$
($\omega'$ the limiting argument for $c'$), \ie, $b_c -f$ goes
to 0 along $c'$.  \enddefinition

Generally, this will be used only for functions 
$f \in \bDtx(M)$, though the definition can apply more broadly.

Comparing $b_c$ with such a function $f$ is particularly simple
along the direction $c$ itself, as we just need to check that
$f$ fulfills the role of $b_c$ in Lemma 3.1; and when this
asymptotic relationship obtains, $b_c$ is actually bounded by $f$:

\proclaim{Lemma 3.2} For any $c \in \Cz(M)$ and $f \in \bDtx(M)$,
the following are equivalent:
\roster
\item $c$ is $c$-asymptotically like $f$; 
\item $\dlim_{s \to \omega}(f(c(s)) -s) = 0$;
\item for some sequence $\{s_k\}$ increasing to $\omega$,
$\dlim_{k \to \infty} (f(c(s_k)) - s_k) = 0$.
\endroster
Furthermore, when any of these obtain, $b_c \le f$ on $M$.
\endproclaim

\demo{Proof} In light of Lemma 3.1, all we need to show for the
first part is that (3) implies (2).  Let $f_c = f \circ c$, a
real-valued function defined on $[\alpha, \omega)$, the domain of
$c$.  Note that since $f \in \Lp(M)$ and $c$ is unit-speed, $f_c$ is
Lipschitz-1 also.  Let $\delta :[\alpha,\omega) \to \R$ be defined
by $ \delta(s) = f_c(s) - s$.  We know $\delta$ goes to 0 on the
sequence $\{s_k\}$, and we wish to show that $\delta$ goes to 0
generally.  Because $\{s_k\}$ goes out to $\omega$, this follows
directly from showing that $\delta$ is monotonic decreasing:

Suppose $t \ge s$.  Then $f_c(t) - f_c(s) \le t - s$ (since
$f_c$ is Lipschitz-1), so $f_c(t) \le f_c(s) + t - s = \delta(s) +
t$.  Therefore, $f_c(t) -t \le \delta(s)$, or $\delta(t) \le
\delta(s)$.  This finishes the first part.

For the second part, suppose that (2) holds and that $f$ is the
pointwise limit of a sequence $\{d^{t_n}_{x_n}\}$ in $\Dtx(M)$. 
Let $x$ be an arbitrary point in $M$.  We will show that for any
$\epsilon > 0$, for $s$ sufficiently large, $f(x) \ge s -
d(c(s),x) - \epsilon$; that will establish that $f(x) \ge b_c(x)$.

For $n$ sufficiently large (depending on $x$ and $\epsilon$),
$|f(x) - d^{t_n}_{x_n}(x)| < \frac{\epsilon}3$.  Therefore, we have, for
any $s < \omega$,
$$\align
f(x) & \ge d^{t_n}_{x_n}(x) - \frac{\epsilon}3 \\
& = t_n - d(x_n,x) - \frac{\epsilon}3 \\
& \ge t_n - d(x_n, c(s)) - d(c(s),x) - \frac{\epsilon}3 \\
& = d^{t_n}_{x_n}(c(s)) - d(c(s),x) - \frac{\epsilon}3 \\
& = d^{t_n}_{x_n}(c(s)) - f(c(s)) + f(c(s)) - s + s - d(c(s),x) -
\frac{\epsilon}3. \tag{3.1}
\endalign$$
From our hypothesis of (2), if we choose $s$ 
sufficiently large, then $|f(c(s)) - s| < \frac{\epsilon}3$.  For any
fixed such $s$, we can require that $d^{t_n}_{x_n}$ be close to
$f$ at both $x$ and $c(s)$, \ie, that for $n$ sufficiently large,
we also have, $|d^{t_n}_{x_n}(c(s)) - f(c(s))| < \frac{\epsilon}3$. 
Putting these inequalities into (3.1) results in 
$$f(x) \ge s-d(c(s),x) - \epsilon$$
as desired. \qed   
\enddemo

In Lemma 3.2, the functions $f$ we considered for comparison with
Busemann functions were specified as those in the
$\Lp(M)$-closure of $\Dtx(M)$, i.e., the function-space closure of
$V$.  Of course, that also includes anything in the closure of
$V^+_0$---i.e., $\Dtx(M) \cup \Bu(M)$---as the elements of $\Bu(M)$
are part of the closure of $\Dtx(M)$ in $\Lp(M)$; that is to say,
any function expressible as a limit of elements of $\Dtx(M) \cup
\Bu(M)$ is also expressible as a limit of elements purely of
$\Dtx(M)$.  As a matter of practicality, it is worth noting just
how this can be implemented; this is the content of the next lemma.

Standard techniques, such as covering a
compact set with open balls, easily show that among Lipschitz-1
functions, uniform convergence on compact subsets is equivalent to
pointwise convergence (and uniform convergence on compact subsets
is equivalent to convergence in the compact-open topology, even for
merely continuous functions).  This is useful for showing the
following.

\proclaim{Lemma 3.3} Let $\{c_n\}$ be  a sequence of
curves in $\Cz(M)$.  Then there is a sequence of numbers
$\{s_n\}$ such that $\{b_{c_n}\}$ converges pointwise to
some function $f$ if and only if $\{d^{s_n}_{c_n(s_n)}\}$
converges pointwise to $f$.  Furthermore, the same is
true with respect to the future-chronological topology,
\ie, $\L(\{b_{c_n}\}) = \L(\{d^{s_n}_{c_n(s_n)}\})$.
\endproclaim

\demo{Proof} Let $\{K_n\}$ be an exhaustion of  $M$ by
compact subsets, $K_n \subset \text{interior}(K_{n+1})$. 
Since each $b_{c_n}$ is the pointwise limit of
$\{d^s_{c_n(s)}\}$ as $s$ goes to $\omega_n$ (the
limiting argument for $c_n$), we also have uniform
convergence on $K_n$:  For each $n$, there is some
$s_n < \omega_n$ such that $|b_{c_n} - d^{s_n}_{c_n(s_n)}| < 1/n$
on $K_n$.  (The same is true with $s_n$ replaced by any
$s$ with $s_n \le s < \omega_n$, but we have no particular need
for that.)  Let $\delta_n = d^{s_n}_{c_n(s_n)}$.

Suppose that the sequence $\{b_{c_n}\}$ converges
pointwise to $f$.  Then by uniform convergence on each
$K_k$, we have for any $\epsilon > 0$, there is some
$N^k_\epsilon$ such that for any $n \ge N^k_\epsilon$, 
$$|f - b_{c_n}| < \frac{\epsilon}2 \quad \text{ on } K_k.$$
We also have for all $n > \frac2{\epsilon}$, 
$$|b_{c_n} - \delta_n| < \frac{\epsilon}2 \quad 
\text{ on }K_n.$$
Therefore, for all $n> \max\{N^k_\epsilon,
\frac2{\epsilon}, k\}$,
$$|f - \delta_n| < \epsilon \quad \text{ on }
K_k.$$   
This shows $\{\delta_n\}$ converges to
$f$ uniformly on compact subsets, so the convergence is
also pointwise.

Suppose $\{\delta_n\}$ converges pointwise to
$f$.  Then we proceed similarly:  For each $k$, for every
$\epsilon > 0$, there is some $M^k_\epsilon$ so that for
every $n \ge M^k_\epsilon$, $|f - \delta_n| <
\frac{\epsilon}2$ on $K_k$.  Then for
$n > \max\{M^k_\epsilon, \frac2{\epsilon}, k\}$, 
$|f-b_{c_n}| < \epsilon$ on $K_k$, and $\{b_{c_n}\}$
converges to $f$.

For future-chronological limits, it works similarly:

Suppose $f \in \L(\{b_{c_n}\})$; that means $f \le
\liminf_{n \to \infty} b_{c_n}$ and $f$ is the only
function $g \in \IP(M)$ such that $f \le g \le 
\limsup_{n \to \infty} b_{c_n}$.  We use the fact
that for any $x  \in M$, for some integer $I_x$, for
all $n \ge I_x$,  $x \in K_n$.   

We know that for all $x \in M$, for all  $\epsilon >
0$, for some $J_\epsilon^x$, for all $n \ge
J_\epsilon^x$, 
$$f(x) \le b_{c_n}(x) + \frac\epsilon2.$$  
We also know for $n \ge \max\{I_x,\frac2\epsilon\}$,
$$|b_{c_n}(x) - \delta_n(x)|< \frac\epsilon2.$$
Then for all $n >
\max\{I_x,J_\epsilon^x,\frac2\epsilon\}$,
$f(x) \le \delta_n(x) + \epsilon$; thus, $f \le
\liminf \delta_n$.  

For any $g \le \limsup \delta_n$, we know that for
all $x \in M$, for all $\epsilon > 0$, there is a
subsequence $\sigma$ of the integers such that for
all $k$, 
$$g(x) \le\delta_{\sigma(k)}(x) + \frac\epsilon2.$$  
Then for $k$ so large that $\sigma(k) \ge \max\{I_x,
\frac2\epsilon\}$, 
$$|b_{c_{\sigma(k)}}(x) - \delta_{\sigma(k)}(x)| <
\frac\epsilon2,$$ 
so $g(x) \le b_{c_{\sigma(k)}}(x) + \epsilon$; thus,
$g \le \limsup b_{c_n}$.  It follows that if $f \le g
\le \limsup \delta_n$, then $f \le g \le
\limsup b_{c_n}$, so $g = f$; thus, $f \in
\L(\{\delta_n\})$.

Suppose $f \in \L(\{\delta_n\})$.  Then we proceed 
exactly the same way, just exchanging the roles of
$\delta_n$ and $b_{c_n}$, and ending with $f \in
\L(\{b_{c_n}\})$. \qed
\enddemo

\head 4. Asymptotically Ray-like Curves and Rays \endhead

In a Hadamard manifold, the only Busemann functions are those
which are generated by rays; but that may not be the case in a
general manifold.  An example occurs in a manifold similar to
that of the unwrapped grapefruit on a stick of section 2: 
Instead of just one grapefruit, have an infinite number of
them.  In other words:  $M$ has alternating strips of zero and
uniform positive curvature (separated by minuscule regions of
negative curvature).  If the $k$th grapefruit, of radius 1, is
centered at $z = 4k$, then the $k$th flat region is given by
$4k+1 < z < 4k+3$.  Let
$c_k$ be a ray in the $k$th flat region, $c_k(t) = ((1/r)t,
4k+2)$ (where, as before, $r$ is the radius of the stick); then
$b_{c_k}$ has a simple formula for those points
$(\theta,z)$ lying in the flat regions.  With $\kappa(z)$ denoting
which flat region $(\theta,z)$ lies in (\ie, $4\kappa+1 < z <
4\kappa+3$), we have
$$b_{c_k}(\theta,z) = r\theta - (1-r)|k-\kappa(z)|\pi\,.$$
 For points in the round
regions, the formula for this Busemann function involves the
function $f$ from the example in section 2; but we can get a good
picture of what's transpiring here without engaging that level of
detail.  The important point to note here is that for fixed
$\theta$, the $k$th Busemann function peaks for points in the
$k$th flat region, and that is the only way in which these
functions differ from one another.  Thus, $b_{c_{k+1}} + (1-r)\pi$
is identical with $b_{c_k}$ for those points with
$\kappa \le k$ and is greater than it for the remainder.  In other
words, with proper adjusting by additive constants---\ie, using
$\bar c_k = c^{(1-r)k\pi}_k$ in place of
$c_k$, thus generating the Busemann function $b_{\bar c_k} =
b_{c_k} + (1-r)k\pi$---the respective IPs generated by these rays
are sandwiched together very much like the IPs in
$\mk2$ given by
$P_k = I^-(k,k)$.  And just as
$\bigcup_{k=0}^\infty P_k$ is an IP in $\mk2$, the union of the
properly sandwiched IPs generated by $\{\bar c_k\}$ in this space
is another IP.  This is not obvious at first blush; it's easy to
calculate what the limiting function $b_\infty = \dlim_{k \to
\infty} b_{\bar c_k}$ is:
$$b_\infty(\theta,z) = r\theta + (1-r)\kappa(z)\pi\,;$$
but it is not immediately clear that this is a Busemann function,
\ie, that $\Pa(b_\infty)$ is an IP.  It's
clearly not the Busemann function of any geodesic, as those are
all known: the functions $b_{c_k}$ above and the functions coming
from geodesics pitched at an angle to the horizontal (in the flat
regions); the geodesics remaining within a round section have
infinite Busemann fucnctions.  But is there any curve
$c$ with $b_c = b_\infty$?

We will form a broken geodesic $c$ by taking minimal geodesic arcs
between points $\{x_n\}$, where $x_n$ lies in the $n$th flat
region: $x_n = (\theta_n,4n+2)$.  Then we must have $c(t_n) =
x_n$ where $t_n = \sum_{i=1}^n d(x_{i-1},x_i)$.  This yields
$$\align
b_c(\theta,z) = \kappa(z)\pi & + \lim_{n\to\infty}
\biggl(\sum_{i=1}^n\sqrt{r^2(\theta_i-\theta_{i-1}-\pi)^2 + 4} \\
& -  \sqrt{r^2(\theta_n-n\pi-(\theta-k\pi))^2 +
(2n-z+z_0+2(k-k_0))^2}\biggr).
\endalign$$
This is not necessarily finite; but if we have $\{\theta_n\}$
increase sufficiently quickly, it will be.  Specifically, we need
$\dlim_{n \to \infty} \theta_n/n = \infty$ and $\sum_{i=1}^\infty
1/(\theta_i - \theta_{i-1} - \pi) < \infty$ (example: $\theta_n =
n^3$).  With that provision, we have
$$b_c(\theta,z) = r(\theta-\theta_0) + (1-r)\kappa(z)\pi + C,$$
where $C$ is a positive constant bounded above by
$\sum_{i=1}^\infty 2/(r(\theta_i -\theta_{i-1} -\pi))$.  (Reaching
this simplification requires use of the identity
$\sqrt{a^2 + b} =  a + \dfrac{b}{\sqrt{a^2+b}+a}$.)  Thus,
$b_\infty$ is in the same equivalence class as the Busemann
function for $c$.

This section is devoted to exploring just how the Busemann
function of an arbitrary asymptotically ray-like curve $c$---which
may not be the Busemann function of any geodesic---is related to
the Busemann functions of geodesics, specifically, of rays related
to $c$.  It is applicable to complete manifolds (though results
can be obtained for incomplete manifolds which are appropriately
convex).

Our first result says that although the Busemann function from a
curve $c \in \Cz(M)$ may not come from a geodesic, there is always
a geodesic $\gamma$ whose Busemann function is in some sense an
approximation of that of $c$, in that $b_\gamma$ is bounded above
by $b_c$ and the two functions are equal along $\gamma$.  What is
particularly interesting is that if the two are also equal along
$c$ then they are equal everywhere.

\proclaim{Proposition 4.1} Suppose $M$ is complete.  Let $c$ be
any asymptotically ray-like curve and
$x_0$ any point in $M$.  Then there is an endless unit-speed ray
$\gamma$ in $M$, starting at $x_0$, such that 
\roster
\item $b_\gamma$ and $b_c$ are equal on $\gamma$, and
\item $b_\gamma  \le b_c$ on $M$.
\endroster
Furthermore,  for any sequence $\{s_n\}$ of numbers approaching
$\infty$,
\roster
\item[3] if $\dlim_{n \to\infty}(b_\gamma(c(s_n)) - b_c(c(s_n))) =
0$ (\ie, $b_\gamma - b_c$ goes to 0 along a cofinal sequence on
$c$), then $b_\gamma = b_c$ on all of
$M$. 
\endroster
\endproclaim

\demo{Proof} For each $n$, let $\gamma_n$ be a unit-speed
minimal geodesic segment from $x_0$ to $c(n)$.  We will parametrize
$\gamma_n$ on $[\alpha, T_n]$, where $\alpha = b_c(x_0)$, so that
$T_n - \alpha = d(x_0,c(n))$, or 
$$b_c(x_0) = T_n - d(x_0,c(n)).$$
We also know
$$\align 
b_c(x_0) & = \lim_{n \to \infty}(n-d(x_0,c(n))) \\
& = \lim_{n \to \infty}(n-T_n + b_c(x_0))
\endalign$$
whence
$$0 = \lim_{n \to \infty}(n-T_n).$$
We thus know that $\{T_n\}$ goes out to infinity.

The unit vectors $\{\dot\gamma_n(\alpha)\}$ (all at the point
$x_0$) have a vector $v$ which is a limit-point of that sequence,
and we will consider the sequence $\{\gamma_n\}$ replaced by a
subsequence with $\{\dot\gamma_n(\alpha)\}$ converging to $v$.
Let $\gamma: [\alpha, \infty) \to M$ be the geodesic with
$\dot\gamma(\alpha) = v$.  Since every arc of $\gamma$ (say, on
$[\alpha, s]$) is the limit of arcs in $\{\gamma_n\}$ (with $T_n >
s$), $\gamma$ is a ray.

Fix any number $s > b_c(x_0)$.  We will show $b_c(\gamma(s)) = s$;
since $\gamma$ is a ray, $b_\gamma(\gamma(s)) = s$, thus proving
(1).

Consider the triangle of points $\gamma_n(s)$, $\gamma(s)$, and
$c(n)$; note that the distance between the first two, $\delta^s_n =
d(\gamma_n(s), \gamma(s))$, goes to zero as $n$ goes to infinity
(since $\{\gamma_n\}$ approaches $\gamma$ on compact subsets), and
the distance between the first and third is
$d(\gamma_n(s), c(n)) = T_n-s$ (since $c(n) = \gamma_n(T_n)$). 
From the triangle inequality we have 
$|T_n - s - d(\gamma(s),c(n))| \le \delta^s_n$.
This gives us $\lim_{n \to \infty}(T_n - s - d(\gamma(s),c(n))) =
0$, or $\lim_{n \to \infty}(T_n - d(\gamma(s),c(n))) = s$.  Since,
as shown previously, $\lim_{n \to \infty}(T_n - n) = 0$, this
establishes
$\lim_{n \to \infty}(n - d(\gamma(s),c(n))) = s$ or
$b_c(\gamma(s)) = s$, as desired.

Thus we have shown that $b_\gamma = b_c$ along $\gamma$, i.e,
that (1) is true.  In particular, we have that $\gamma$ is
$\gamma$-asymptotically like $b_c$.  By Lemma 3.2 (using $f =
b_c$) we conclude that $b_\gamma \le b_c$ on $M$, \ie, (2) is
true.  

Suppose now that $b_\gamma - b_c$ goes to zero along a cofinal
sequence $\{s_k\}$ on $c$; since, by Lemma 3.1, $b_c(c(s)) -s$
goes to zero as $s$ goes to infinity, this implies
$b_\gamma(c(s_k)) - s_k$ goes to zero.  Again employing Lemma
3.2 (this time with $f = b_\gamma$), we conclude $b_c \le
b_\gamma$ on $M$.  It follows that $b_\gamma = b_c$ on $M$,
\ie, (3) is true. \qed
\enddemo

Although the Busemann function for a curve $c$ may not be the
Busemann function for any geodesic, we can none the less
approximate it by the Busemann functions of geodesics, in that
it is the pointwise limit of such functions.

\proclaim{Proposition 4.2} Suppose $M$ is complete.  Let $c$ be
any asymptotically ray-like curve in $M$.  Then
there is a sequence of rays $\{\gamma_n\}$ such that $b_c =
\lim_{n\to\infty}b_{\gamma_n}$ (pointwise limit, hence, uniform
convergence on compact subsets).  Also, for all $n$,
$b_{\gamma_n} \le b_c$; thus, $b_c =
\sup_n b_{\gamma_n}$.
\endproclaim

\demo{Proof}For each $n$, apply Proposition 4.1 to the curve $c$
and the point $c(n)$ to obtain the ray $\gamma_n$.  We have
$b_{\gamma_n} \le b_c$ on $M$ and $b_{\gamma_n} = b_c$ along
$\gamma_n$.  In particular, since $\gamma_n$ starts at $c(n)$,
we have
$$b_{\gamma_n}(c(n)) = b_c(c(n)).$$

Consider an arbitrary point $x \in M$.  Since $b_{\gamma_n} \in
\Lp(M)$, we have
$$\align
b_{\gamma_n}(x) & \ge b_{\gamma_n}(c(n)) - d(c(n),x) \\
 & = b_c(c(n)) - d(c(n),x). \tag{4.1}
\endalign$$

Note that for all $s$, $b_c(c(s)) \ge s$: We know
$b_c(c(s)) = \lim_{t\to\infty}(t - d(c(t),c(s)))$, and $t\mapsto t
- d(c(t),c(s))$ is an increasing function.  Thus, for all
$t$, $b_c(c(s)) \ge  t - d(c(t),c(s))) \ge t - |t-s|$; choosing
$t \ge s$, we get $b_c(c(s)) \ge s$.  In particular, we have
$b_c(c(n)) \ge n$.  Thus, from (4.1) plus $b_c \ge
b_{\gamma_n}$ we have
$$b_c(x) \ge b_{\gamma_n}(x) \ge n - d(c(n),x).$$
Since $b_c(x) = \lim_{n\to\infty}(n - d(c(n),x))$, we end up
with 
$$b_c(x) =  \lim_{n\to\infty}b_{\gamma_n}(x).$$
\qed
\enddemo

\head 5. The Limit-Operator $\L$ and Quasi-Compactness 
\endhead

In this section we will present a thorough analysis of the
limit-operator $\L$ defining the future-chronological topology 
in $\V$.  This will include what are likely the most important
results of this paper: that the function-space and
future-chronological topologies coincide if and only if the
latter is Hausdorff, and that $\M$ is always compact in the
chronological topology.  In fact, the future-chronological
topology induces a quasi-compactness on any future-complete
chronological set, translating into a very useful compactness
feature for spacetimes.

We will begin with a way of characterizing $\L$ in a
general chronological set $X$.  Much of what we prove in this
section is not restricted to standard static spacetimes, or 
even to spacetimes generally, but is applicable more widely,
such as to spacetimes with boundaries.  As in section 2, for
simplicity of presentation we will restrict our attention to
those chronological sets which are {\it past-regular\/}, \ie,
in which the past of every point is indecomposable; as
previously mentioned, this includes both spacetimes and
future-completions of spacetimes, thus everything considered in
this paper.

Let $\{A_n\}$ be a sequence of subsets in a set $X$.  We
will define $\LI(\{A_n\})$  (meaning
$\displaystyle\liminf_{n\to\infty}\{A_n\}$) and
$\LS(\{A_n\})$ (meaning
$\displaystyle\limsup_{n\to\infty}\{A_n\}$) as follows:
$$\align
\LI(\{A_n\}) & = \bigcup_{n=1}^\infty\bigcap_{k = n}^\infty A_k, \\ 
\LS(\{A_n\}) & = \bigcap_{n=1}^\infty\bigcup_{k = n}^\infty A_k.
\endalign$$
It is evident that these constructions can also be
characterized as 
$$\align
\LI(\{A_n\}) & = \{x \in X \st x \in A_n \text{ for all } n 
\text{ sufficiently large}\}, \\ 
\LS(\{A_n\}) & = \{x \in X \st x \in A_n
\text{ for infinitely many }n\}.
\endalign$$

These constructions make for an easy way to express the
future-chronological limit-operator $\L$ in a chronological set:

\proclaim{Proposition 5.1} Let $X$ be a past-regular 
chronological set, and let $\sigma = \{\alpha_n\}$ be any
sequence in $X$.  Then for any $\alpha \in X$,  
$\alpha
\in \L(\sigma)$ if and only if
\roster
\item $I^-(\alpha) \subset \LI(\{I^-(\alpha_n)\})$ and
\item $I^-(\alpha)$ is a maximal IP in $\LS(\{I^-(\alpha_n)\})$.
\endroster
In particular, if $\LI(\{I^-(\alpha_n)\}) = 
\LS(\{I^-(\alpha_n)\}) = R$, then $\alpha \in \L(\sigma)$ if and
only if $I^-(\alpha)$ is a maximal IP in $R$.

In all the above, one may replace $\LI(\{I^-(\alpha_n)\})$
and $\LS(\{I^-(\alpha_n)\})$ by, respectively,
$I^-[\LI(\{I^-(\alpha_n)\})]$ and
$I^-[\LS(\{I^-(\alpha_n)\})]$.
\endproclaim

\demo{Proof} Conditions (1) and (2) above are just
restatements of the two conditions given in the definition
in section 2:  Condition (1) in section 2 amounts to
saying that for all
$\beta \in I^-(\alpha)$, $\beta
\in I^-(\alpha_n)$ for $n$ sufficiently large.  Condition 
(2) in section 2 says that if $P$ is any IP properly
containing $I^-(\alpha)$, then $P$ contains a point $\beta$
with $\beta \not\in I^-(\alpha_n)$ for $n$ sufficiently
large; that is the same as saying that $P$ is not contained
in $\LS(\{\alpha_n\})$.

For any set $A$ and any past set $P$, $P \subset A$ if and
only if $P \subset I^-[A]$; this allows the replacement in
the last statement of the proposition.
\qed \enddemo

Note that $\LI(\{I^-(\alpha_n)\})$ and
$\LS(\{I^-(\alpha_n)\})$ generally are not past sets.  For
instance, in $\mk2$, for each $n$, let $\alpha_n$ be the point
with $t = 0$ and $x = 1/n$; then $\LI(\{I^-(\alpha_n)\}) =
\LS(\{I^-(\alpha_n)\}) = \{(t,x)
\st t<-|x| \text{ or } t = -x < 0\}$: a past set plus some
of its boundary.

On the other hand, the union of a future chain of IPs is always an
IP (this will be important for an application of the proposition
above):

\proclaim{Lemma 5.2} Let $\{P_\iota \,|\, \iota \in I\}$ be a
collection of IPs indexed by a well-ordered set $I$ such
that for all $\iota$ and $\kappa$ in $I$, $\iota \le \kappa$
implies $P_\iota \subset P_\kappa$.  Then $\bigcup_\iota 
P_\iota$ is an IP.
\endproclaim

\demo{Proof} Let $P = \bigcup_\iota P_\iota$.  It is easy to
show that $P$ is a past set (\ie, that $I^-[P] = P$).  Then to
show $P$ is an IP, we can use the alternate characterization of
IPs from Theorem 2 of \cite{H1}: A past set $A$ is an IP if and
only if for all $\alpha$ and $\beta$ in $A$, there is some
$\gamma \in A$ with $\alpha \ll \gamma$ and $\beta \ll \gamma$. 
This is easily applied here:  If $\alpha$ and $\beta$ are in
$P$, then $\alpha \in P_\iota$ and $\beta \in P_\kappa$ for some
$\iota$ and $\kappa$; either $\iota \le \kappa$ or $\kappa \le
\iota$, and this characterization does the rest. \qed 
\enddemo

As a direct consequence of Proposition 5.1, we can
characterize a class of sequences which always have 
future-chronological limits. 

\proclaim{Proposition 5.3} Let $X$ be a past-regular,
future-complete chronological set and let $\sigma =
\{\alpha_n\}$ be a sequence in $X$.  If
$\LI(\{I^-(\alpha_n)\}) = \LS(\{I^-(\alpha_n)\}) = R$ is
non-empty, then $\L(\sigma) \neq \emptyset$. 
(Alternatively:  If $I^-[\LI(\{I^-(\alpha_n)\})] =
I^-[\LS(\{I^-(\alpha_n)\})]$ is non-empty, then the same.)

Furthermore, for any $\alpha \in R$,
$\L(\sigma)$ contains an element $\beta$ with
$I^-(\beta) \supset I^-(\alpha)$.
\endproclaim

\demo{Proof} There is some $\alpha_0 \in R$.  Clearly, for any
$\alpha \ll \alpha_0$, $\alpha \in R$ also, \ie, $I^-(\alpha_0)$
is an IP contained in $R$.  Let $\Cal P$ be the set of all IPs
contained in $R$, and consider $\Cal P$ as a partially ordered
set under inclusion.  We know $\Cal P$ is non-empty.  If
$\{P_\iota \st \iota \in I\}$ is any chain in this poset
($I$ a well-ordered index set with $P_\iota \subset
P_\kappa$ for $\iota \le \kappa$), then by Lemma 5.2,
$Q = \bigcup_\iota P_\iota$ is an IP; therefore, $Q \in \Cal
P$.  Thus, every chain in  $\Cal P$ has an upper bound; by
Zorn's lemma there is a maximal IP $P_\infty \in \Cal P$.  

Let $c$ be a future chain generating $P_\infty$.  Since
$X$ is future-complete, $c$ has a future-limit
$\beta$; then $P_\infty = I^-(\beta)$. By
Proposition 5.1,
$\beta \in \L(\sigma)$.  

For any $\alpha \in R$ we can restrict attention to the poset
$\Cal P_\alpha$ consisting of all IPs in $R$ which contain
$I^-(\alpha)$.  We can thus assure that $P_\infty$ contains
$I^-(\alpha)$. \qed
\enddemo

We can thus specify a class of sequences in a spacetime which 
have a future-chronological limit (if not in the spacetime itself,
then in its future-completion):

\proclaim{Corollary 5.4} Let $U$ be a strongly causal 
spacetime and let $\sigma = \{p_n\}$ be a sequence of
events in $U$.   Suppose $\sigma$ has these two
properties: 
\roster
\item There is some $q_0$ in the common past of infinitely many
$p_n$.
\item If $q \ll p_n$
for infinitely many $n$, then $q \ll p_n$ for all $n$ 
sufficiently large.
\endroster
Then $\sigma$ has a future-chronological limit $\alpha$ in
$U^+$ with $I^-(\alpha) \supset I^-(q_0)$.  
\endproclaim

\demo{Proof}  Apply Proposition 5.3 to $U^+$. \qed \enddemo

Let us apply Propositions 5.1 and 5.3 to our standard
static spacetime $V = \mk1 \times M$.  As in section 2,
we will identify $\V$ with $\IP(M)$ (with $V$ represented
by $\Dtx(M)$). 

\proclaim{Proposition 5.5} Let $V = \mk1 \times M$ be a
conformal equivalent of a standard static spacetime.  Suppose
$\{f_n\}$ is a sequence of elements in $\IP(M)$ (representing
elements of $\V$) such that $\phi = \lim_n f_n$ exists as a
pointwise limit and $\phi \neq -\infty$ (though $\phi = \infty$ is
allowed).  Then $\{f_n\}$ has a future-chronological
limit in $\V$.  In particular, for any $x \in M$
and $t < \phi(x)$, $\L(\{f_n\})$ contains an element
$f_\infty \ge d^t_x$. 

Furthermore, for any $f \in \IP(M)$, $f \in \L(\{f_n\})$
if and only if
\roster
\item $f \le \phi$ and
\item for any $g \in \IP(M)$, $f \le g \le \phi$ implies
$g = f$.
\endroster
In particular, if $\phi \in \IP(M)$, then that is the only
future-chronological limit of the sequence: $\hat L(\{f_n\}) =
\{\phi\}$. 
\endproclaim

\demo{Proof} With $\{f_n\}$ converging pointwise to 
$\phi$, it's easy to show $\Pa(\phi) \subset
\LI(\{\Pa(f_n)\})$ and $\LS(\{\Pa(f_n)\}) \subset
\text{closure}(\Pa(\phi))$; therefore,
$\LI(\{\Pa(f_n)\})$ and $\LS(\{\Pa(f_n)\})$ differ, at
most, by some part of $\partial(\Pa(\phi))$, \ie, points
of the form $(\phi(x), x)$.  In any case, the two have
the same interior, hence, the same past, so Proposition
5.3 applies with the alternative condition.

Since $\phi \neq -\infty$, $\LI(\{\Pa(f_n)\})$ is non-empty
(for any $x \in M$ and $t < \phi(x)$, $(t,x) \in
\LI(\{\Pa(f_n)\})$).  Therefore, we can apply Proposition
5.3 and obtain a future-chronological limit; in
particular, for any $t < \phi(x)$, since $(t,x) \in
\LI(\{\Pa(f_n)\})$, there is a future-chronological limit
$f_\infty$ with $\Pa(f_\infty) \supset I^-((t,x))$, \ie,
$f_\infty \ge d^t_x$.  Proposition 5.1 yields the
remainder. 

In case $\phi \in \IP(M)$, then $\phi$ clearly satisfies the
conditions on $f$ in (1) and (2), so $\phi \in
\hat L(\{f_n\})$.  Furthermore, if $f$ satisfies (1) and (2),
then letting $g = \phi$ shows that $f = \phi$. \qed
\enddemo

It is important to realize that even though a sequence of
functions $\{f_n\}$ in $\IP(M)$ has a pointwise limit 
$\phi$, $\phi$ itself may not represent any point in
$V^+$; this was the case for sequences of points in the 
unwrapped grapefruit of section 2.  In fact, when this
happens, $V^+$ is necessarily non-Hausdorff in the
future-chronological topology:

\proclaim{Proposition 5.6}  Suppose $\{f_n\}$ is a 
sequence of elements in $\IP(M)$ with a pointwise limit
$\phi$.  If $\L(\{f_n\})$ has an element $f \neq \phi$,
then it has at least two elements. \endproclaim

\demo{Proof}  We know $\phi \neq - \infty$, for otherwise
$\LS(\{\Pa(f_n)\})$ would be empty, but $f \in \L(\{f_n\})$. 
Using Proposition 5.5, we know $f \le \phi$.  With $f \neq
\phi$, there is some $x_0 \in M$  and $t_0 \in \R$ with $f(x_0)
< t_0 < \phi(x_0  )$.  Note that $d^{t_0}_{x_0} < \phi$:  For
any $x$, $d^{t_0}_{x_0}(x) = t_0 - d(x,x_0) < \phi(x_0) -
d(x,x_0) \le \phi(x)$ (since $\phi$ is Lipschitz-1).

Let us define 
$$F = \{g \in \IP(M) \st g \le \phi \text{ and } g(x_0) 
\ge t_0\},$$ 
considered as a partially ordered set under $\le$ (as
applied to functions).  We have $d^{t_0}_{x_0} \in F$, so
it is non-empty.  For any chain $\{g_\iota\}$ in $F$, $g
= \sup_\iota g_\iota$ is in $F$ (as $\Pa(g) =
\bigcup_\iota \Pa(g_\iota)$, Lemma 5.2 shows $g \in \IP(M)$),
and it is clearly an upper bound for the chain.  Therefore, by
Zorn's lemma, $F$ contains a maximal function $g_\infty$.  For
any $h \in \IP(M)$ with $g_\infty \le h \le \phi$, we have
$h \in F$; therefore, $g_\infty$ being maximal, we must
have $h = g_\infty$.  Thus, by Proposition 5.5, $g_\infty
\in \L(\{f_n\})$.  But since $g_\infty(x_0) > f(x_0)$,  we
know $g_\infty \neq f$. \qed
\enddemo 

It is less easy to express this idea in other contexts.  For any
strongly causal spacetime $U$ we could say that if
$\sigma = \{p_n\}$ is a sequence of events with
$\LI(\{I^-(p_n)\}) = \LS(\{I^-(p_n)\}) = R$ but with
$\L(\sigma)$ containing an element $P$ with past different
from interior($R$), then $\L(\sigma)$ has at least two
elements ($\L$ being understood to take values in
$U^+$).  For a past-regular, future-complete chronological 
set $X$, if $\sigma = \{\alpha_n\}$ is a sequence with
$\LI(\{I^-(\alpha_n)\}) =  \LS(\{I^-(\alpha_n)\}) = R$ but 
with $\L(\sigma)$ containing a point $\alpha$ with
$I^-(\alpha) \neq I^-[R]$, then
$\L(\sigma)$ contains at least two points.  The proofs are
essentially identical to that of Proposition 5.6.  (For the
latter, suppose $\beta_1 \ll \beta_2 \in R$ with $\beta_1 \not
\in I^-(\alpha)$, and let $F = \{\text{IPs } P \text{ in } X \st 
I^-(\gamma) \subset P \subset R\}$ for some $\gamma$ with
$\beta_1 \ll \gamma \ll \beta_2$; $F$ has a maximal element
$I^-(\delta)$, and $\delta \neq \alpha$.)

Proposition 5.3 leads to a characterization
of the behavior of $\L$ in non-Hausdorff situations
exactly reflective of what happens in the example space
$\tilde G$ of section 2 (the unwrapped grapefruit
on a stick), at least in the simple context of the $\liminf$ and
$\limsup$ of the pasts of the sequence being the same.  Recall
that in $\tilde G$, the sequence of points $\sigma =
\{(t_n,x_n)\}$ for $t_n = rn + (1-r)\pi/2$ and $x_n = (n,z_0)$
(with any choice of $|z_0| < r'$) has two elements in its
future-chronological limit, $\hat L(\sigma) =
\{b^+,b^-\}$ (in the notation of section 2, $b^\pm = b_{c^\pm}
\pm f(z_0)$).   Let $\delta_{z_0}$ be the function which gives
the boundary of the past of this limit set, \ie, $I^-[\hat
L(\sigma)] = \Pa(\delta_{z_0})$; this is calculated from the
pasts of the points of the sequence as $\delta_{z_0} = \lim
d^{t_n}_{x_n}$.  Then we also know that $\delta_{z_0} =
\max\{b^+,b^-\}$.  In other words:  The multiple elements of
the future-chronological limit of the sequence have a combined
past which is the limit (\ie, both the LI and LS) of the pasts
of the sequence.  This is true very generally.

\proclaim{Proposition 5.7} Let $X$ be a past-regular, 
future-complete chronological set.  Let $\sigma$ be any sequence
of points in $X$ for which 
$\LI(\{I^-(\sigma(n))\}) = \LS(\{I^-(\sigma(n))\}) = R
\neq
\emptyset$.  Then $I^-[\L(\sigma)] = I^-[R]$.
\endproclaim

\demo{Proof} By Proposition 5.1, $\alpha \in \L(\sigma)$ if and
only if $I^-(\alpha)$ is a maximal IP in $R$.

Suppose $\gamma \in I^-[\L(\sigma)]$, \ie, $\gamma \ll \alpha$ 
for some $\alpha \in \L(\sigma)$; there is some $\beta$ with
$\gamma \ll \beta \ll \alpha$.  Then $I^-(\alpha)
\subset R$, so $\beta \in R$; then $\gamma \in
I^-[R]$.

Suppose $\gamma \in I^-[R]$, \ie, $\gamma \ll \alpha$ for some
$\alpha \in \LI(\{I^-(\sigma(n))\})$.  By Proposition 5.3, 
$\L(\sigma)$ contains an element $\beta$ with
$I^-(\beta) \supset I^-(\alpha)$.  Then $\gamma \in
I^-(\beta)$, so $\gamma \in I^-[\L(\sigma)]$. \qed
\enddemo

Let's apply this to a spacetime:

\proclaim{Corollary 5.8} Let $U$ be a strongly causal
spacetime.  Suppose $\sigma$ is a sequence of events in $U$ for
which any event observed by infinitely many of the elements of
$\sigma$ is observed by almost all of them (\ie, all but
finitely many of them); let $R$ denote this set of observed
events, and assume this is non-empty.  Then the past of
$\L(\sigma)$ is the past of $R$. \qed
\endproclaim

In a standard static spacetime:

\proclaim{Corollary 5.9} Let $V = \mk1 \times M$ be a conformal
equivalent of a standard static spacetime.  Suppose $\{f_n\}$ is
a  sequence of elements in $\IP(M)$ with a pointwise limit
$\phi \neq - \infty$.  Then $\phi = \sup \L(\{f_n\})$; in
particular, if $\L(\{f_n\})$ is finite in size,
$\L(\{f_n\}) = \{g_1, \dots, g_m\}$, then $\phi =
\displaystyle \max_{i \le m} g_i$. \endproclaim

\demo{Proof}  The sets $\LI(\{\Pa(f_n)\})$ and
$\LS(\{\Pa(f_n)\})$ have the same boundary precisely when
$\{f_n\}$ has a pointwise limit $\phi$, and that boundary is
given by $\phi$.  Thus, $\Pa(\phi)$ is the interior---\ie, the
past---of $R$ in Corollary 5.8.  For any collection $Q$ of
functions in $\IP(M)$, the boundary of $I^-[Q]$ is given by
$\sup_{f \in Q} f$.  Thus, Corollary 5.8 identifies $\phi$ as
$\sup \L(\{f_n\})$.
\qed
\enddemo

The key to the main compactness result is a very minor extension
of Proposition 5.3:  We may restrict attention to intersection
with a countable dense set.  We are speaking here of a set $D$
which is chronologically dense, \ie, for any $x \ll y$, there is
an element $d \in D$ with $x \ll d \ll y$.  The existence of a
countable, chronologically dense subset is one of the axioms of 
a chronological set (see \cite{H1}); it is what makes it
possible to do almost as much in such a general setting as one
can do in a spacetime.

\proclaim{Lemma 5.10} Let $X$ be a past-regular, future-complete
chronological set, and let $\sigma = \{\alpha_n\}$ be any
sequence in $X$.  Let $D$ be any chronologically dense
subset of $X$.  If $\LI(\{I^-(\alpha_n)\}) \cap D =
\LS(\{I^-(\alpha_n)\}) \cap D \neq \emptyset$, then $\L(\sigma) \neq \emptyset$; in particular, for any $\alpha \in
\LI(\{I^-(\alpha_n)\})$, $\L(\sigma)$ contains an element
$\beta$ with $I^-(\beta) \supset I^-(\alpha)$.
\endproclaim

\demo{Proof} Let $I = \LI(\{I^-(\alpha_n)\})$ and $S =
\LS(\{I^-(\alpha_n)\})$; we have $\alpha
\in I$. As in the proof of Proposition 5.3, we can demonstrate 
the existence of an IP $P_\infty$, lying in $I$ and containing
$I^-(\alpha)$, which is  maximal for those properties (let $\Cal
P$ be the collection of such IPs; invoking Lemma 5.2, we see
every inclusion-chain in $\Cal P$ has an upper bound); this also
gives us that $P_\infty$ is a maximal IP in $I$.  There is some
$\beta \in X$ with $I^-(\beta) = P_\infty$.  All we need to show
is that $P_\infty$ is maximal in $S$, for then Proposition 5.1
gives us $\beta \in \L(\sigma)$.

Suppose there were some $\gamma \in X$ with $I^-(\beta) \subsetneq
I^-(\gamma) \subset S$.  Since $P_\infty$ is maximal in $I$, we
have $I^-(\gamma) \not\subset I$; thus, there is some $\gamma'
\ll \gamma$ with $\gamma' \not\in I$.  There is some element
$\delta \in D$ with $\gamma' \ll \delta \ll \gamma$; then $\delta
\not\in I$ also.  That means $\delta \not\in I \cap D$, so
$\delta \not\in S \cap D$ also; on the other hand, $\delta \in
I^-(\gamma)$, so $\delta$ must be in $S$. \qed
\enddemo

With this lemma in hand, we can use a  diagonal argument to
establish our basic quasi-compactness result.

\proclaim{Theorem 5.11} Let $X$ be a past-regular,
future-complete chronological set.  Let $\sigma$ be any
sequence of points in $X$ for which
$\LS(\{I^-(\sigma(n))\})$ has a non-empty past.  Then
there is a subsequence $\sigma^\infty \subset \sigma$ with
$\L(\sigma^\infty) \neq \emptyset$; in particular, for any
$\alpha \in \LI(\{I^-(\sigma(n))\})$,
$\L(\sigma^\infty)$ contains an element
$\beta$ with $I^-(\beta) \supset I^-(\alpha)$.
\endproclaim

\demo{Proof}  Let $D = \{d_n\}$ be a countable, chronologically
dense set in $X$.  We will construct a sequence of nested
subsequences of the positive integers, $\cdots \subset \tau^{n}
\subset \tau^{n-1} \subset \cdots \subset \tau^{1} \subset 
\tau^0 = (1,2,3,\dots)$.  First some notation: 
For any sequence $\bar\sigma$ of points in $X$, let
$\LS_{I^- }\{\bar\sigma\} =
\LS(\{I^-(\bar\sigma(n))\})$; for any subsequence
$\tau$ of $(1,2,3,\dots)$, let
$\sigma\circ\tau$ denote the sequence of points
$(\sigma(\tau(1)), \sigma(\tau(2)), \dots)$ and let $\text
S_\sigma[\tau] = \LS_{I^-}\{\sigma\circ\tau\}$.  Similarly
for $\LI_{I^-}\{\bar\sigma\}$ and $\text I_\sigma[\tau]$.  Note
that for $\tau' \subset \tau$, we have $\text S_\sigma[\tau']
\subset \text S_\sigma[\tau]$.  Also, $\text S_\sigma[\tau^0] =
\LS_{I^-}\{\sigma\}$.

Suppose we have defined $\tau^{n-1}$.  Consider the element
$d_n$:  If $d_n \not\in \text S_\sigma[\tau^{n-1}]$---\ie, if
$d_n \ll \sigma(\tau^{n-1}(k))$ for only finitely many 
$k$---then let $\tau^{n} = \tau^{n-1}$.  If, however, $d_n \in
\text S_\sigma[\tau^{n-1}]$, then let $\tau^{n}$ be that
subsequence of $\tau^{n-1}$ containing all numbers $k$ such
that $d_n \ll \sigma(\tau^{n-1}(k))$ (which occurs for
infinitely many $k$).  Note that this implies that if
$d_n \in \text S_\sigma[\tau^{n-1}]$ then for all $k$, $d_n
\ll \sigma(\tau^n(k))$ (a stronger statement than $d_n
\in \text I_\sigma[\tau^n]$).  

Having defined all the subsequences $\{\tau^n\}$, we have, for
all $k \le j$,  $\text  S_\sigma[\tau^k]  \supset \text
S_\sigma[\tau^j]$.  Therefore, for all $n$, if for any $j \ge
n-1$, $d_n \in \text S_\sigma[\tau^j]$---that is to say, $d_n$
is in the past of infinitely many of the elements of $\tau^j$---
then $d_n$ is also in $S_\sigma[\tau^{n-1}]$ and, consequently,
for all $m \ge n$ and for all $k,\; d_n  \ll
\sigma(\tau^m(k))$.  In summary:
$$\align
d_n \in \text S_\sigma[\tau^j] & \text{ for some } j \ge n -1
\implies \\
d_n  \ll \sigma(\tau^m(k)) & \text{ for all } m \ge n \text{ and
for all } k.
\endalign$$

Now let $\tau^\infty$ be the sequence defined by
$\tau^\infty(k) = \tau^k(k)$.  Consider any $d_n \in \text
S_\sigma[\tau^\infty]$:  For infinitely many $k$, $d_n \ll
\sigma(\tau^k(k))$, \ie, there is a strictly increasing sequence
of integers $\{k_i\}$ with $d_n \ll \sigma(\tau^{k_i}(k_i))$ for
all $i$.  Pick some $j$ with $k_j \ge n-1$.  For all $i \ge
j$, we have $\tau^{k_i}$ is a subsequence of $\tau^{k_j}$, so 
for some $m_i$, $\tau^{k_i}(k_i) = \tau^{k_j}(m_i)$.  Then
$d_n \ll \sigma(\tau^{k_j}(m_i))$ for all $i \ge j$, an infinite
collection $\{m_i\}$.  Thus, $d_n \in \text S_\sigma[\tau^{k_j}]$. 
It follows that for all $m \ge n$ and all $k$, $d_n \ll
\sigma(\tau^m(k))$.  In particular, we have $d_n \ll
\sigma(\tau^k(k))$ for all $k \ge n$, which gives us $d_n \in
\text I_\sigma[\tau^\infty]$.  Therefore, $\text
I_\sigma[\tau^\infty] \cap D = \text S_\sigma[\tau^\infty] \cap
D$.  

Let $\sigma^\infty = \sigma\circ\tau^\infty$.  We have 
$\LI_{I^-}\{\sigma^\infty\} \cap D =
\LS_{I^-}\{\sigma^\infty\} \cap D$; we need to know
this is non-empty.  We know
$\LS_{I^-}\{\sigma\}$ contains some element $\gamma$ with some
$\gamma' \ll \gamma$.  There is some element $\delta$ of $D$
with $\gamma' \ll \delta \ll \gamma$.  All of $I^-(\gamma)$ lies
in $\LS_{I^-}\{\sigma\}$, so $\delta \in \LS_{I^-}\{\sigma\}$,
\ie, $\LS_{I^-}\{\sigma\} \cap D$ is non-empty.  Let $n_0 =
\min\{n \st d_n \in \LS_{I^-}\{\sigma\}\}$.  Then for all $i <
n_0$, $\tau^i = \tau^0$, and $d_{n_0} \in
I^-((\sigma\circ\tau^{n_0})(k))$ for all $k$.  For all larger
$n$, $\sigma\circ\tau^n$ is a subsequence of
$\sigma\circ\tau^{n_0}$, so $d_{n_0}$ is also in every
$I^-((\sigma\circ\tau^n)(k))$.  In particular, $d_{n_0} \in
I^-((\sigma\circ\tau^n)(n))$ for all $n \ge n_0$; it follows 
that $d_{n_0} \in \LS_{I^-}\{\sigma\circ\tau^\infty\}$. 
Therefore, $\LS_{I^-}\{\sigma^\infty\} \cap D$ is non-empty.  

Lemma 5.10 now applies to $\sigma^\infty$; this immediately
gives us that $\L(\sigma^\infty)$ is non-empty.  Now consider
any $\alpha \in \LI_{I^-}\{\sigma\}$; since $\sigma^\infty$ is a
subsequence of $\sigma$, we also have $\alpha \in
\LI_{I^-}\{\sigma^\infty\}$.  Therefore, Lemma 5.10 gives us an
element $\beta \in \L(\sigma^\infty)$ with $I^-(\beta)
\supset I^-(\alpha)$. \qed
\enddemo

This has immediate application to spacetimes:

\proclaim{Corollary 5.12}  Let $U$ be a strongly causal
spacetime, and let $\sigma$ be any sequence of events in $U$. 
Suppose there is an event in $U$ which is observed by an
infinite number of the elements of $\sigma$; then there is a
subsequence $\sigma^\infty \subset \sigma$ with
$\L(\sigma^\infty) \neq \emptyset$ (where $\L$ takes values
possibly in $\FB(U)$).  Moreover, if there is some event $x$
observed by all but finitely many of the elements of $\sigma$,
then $\L(\sigma^\infty)$ contains an element $\beta$ whose past
contains the past of $x$.

In other words:  A sequence $\sigma$ of events contains a
subsequence $\sigma^\infty$ with a future-chronological limit if
and only if some event in $U$ is in the common past of some
subsequence of $\sigma$; and the combined observations of all 
the future-chronological limits of $\sigma^\infty$ include
anything observed by almost all the elements of $\sigma$.
\endproclaim

\demo{Proof} Just apply Theorem 5.11 to $U^+$. \qed 
\enddemo

Specializing to standard static spacetimes, we have this:

\proclaim{Corollary 5.13}  Let $V = \mk1\times M$ be a conformal
equivalent of a standard static spacetime, and let $\{f_n\}$ be a
sequence of elements in $\IP(M)$; let $f^+ = \limsup f_n$ and $f^-
= \liminf f_n$.  If $f^+ \neq -\infty$, then there is a
subsequence $\{f_{n_k}\}$  with a future-chronological limit in
$\V$; moreover, this sequence has the property that for any $x \in
M$ and $t < f^-(x)$, there is a future-chronological limit
$f_\infty$ of $\{f_{n_k}\}$ with $f_\infty \ge d^t_x$.  That
future-chronological limit cannot be
$i^+$ unless $f^+ = \infty$, which can happen only when there is a
subsequence of the events represented by $\{f_n\}$ with no events
in its common future.  Furthermore, the subsequence $\{f_{n_k}\}$
has a pointwise limit.
\endproclaim

\demo{Proof} First note that $f^+$ is either everywhere finite 
or everywhere infinite (and the same for $f^-$). Note also that
$\Pa(f^+)$ is the interior of $\LS(\{\Pa(f_n)\})$ and $\Pa(f^-)$
the interior of $\LI(\{\Pa(f_n)\})$.  Then Theorem 5.11 applies
to $\V$.  

The statement about $i^+$ follows from noting that $f_\infty
\le f^+$, as $\Pa(f_\infty)$ must lie inside $\LS(\{\Pa(f_n)\})$.

The final statement follows from noting that the subsequence
created in Theorem 5.11 has $\LI(\{\Pa(f_{n_k})\}) \cap D =
\LS(\{\Pa(f_{n_k})\}) \cap D$ for a countable dense set $D$ (in
$V$, chronologically dense is the same as topologically dense). 
Let $D = \{(t_m,x_m)\}$.  Consider the sets 
$$\align
A & = \{m \st \text{for infinitely many }k,\;  t_m < f_{n_k}(x_m)\}
\text { and} \\ 
B & = \{m \st \text{for almost all }k,\;  t_m <
f_{n_k}(x_m)\}.
\endalign$$ 
Define $\bar f^+ =
\limsup f_{n_k}$ and $\bar f^- = \liminf f_{n_k}$, and let
$$\align
A^+ & = \{m \st t_m < \bar f^+(x_m)\} \text{ and} \\ 
B^- & = \{m \st t_m \le\bar f^-(x_m)\}.
\endalign$$
Note that $A^+ \subset A$ and
$B \subset B^-$.  The information from Theorem 5.11 is that $A =
B$.  It follows that
$A^+ \subset B^-$, \ie, 
$$\text{for all }m, \;\; t_m < \bar f^+(x_m) \,\text{ implies }\, 
t_m \le \bar f^-(x_m).$$
We know $\bar f^+$ and $\bar f^-$ are Lipschitz-1, hence, continuous;
so if they do not coincide, then there is some open set $U$ and some
numbers $t^- < t^+$ such than on $U$, $\bar f^- < t^-$ and $t^+
< \bar f^+$.  There are infinitely many $m$ (though only one
is needed) such that
$(t_m,x_m) \in (t^-,t^+) \times U$.  For each such $m$, $\bar
f^-(x_m) < t^- < t_m < t^+ < \bar f^+(x_m)$, contradicting the
implication displayed above.  It follows that $\bar f^- = \bar
f^+$, which means that
$\{f_{n_k}\}$ converges pointwise. \qed
\enddemo

While Corollary 5.13 provides only a quasi-compactness 
for $\V$ (sequences have convergent subsequences only if
they don't converge to $-\infty$), it is good enough to
provide full compactness for $\M$:

\proclaim{Theorem 5.14} Let $M$ be any Riemannian
manifold.  Then $\M$ is compact in the chronological
topology. \endproclaim

\demo{Proof}  We represent $\M$ as $\V_0/\R$ ($V = \mk1
\times M$), which we treat as $\IP_0(M)/\R$.  Let $\sigma
= \{[f_n]\}$ be any sequence in $\IP_0(M)/\R$.  Pick a
point $x_0 \in M$.  For each $n$, let $\bar f_n = f_n -
f_n(x_0)$; then $[\bar f_n] = [f_n]$, so these functions
represent the same sequence $\sigma$ in $\M$.  By
Corollary 5.13, there is a subsequence $\{\bar
f_{n_k}\}$ with a future-chronological limit $\bar
f_\infty$ in $\V_0$ ($\{\bar f_n(x_0)\}$ being
obviously bounded below).  Since $\bar
f_\infty$ is a limit of the sequence $\{\bar f_{n_k}\}$
in the future-chronological topology on $\V_0$, $[\bar
f_\infty]$ is a limit of the sequence $\{[\bar f_{n_k}]\}
= \{[f_{n_k}]\}$ in the chronological topology on $\M$. 
(Note that we are using the fact that $\pi: \V_0 \to \M$
is continuous in the quotient topology; we are not
assuming that $e_{x_0} : \M \to \V_0$ is continuous.) 

Since every sequence in $\M$ has a subsequence with a
limit, it is compact.\qed
\enddemo

(This result in purely Riemannian geometry can be established
without recourse to chronological sets by proving a result for
separable metric spaces:  Any sequence of Lipschitz-1 functions
which is bounded below at one point has a pointwise-convergent
subsequence with convergence to $\infty$ being allowed.   A
diagonal procedure is used, employing a countable dense set in
the metric space.) 

Corollary 5.13 also establishes that it is precisely the
non-Hausdorff character of the future-chronological
topology that distinguishes it from the function-space
topology.  Note that it is the behavior of
$\L$---whether or not, for a sequence $\sigma$ in $\V$,
$\L(\sigma)$ can have more than one element---that
determines whether or not the future-chronological
topology is the same as the function-space topology: 
For if $\L(\sigma)$ never has more than one element,
then, by Proposition 5.6, it is the same as the
pointwise limit.

\proclaim{Theorem 5.15} Let $V$ be a standard static
spacetime.  The future-chronological topology on $\V$ is
identical to the function-space topology if and only if
the former is Hausdorff. \endproclaim

\demo{Proof}  The compact-open topology on a real-valued
function space is always Hausdorff (assuming points are
closed in the domain space):  If $f \neq g$, then there 
is some $x$ in the domain with, say, $f(x) < g(x)$. 
Pick a number $a$ with $f(x) < a < g(x)$. Then $\{h \st
h[\{x\}] \subset (-\infty,a)\}$ and $\{h \st h[\{x\}]
\subset (a,\infty)\}$ are disjoint neighborhoods of $f$
and $g$ respectively.

Assume that the future-chronological topology on $\V$
is Hausdorff.  We will show that any sequence
$\{f_n\}$ in $\IP(M)$ ($V = \mk1 \times M$) can have only
the pointwise limit of $\{f_n\}$ as a 
future-chronological limit.  

Suppose we have $f \in \L(\{f_n\})$.  Let $f^- =
\liminf f_n$ and $f^+ = \limsup f_n$.  We first show $f^-
= f^+$:

Suppose the two differ on some point $x$. Let $\{f'_m\}$
be a subsequence of $\{f_n\}$ with $\{f'_m(x)\}$
converging to $f^+(x)$, and let $f'{}^- = \liminf
f'_m$.  As $f \le f^-$, we know $\{f'_m(x)\}$ is bounded
below by $f(x)$.    Pick some $t$ with $f^-(x) < t <
f^+(x)$; note that $t < f'{}^-(x)$.   Then by
Corollary 5.13, there is a subsequence
$\{f'_{m_k}\}$ with a future-chronological limit
$f'_\infty \ge d^t_x$.  Then
$$f(x) \le f^-(x) < t = d^t_x(x) \le f'_\infty(x).$$
It follows that $f \neq f'_\infty$.  Then, since
$\L(\{f_n\}) \subset \L(\{f'_{m_k}\})$, we have $f$ and
$f'_\infty$ as distinct elements of $\L(\{f'_{m_k}\})$, 
contradicting Hausdorffness.  

Therefore, $\{f_n\}$ converges pointwise to some $\phi$
(possibly $\infty$).  Then, by Proposition 5.6, $\phi = 
f$ must be the only element of $\L(\{f_n\})$, if $\V$ is
Hausdorff in the future-chronological topology.

We have shown that under the assumption of Hausdorffness
on the future-chronological topology, the only
future-chronological limits are also function-space
limits; it follows that the two topologies are identical.
\qed
\enddemo

\head 6. When the Function-Space and Future-Chronological
Topologies Coincide \endhead

We conclude this paper with an examination of standard static
spacetimes $\mk1 \times M$ for which it can be shown that the
future-chronological topology on the future causal boundary is
as nice as can be desired: a cone on the Busemann boundary, with
only geodesics being needed for Busemann functions.  We will show
that this includes a large class of spacetimes encompassing
various classical examples, such as external Schwarzschild or parts
of Schwarzschild--de Sitter or Reissner-Nordstr\"om, addressed in a
discussion subsection.

We need first a couple of definitions:

\definition{Divergence of a sequence of points} Let $\sigma$ be a
sequence of points in $M$.  Then $\sigma$ is {\it
divergent\/} if there is no convergent subsequence (\ie,
$\sigma' \subset \sigma$ such that $\dlim_{n \to \infty}
\sigma'(n)$ exists in $M$).
\enddefinition

\definition{Convergence of a sequence of geodesic segments}
Suppose
$\{\gamma_n\}$ is a sequence of unit-speed geodesic segments in
$M$, $\gamma_n$ defined on
$[0,L_n]$.  We will say that $\{\gamma_n\}$ converges to the
unit speed geodesic
$\gamma_\infty$, defined on $[0,\omega)$, if
\roster
\item $\{\dot\gamma_n(0)\}$ converges to
$\dot\gamma_\infty(0)$,  
\item $\omega = \lim_{n \to \infty}\{L_n\}$, and
\item $[0,\omega)$ is the maximal domain for $\gamma_\infty$.
\endroster
Note that these three conditions imply that for all $t < \omega$, 
$\{\gamma_n(t)\}$ (defined for $n$ sufficiently large) converges to
$\gamma_\infty(t)$. 

The point of this is to define convergence of
geodesics even in a non-complete manifold, where geodesics of
maximal length cannot be assumed to be infinite in length.  Since a
spacetime of interest is typically only conformal to a standard
static spacetime, and the conformal factor may easily result in
the Riemannian factor being rendered non-complete, we must retain
an interest in non-complete manifolds.
\enddefinition

The following proposition is a condition that allows us to conclude
that the Busemann boundary has the simplest of forms:  It has the
function-space topology, and it uses only geodesics---and only a
restricted class of limit-geodesics, at that.  It appears to be an
unnaturally formidable condition; but Theorem 6.2 will show that is
not the case, that it applies to a substantial class of manifolds.

\proclaim{Proposition 6.1} Suppose distances in $M$ are realized by
geodesics and that $M$ obeys the following property:

For every  divergent sequence $\{x_n\}$ in $M$, for some
subsequence of that sequence (which we again call $\{x_n\}$),
for some point $z$, for some choice of minimal unit-speed
geodesic segments $\gamma_n$ from $z$ to $x_n$ (with
$\gamma_n(0) = z$), the geodesic segments converge to a
geodesic $\gamma_\infty:[0,\omega) \to M$ such that for all $x \in
M$,
$$\lim_{t \to \omega}\bigl(d(z,\gamma_\infty(t)) -
d(x,\gamma_\infty(t))\bigr) = 
\lim_{n \to \infty}\bigl(d(z,x_n) -
d(x,x_n)\bigr) \tag*$$
(wherein $\infty$ is admissable as a limit);
or, more generally, for some sequence $\{t_n\}$ approaching
$\omega$, for all $x \in M$,
$$\lim_{n \to \infty}\bigl(d(z,\gamma_\infty(t_n)) -
d(x,\gamma_\infty(t_n))\bigr) = 
\lim_{n \to \infty}\bigl(d(z,x_n) -
d(x,x_n)\bigr). \tag{**}$$
(The left side of either \rom{(*)} or \rom{(**)} always exists with
$\infty$ admitted as a value; these hypotheses are conditions upon
the right sides.)

Then 
\roster
\item the chronological topology on the Busemann boundary of
$M$, $\partial_B(M)$, is the same as the function-space
topology, and
\item only geodesics need be considered for $\partial_B(M)$;
moreover, only geodesics of the form $\gamma_\infty$.
\endroster
Thus, the future causal boundary for $V =
\mk1\times M$ (in the future-chronological topology) is a cone on
$\partial_B(M)$, and $\partial_B(M)$ may be identified with
equivalence classes of Busemann functions of geodesics of the form
$\gamma_\infty$ above.
\endproclaim

\demo{Proof}  By Theorem 5.15, to show the chronological
topology the same as the function-space topology, it suffices
to show the former is Hausdorff.  By Proposition 5.5, to show
Hausdorffness, it suffices to show that for any sequence  of
functions $\{f_n\}$ in $\IP(M)$ for which there is a
future-chronological limit, there is a subsequence with a
pointwise limit $\phi$ such that $\phi \in \IP(M)$
also.  By Corollary 5.13 (last statement), we may restrict
ourselves to consideration of sequences which have a pointwise
limit.  By Lemma 3.3, it is sufficient to consider the functions
$f_n$ as being of the form
$a_n - d(x_n,-)$ for some numbers $\{a_n\}$ and points
$\{x_n\}$.  

So let us consider a sequence of functions $\{d_n\}$ with $d_n
= a_n - d(x_n,-)$ such that $\{d_n\}$ has the pointwise limit
$\phi$; we can assume $\phi$ is finite, since $\infty$ is
always in $\Cal I \Cal P(M)$.  Let $x_0$ be some point in
$M$.  Let $a'_n = d(x_n,x_0)$ and $d'_n = a'_n - d(x_n,-) = d_n +
a'_n - a_n$.  Since $\phi(x_0) = \lim d_n(x_0) = \lim (a_n -
a'_n)$, we have $\{d'_n\}$ converges pointwise to $\phi -
\phi(x_0)$, which represents the same element as $\phi$ in
$\partial_B(M)$.  Thus, we may assume that our functions $d_n$
are of the form of $d'_n$, i.e., $d_n = d(x_n,x_0) - d(x_n,-)$.

By restricting to an appropriate subsequence, we may assume that
for $\gamma_n$ a minimizing unit-speed geodesic from some point
$z$ to $x_n$, the geodesic segments $\{\gamma_n\}$ converge to a
unit-speed geodesic $\gamma_\infty : [0, \omega) \to M$.  Clearly,
$\gamma_\infty$ is a ray, i.e., it is minimizing on its maximal
domain, $[0,\omega)$; consequently, its Busemann function
$b_{\gamma_\infty}$ is finite.  We will see, in fact, that the
hypothesis (*) yields (with choice of $x_0 = z$)  
$b_{\gamma_\infty} = \phi$.  This will then establish both
conclusion (1) (since $\phi$ is then in
$\Cal I \Cal P(M)$) and conclusion (2) (since $\gamma_\infty$ is 
a geodesic).

We know for all $x$, $\phi(x) =  \lim d_n(x) = \lim 
(d(x_n,x_0)-d(x_n,x))$.  We have freedom to choose the arbitrary
point $x_0$; letting $x_0 = z$ then gives us $\phi(x)$ as the
right side of (*).  Since
$\gamma_\infty$ is a ray commencing at $z$, for all $t < \omega$,
$d(z,\gamma_\infty(t)) = t$; thus,
$b_{\gamma_\infty}(x)$ is the left side of (*) (which is why that
limit always exists). 

Note that the left side of (**) is $\lim_{n
\to\infty}(t_n - d(x,\gamma_\infty(t_n)))$, which is necessarily
$b_{\gamma_\infty}(x)$.  Thus, (**) is also a sufficient
condition. \qed
\enddemo

Before looking at an application of this proposition, we should
perhaps look at examples where it does not apply.  If $M$ is the
unwrapped grapefruit on a stick---$\tilde G$ of section 2---and
$\{x_n\}$ is a sequence of points running out to infinity along the
unwrapped equator, then, as $M$ is complete, for any $z$, there can
be found a minimizing geodesic $\gamma_n$ from $z$ to $x_n$.  If
$z$ is in the upper flat region, for instance, then $\{\gamma_n\}$
converges to a geodesic $\gamma_\infty$ which is the horizontal
geodesic starting at $z$.  But (*) is not satisfied, as the left
side is $b_{\gamma_\infty}$ (or $b_{c^+}$), while the right side
is $\infty$ (since the unwrapped equator is not asymptotically
ray-like).  If $z$ is in the round sector, then the convergence
is a bit more interesting, as the geodesic segments $\gamma_n$
never enter the flat region; they stay in the round region and,
for a long distance, in the tiny negative-curvature region
between the flat and round regions.  They still converge (or, in
the case of $z$ lying right on the unwrapped equator, some
subsequence of them converge) to a geodesic $\gamma_\infty$; but
in this case $\gamma_\infty$ is a somewhat remarkable geodesic
that is uniquely characterized as starting from $z$, entering the
negative-curvature region (either the upper or lower one) and
never leaving it.  However, the Busemann function for this
geodesic is exactly the same (up to a constant) as that for $c^+$
(or $c^-$, for the lower negative-curvature region).  Thus, (*) is
never satisfied for $\{x_n\}$, and the Busemann boundary is
non-Hausdorff:  $\{d_n\}$ has a limit $\phi$ which is not in
$\Cal I \Cal P(M)$, but is the max of $\{b_{c^+},b_{c^-}\}$ (up
to constants), and $[b_{c^+}]$ and $[b_{c^-}]$ are both
chronological limits of $\{[d_n]\}$, i.e. of $\{x_n\}$.

If, instead of a single grapefruit, we unwrap an infinite string
of them, all on a single stick, then we obtain a manifold $M$
which violates not conclusion (1) of Proposition 6.1, but
conclusion (2).  Let $x_n$ be on the unwrapped equator of the 
$n$th grapefruit, at a distance $\theta_n = n^3$ out from $\theta
= 0$.  If we pick $z_k$ in the $k$th flat region, then the
geodesic segments $\{\gamma^k_n\}$ converge, as before, to a
geodesic $\gamma^k_\infty = c_k$ which is a horizontal geodesic
starting at $z_k$.  The left-hand side of (*) is now $b_{c_k}$. 
The right-hand side, instead of being infinite, is equal to
$\max\{b_{c_n}\}$; this is $b_c$ for $c$ the broken geodesic
connecting the points $\{x_n\}$, a finite Busemann function but
not the Busemann function of any geodesic.  

In spite of the non-geodesic nature of the element $[b_c]$ of the
Busemann boundary for $M$, $\partial_B(M)$ is, none the less,
Hausdorff.  This suggests a modification of Proposition 6.1:
Suppose $M$ has distances realized by geodesics and every
divergent sequence $\{x_n\}$ has a subsequence (call it again
$\{x_n\}$) so that for some choice of minimizing unit-speed geodesic
segments $\gamma^m_n$ from $x_m$ to $x_n$ (with $\gamma^m_n(0) =
x_m$), for each fixed $m$, the geodesic segments $\{\gamma^m_n \,|\,
n \ge m\}$ converge to a geodesic $\gamma^m_\infty$ with Busemann
function $b_{\gamma^m_\infty} = b_m$ such that the sequence of
functions $\{b_m - b_m(x_0)\}$ is monotonic increasing and converges
pointwise to $\lim_{n\to\infty} (d(x_0,x_n) - d(-,x_0))$; then the
chronological topology on $\partial_B(M)$ is the same as the
function-space topology.  (Proof sketch: Let $\phi$ be the
pointwise limit of $\{d_n\}$ for $d_n =d(x_0,x_n) - d(-,x_0)$.  By
Lemma 5.2, the pointwise limit of a monotonically increasing
sequence of elements of $\IP(M)$ is in $\IP(M)$; apply this to
$\phi$. Then, as in Proposition 6.1, it follows
that $\partial_B(M)$ is Hausdorff.)

The motivation for Proposition 6.1 is its application to external
Schwarzschild; this has as manifold $\R \times (2m,\infty)
\times \sph 2$ and as metric
$ds^2 = -\(1-\dfrac{2m}r\)dt^2 + \(1-\dfrac{2m}r\)^{-1}dr^2
+ r^2 d\sph 2$, where $d\sph 2$ denotes the round metric on the
sphere.  This is conformal to the metric $-dt^2 + h = -dt^2  + 
\(1-\dfrac{2m}r\)^{-2} dr^2 +
r^2\(1-\dfrac{2m}r\)^{-1} d\sph 2$.  Since the future
causal boundary is conformally invariant, we can use this latter
metric.   Thus we have external Schwarzschild is conformal to the
standard static spacetime $\mk1 \times M$, where $M$ is
topologically $(2m, \infty) \times \sph2$.  It will be much
easier to treat $M$ as a warped product by defining $\rho(r) =
\int (1-\frac{2m}r)^{-1} dr = r + 2m\ln(r-2m)$,
yielding the metric $h$ on $M$ as $d\rho^2 +
r(\rho)^2(1-\frac{2m}{r(\rho)})^{-1}d\sph 2$, a
warped product metric on $(-\infty,\infty) \times \sph 2$. 
Note that as $\rho$ goes to $\infty$, so does the warping
function $r^2(1-\frac{2m}r)^{-1}$; and as $\rho$ goes to $-\infty$,
$r$ goes to $2m$ and, again, the warping function goes to
$\infty$.  More specifically, the warping function is monotonic
increasing on $[\rho_0, \infty)$ and monotonic decreasing on
$(-\infty,\rho_0]$, where $r(\rho_0) = 3m$.  This, in fact, is
sufficient to allow us to conclude that the topology of the
future causal boundary is a null cone on two copies of $\sph 2$,
as the next theorem establishes.    

\proclaim{Theorem 6.2}  Let $M$ be a Riemannian manifold
expressible as a warped product with a compact manifold: $M =
(\alpha,\omega) \times_a K$, where $(\alpha,\omega)$ is a
possibly infinite interval in $\R$, $a: (\alpha,\omega) \to \R$ is
a positive function, and $K$ is compact, and the metric on $M$ is
given by $h = d\rho^2 + a(\rho)^2j_K$ ($j_K$ being the
metric on $K$). 

Suppose that for some $\rho_0 \in (\alpha,\omega)$, $a(\rho)$ is
monotonic decreasing for $\rho  \in (\alpha,\rho_0]$ and monotonic
increasing for $\rho \in [\rho_0,\omega)$.  Then $M$ satisfies
the hypotheses of Proposition 6.1, and $\BB(M)$ consists
of spaces $B_\alpha$ and $B_\omega$, attached respectively at
$\{\alpha\} \times K$ and $\{\omega\} \times K$, with each
$B_\iota$ ( $\iota = \alpha,\,\omega$) either $K$ or a single point
$*$:  $B_\iota \cong K$ if
$|\int_{\rho_0}^\iota 1/a(\rho)^2
\,d\rho| < \infty$, and $B_\iota = *$ if $|\int_{\rho_0}^\iota
1/a(\rho)^2 \,d\rho| = \infty$. 

Thus, if $\int_\alpha^{\rho_0} 1/a(\rho)^2 \,d\rho$ and
$\int_{\rho_0}^\omega 1/a(\rho)^2 \,d\rho$ are both finite, the
future causal boundary (using the future chronological
topology) for $V = \mk1\times (\alpha,\omega) \times_a K$ consists
of conjoined cones on two copies of $K$, one for each endpoint of
the interval $(\alpha,\omega)$; the cone-elements corresponding to
each endpoint are null if that endpoint is infinite, timelike if the
endpoint is finite. 
\endproclaim

\demo{Proof} This is a quite lengthy proof, as there is much to be
established.  Part I is the proof that condition (**) of
Proposition 6.1 is met; this requires a close consideration of the
form that geodesics in $M$ take.  Part II is the proof that the
spaces $B_\iota$ have the form claimed: IIa for the infinite
integral and IIb for the finite integral (the latter requiring the
establishment of a homeomorphism with $K$).

\remark
{\bf Part I: Applying Proposition 6.1}\endremark

We need to establish what the distance function on
$M$ is.   Let $M^- = (\alpha,\rho_0) \times K$ and $M^+ =
(\rho_0,\omega) \times K$; we will look at how geodesics either
stay in the one half or the other of $M$ or how they move from
one half to the other.

Let $\gamma(s) = (\rho(s),c(s))$ be a geodesic in $M$; then the
geodesic equation in $M$ gives us
$$\ddot \rho = aa'|\dot c|^2 \quad\text{and}\quad
\ddot c = -2\dot \rho \frac{a'}{a}\dot c,$$
where $\,\dot{}\,$ denotes $d/ds$, $\,{}'\,$ denotes $d/d\rho$,
and $|\,|$ denotes the norm in the metric on $K$.  Thus, $c$ is a
pregeodesic (reparametrization of an actual geodesic) in $K$, and
we have the following integrations (taking $\gamma$ to be
unit-speed):
$$\dot \rho^2 = 1 - \frac{a_0^2\sin^2\theta_0}{a^2}
\quad\text{and}\quad |\dot c| = \frac{a_0\sin\theta_0}{a^2},$$
where $a_0 = a(\rho(0))$ and $\theta_0$ is the angle between
$\dot\gamma(0)$ and the radial vector field $\bold R$ in $M$ given
by $(d/d\rho)(\rho,p)$ for fixed $p \in K$.

Note in particular that in the regime of $a$ being increasing,
$\ddot \rho$ is positive, so if $\dot \rho$ starts off positive, it
remains positive.  Therefore, if $\rho(0) \ge \rho_0$ and
$\theta_0 \le \pi/2$---and at least one of those inequalities is
strict---then $\rho(s)$ will be a strictly increasing function of
$s$, and we can reparametrize $\gamma$ by $\rho$, which greatly
simplifies the analysis.  If $\gamma$ goes from $x_1 =
(\rho_1,p_1)$ to $x_2 = (\rho_2,p_2)$ and if we know $\rho(s)$ is
strictly increasing, then let $\bar \gamma :[\rho_1,\rho_2] \to
M$ be the parametrization of $\gamma$ by $\rho$, i.e,
$\bar\gamma(\rho) =  (\rho,\bar c(\rho))$, where $\bar c: [\rho_1,
\rho_2] \to K$ is a reparametrization of $c$; then
$$\align
L(\gamma) & = L(\bar\gamma) \\
& = \int_{\rho_1}^{\rho_2} 
\left|\frac{d\bar\gamma}{d\rho}\right|\,d\rho \\
& = \int_{\rho_1}^{\rho_2}
\left|\frac{ds}{d\rho}\frac{d\gamma}{ds}\right|\,d\rho \\
& = \int_{\rho_1}^{\rho_2} 
\frac1{\left|\frac{d\rho}{ds}\right|}\,d\rho \\
& = \int_{\rho_1}^{\rho_2}
\frac{d\rho}{\sqrt{1-\left( \frac{a_1\sin\theta_0}{a(\rho)}
\right)^2}}\;,
\endalign$$
where $a_1 = a(\rho_1)$.  For any numbers $\rho_1$, $\rho_2$, and
$\theta_0$ with $\rho_0 \le \rho_1 \le \rho_2$, we can define
$$J_{\rho_1}^{\rho_2}(\theta_0) = 
\int_{\rho_1}^{\rho_2}
\frac{d\rho}{\sqrt{1-\left( \frac{a_1\sin\theta_0}{a(\rho)}
\right)^2}}\;;$$
then if we know $\gamma$ is minimizing, we have
$d(x_1,x_2) = J_{\rho_1}^{\rho_2}(\theta_0)$.

If, on the other hand, $\theta_0 > \pi/2$, then $\rho(s)$ is
decreasing in $s$ until $\rho(s)$ reaches a value $\rhm$ defined
by  $a(\rhm) = a_1\sin\theta_0$, at which point it becomes
increasing (if there is no such $\rhm$---\ie, if $\sin\theta_0 <
a(\rho_0)/a_1$ for $\theta_0 > \pi/2$---then either
$\gamma$ is asymptotic to the $\rho =\rho_0$ curve or it leaves
$M^+$ and enters $M^-$, never returning to $M^+$).  We can still
parametrize $\gamma$ by $\rho$, so long as we are careful to do
so in two segments, with domains $[\rhm,\rho_1]$ and
$[\rhm,\rho_2]$.  Then we obtain
$$L(\gamma) = \int_{\rhm}^{\rho_1}
\frac{d\rho}{\sqrt{1-\left( \frac{a_1\sin\theta_0}{a(\rho)}
\right)^2}} +
\int_{\rhm}^{\rho_2}
\frac{d\rho}{\sqrt{1-\left( \frac{a_1\sin\theta_0}{a(\rho)}
\right)^2}}\;.$$
In this case, if $\gamma$ is minimizing, then $d(x_1,x_2) =
J_{\rhm}^{\rho_1}(\theta_0) + J_{\rhm}^{\rho_2}(\theta_0)$.

In case $\rho_1 < \rho_0 < \rho_2$, we have $\ddot \rho \le 0$
for $\rho \le \rho_0$ and $\ddot \rho \ge 0$ for $\rho \ge
\rho_0$.  This means that $\rho$ is strictly increasing from
$\rho_1$ to $\rho_2$: If $\rho$ starts to decrease while $\rho <
\rho_0$, it never can increase again, so it stays below $\rho_0$;
but it must reach $\rho_2$.  And since it is increasing when it
passes $\rho_0$, it cannot decrease after that point.  Thus, we
can again use $\rho$ for the parametrization of $\gamma$; if
$\gamma$ is minimizing, a similar expression obtains for
$d(x_1,x_2)$.

Note that we have established that all geodesics bend towards
$\rho = \rho_0$, in the following sense:  Take the case of $\rho_0
\le  \rho_1 \le \rho_2$; in going from
$(\rho_1,p_1)$ to $(\rho_2,p_2)$, a geodesic either will have all
its values of $\rho$ lying between $\rho_1$ and $\rho_2$, or, if it
takes values of $\rho$ outside that range, those other values lie
between $\rho_0$ and $\rho_1$.  The case of $\rho_1 \le \rho_2 \le 
\rho_0$ is analogous:  $\rho$-values for the geodesic all lie
between $\rho_1$ and $\rho_2$ or, if they depart from that, take
their additional values between $\rho_2$ and $\rho_0$.  In the case
of $\rho_1 \le \rho_0 \le \rho_2$, the $\rho$-values always fall
between $\rho_1$ and $\rho_2$. It follows that distances in $M$
are realized by geodesics:  

Consider any two points $x_i = (\rho_i,p_i)$, $i = 1,2$.  Pick
$\rho^- < \rho^+$ so that $\rho_0$, $\rho_1$, and $\rho_2$ are all
in $(\rho^-,\rho^+)$. Let
$\bar M = (-\infty,\infty) \times_{\bar a} K$, where $\bar a$ agrees
with $a$ on $(\rho^-,\rho^+)$, is constant off of $(\rho^- -
\epsilon, \rho^+ + \epsilon)$, obeys $\bar a \le a$, and has the
requisite monotonicity properties.  Then $\bar M$ is obviously
complete (differing only on a compact region from the product
metric on $\euc 1\times K$).  Let $M_0 =
(\rho^-,\rho^+) \times_a K$; this is an open submanifold of $M$ and
also of $\bar M$.  The preceding analysis on geodesics in $M$
applies to $\bar M$; thus, if $\bar \gamma$ is a minimizing $\bar
M$-geodesic from $x_1$ to $x_2$, we know that $\bar
\gamma$---bending towards $\rho = \rho_0$---remains within
$M_0$, hence, within $M$.  We can map $M$ into $\bar M$ by $i :
(\rho,p)
\mapsto (\rho,p)$; then $i$ is length-decreasing.  It follows that
$\bar \gamma$ is minimizing in $M$ as well as in $\bar M$, so it is
also a minimizing $M$-geodesic.  Thus, all distances in $M$
are realized by geodesics.

It follows that for any two points $x_1 = (\rho_1,p_1)$ and $x_2
= (\rho_2,p_2)$ with $\rho_1 \le \rho_2$, the distance between
$x_1$ and $x_2$ can be expressed as
$$d(x_1,x_2) = \oint_{\rho_1}^{\rho_2} 
\frac{d\rho}{\sqrt{1-\left( \frac{a(\rho_1)\sin\theta_0}{a(\rho)}
\right)^2}}\;,$$
where $\theta_0$ is the angle made at $x_1$ between the radial
vector field $\bold R$ and a minimizing geodesic from $x_1$ to
$x_2$, and $\oint_{\rho_1}^{\rho_2}$ denotes either
$\int_{\rho_1}^{\rho_2}$ or $\int_{\rhm}^{\rho_1} +
\int_{\rhm}^{\rho_2}$ for the appropriate $\rhm$, depending on
the value of $\theta_0$ (with appropriate readjustment of
nomenclature in case $\rho_2 < \rho_0$).

We must show that the hypotheses of Proposition 6.1 are satisfied
by $M$.  Let $\{x_n\} = \{(\rho_n,p_n)\}$ be a divergent sequence
of points in $M$.  As $K$ is compact, it must be that
$\{\rho_n\}$ has no  accumulation point in $(\alpha,\omega)$; we
may assume that for some subsequence $\{\rho_n\}$ approaches
$\omega$.  Again using that $K$ is compact, we know that for some
subsequence, $\{p_n\}$ converges to a limit $p_\infty$.  
We will, for now, let $z = (\rho_z,p_\infty)$ with $\rho_z$
unspecified.  
For each $n$, let
$\gamma_n$ be a minimizing unit-speed geodesic from $z$ to $x_n$ (so
$\gamma_n(0) = z$).  For some subsequence, the vectors
$\{\dot\gamma_n(0)\}$ converge to a unit-speed vector $v$ at $z$.

\proclaim{Lemma 6.3} The vector $v$ is outward-radial, \ie, $v
= \bold R$. \endproclaim

\demo{Proof of Lemma} 
Let $z_n = (\rho_z,p_n)$, and let $\delta_n$
be the radial geodesic from $z_n$ to $x_n = (\rho_n,p_n)$; clearly,
$L(\delta_n) =\rho_n-\rho_z$ (for $n$ so large that $\rho_n >
\rho_z$).  Let
$\alpha_n$ be the $\rho = \rho_z$ curve from $z$ to $z_n$;
$L(\alpha_n) = a(\rho_z)d_K(p_n,p)$.  Since $\gamma_n$ is
minimizing, we have $L(\gamma_n) \le L(\delta_n) + L(\alpha_n)$. 
We also have
$L(\gamma_n) = \oint_{\rho_z}^{\rho_n}
d\rho/\sqrt{1-\big((a(\rho_z)\sin\theta_n)/a(\rho)\big)^2}$, where
$\theta_n$ is the initial angle of $\gamma_n$ with $\bold R$.  In
the case that $\theta_n > \pi/2$, the two integrals symbolized by
$\oint_{\rho_z}^{\rho_n}$ contain, as a subinterval, the integral
$\int_{\rho_z}^{\rho_n}$.  Thus, in any case, we have
$$\int_{\rho_z}^{\rho_n}\frac{d\rho}{\sqrt{1-\left( 
\frac{a(\rho_z)\sin\theta_n}{a(\rho)}\right)^2}} - (\rho_n-\rho_z)
\le a(\rho_z)d_K(p_n,p_\infty).$$
Combining those two terms on the left into a single integral, we
have
$$\int_{\rho_z}^{\rho_n} \left(\frac1{\sqrt{1-\( 
\frac{a(\rho_z)\sin\theta_n}{a(\rho)}\right)^2}} - 1\)d\rho
\le a(\rho_z)d_K(p_n,p_\infty).$$
Using $\dfrac1{\sqrt{1-x}} \ge 1 + (1/2)x$, we get 
$$\sin^2\theta_n
\int_{\rho_z}^{\rho_n} \frac{d\rho}{a(\rho)^2} \le
\frac2{a(\rho_z)}d_K(p_n,p_\infty).$$
Since $\{p_n\}$ approaches $p_\infty$,
those distances $d_K(p_n,p_\infty)$ approach 0.  On the other side
of the inequality, the integrals are increasing in $n$.  It follows
that $\{\sin \theta_n\}$ must approach 0.  It's not possible that
any subsequence of $\{\theta_n\}$ approach $\pi$ (since that would
have $\gamma_n$ entering $M^-$ and not making it to $x_n$, for $n$
sufficiently large), so $\{\theta_n\}$ must approach 0, i.e.,
$\{\dot\gamma_n(0)\}$ approaches $\bold R$. \qed 
\enddemo

We will choose $\rho_z = \rho_0$, the value for $\rho$ which
minimizes $a(\rho)$. We have the geodesic segments $\{\gamma_n\}$
converging to the geodesic $\gamma_\infty$ defined by
$\dot\gamma_\infty(0) = \bold R$, $\gamma_\infty: [0,\omega-\rho_0)
\to M$.  We will show condition (**) of Proposition 6.1 holds if we
select $t_n = \rho_n-\rho_0$ for each $n$ (so $\gamma_\infty(t_n) =
(\rho_n,p_\infty)$); that is to say, we must show, for all $x \in
M$,
$$\lim_{n \to \infty}\bigl(d(z,\gamma_\infty(t_n)) -
d(x,\gamma_\infty(t_n))\bigr) = 
\lim_{n \to \infty}\bigl(d(z,x_n) -
d(x,x_n)\bigr). \tag **$$
We will do this by showing
$$\lim_{n \to \infty}\bigl(d(z,\gamma_\infty(t_n)) -
d(z,x_n)\bigl) = 0 \tag\dag$$
and
$$\lim_{n \to \infty}\bigl(d(x,\gamma_\infty(t_n)) -
d(x,x_n)\bigl) = 0. \tag\ddag$$
(Showing (\dag) and (\ddag) will suffice, whether the left side of
(**) is finite or infinite.) As can be seen, (\dag) is just a
specialization of (\ddag) to the case of $x = z$; thus (\ddag) is
all we need be concerned with.

We will need the following lemma relating the distance function
$d$ in $M$ to the distance function $d_K$ in $K$.

\proclaim{Lemma 6.4}  Given four points in $K$, $p_1$, $p_2$,
$p'_1$, and $p'_2$ satisfying $d_K(p_1,p_2) \le d_K(p'_1,p'_2)$,
and any two numbers $\rho_1$ and $\rho_2$ in $(\alpha,\omega)$, 
$$d((\rho_1,p_1),(\rho_2,p_2)) \le
d((\rho_1,p'_1),(\rho_2,p'_2)).$$
\endproclaim

\demo{Proof of Lemma} Let $z_i = (\rho_i,p_i)$ and $z'_i =
(\rho_i,p'_i)$  for $i = 1,2$.  We must show $d(z_1,z_2) \le
d(z'_1,z'_2)$.  Let
$\gamma'$ be a minimizing geodesic from $z'_1$ to $z'_2$.  We know
we can parametrize $\gamma'$ by $\rho$, save that we may have to
split it into two segments.  Let us first assume that we do not
need two segments; then, taking $\rho_1 \le \rho_2$, we know that
$\gamma'$ is expressible as $\gamma'(\rho) =
(\rho,\sigma'(\alpha(\rho)))$, where $\sigma'$ is a geodesic in
$K$ from $p'_1$ to $p'_2$ and $\alpha$ is some reparametrization;
if we take $\sigma'$ to be unit-speed, then $\alpha(\rho_1) = 0$
and $\alpha(\rho_2) = L(\sigma')$.  (Note that it is entirely
possible that $\sigma'$ is not a minimizing geodesic, even
though $\gamma'$ is.)  We have $d(z'_1,z'_2) = L(\gamma') =
\int_{\rho_1}^{\rho_2}
\sqrt{1+a(\rho)^2\dot\alpha(\rho)^2} \,d\rho$ (with $\dot\alpha
= d\alpha/d\rho$).

Let $\sigma$ be a minimizing unit-speed geodesic in $K$ from $p_1$
to $p_2$.  We know $L(\sigma) = d_K(p_1,p_2) \le d_K(p'_1,p'_2)
\le L(\sigma')$; let $f = L(\sigma)/L(\sigma')$.  Define the curve
$\gamma:[\rho_1,\rho_2] \to M$ by $\gamma(\rho) = (\rho,
\sigma(f\alpha(\rho)))$; then
$\gamma$ goes from $z_1$ to $z_2$.  We have $d(z_1,z_2) \le
L(\gamma) = \int_{\rho_1}^{\rho_2} \sqrt{1 +
a(\rho)^2 f^2\dot\alpha(\rho)^2}\,d\rho \le L(\gamma') =
d(z'_1,z'_2)$, as desired.

Now suppose that $\gamma'$ is expressed as a portion $\beta'$
parametrized by $\rho$ on $[\rhm',\rho'_1]$ and a portion
$\delta'$ parametrized by $\rho$ on $[\rhm',\rho'_2]$.  We still
have a geodesic $\sigma'$ in $K$ from $p'_1$ to $p'_2$ providing
(after reparametrization) the second coordinate for each of the
segments $\beta'$ and $\delta'$; the only difference  from before
is that the reparametrization in $\beta'$ runs backwards.  The
same technique as above yields a curve $\gamma$ from $z_1$ to
$z_2$, using the same parametrizations by $\rho$ as $\gamma'$
uses, again yielding $L(\gamma) \le L(\gamma')$, whence $d(z_1,z_2)
\le d(z_1',z_2')$.
\qed
\enddemo

We need to show $\lim_{n\to\infty}\bigl(d(x,\gamma_\infty(t_n)) -
d(x,x_n)\bigl)\, = 0$. Represent $x$ as $x = (\rho_x,p_x)$.  

Suppose $\limsup\bigl(d(x,\gamma_\infty(t_n)) -
d(x,x_n)\bigl)\, > 0$, \ie, for some subsequence, for some
$\epsilon > 0$, for all $k$, $d(x,\gamma_\infty(t_{n_k})) \ge
d(x,x_{n_k}) + \epsilon$.  We will define a sequence of points
$\{q_k\}$ approaching $p_x$ with distances from $p_\infty$
mirroring the distances of $p_{n_k}$ from $p_x$; the idea is to be
able to apply Lemma 6.4 to each quartet of points
$\{(q_k,p_\infty),(p_{n_k},p_x)\}$.  If $p_x = p_\infty$, then we
just let each $q_k = p_\infty$ (thus clearly yielding
$d_K(q_k,p_\infty) \le d_K(p_{n_k},p_x)$).  If $p_x
\neq p_\infty$, then we will choose the points $\{q_k\}$ along a
path from $p_\infty$ to $p_x$.  Specifically, let
$\sigma$ be a  minimizing geodesic in
$K$ from $p_\infty$ to $p_x$ with $\sigma(0) = p_\infty$ and
$\sigma(1) = p_x$. Define $q_k = \sigma (s_k)$ where $s_k =
\min\{1,d_K(p_x,p_{n_k})/d_K(p_x,p_\infty)\}$; note $\{s_k\}$
converges to 1.  Then we have $d_K(q_k,p_\infty) = 
s_kd_K(p_x,p_\infty) \le d_K(p_x,p_{n_k})$ for all $k$, while
$\{q_k\}$ converges to
$p_x$.

For all $k$, let $y_k = (\rho_x,q_k)$.  Let $c_k$ be the curve
from $x$ to $y_k$ of constant $\rho$ given by $c_k(s) =
(\rho_x,\sigma(1-s))$ (defined for $s \le 1-s_k$), and let
$\delta_k$ be a minimizing geodesic in $M$ from $y_k$ to
$\gamma_\infty(t_{n_k})$; finally, let
$\beta_k$ be the concatenation of $c_k$ with $\delta_k$, a curve
from $x$ to $\gamma_\infty(t_{n_k})$.  On the one hand, we
have $L(\beta_k) = L(c_k) + d(y_k,\gamma_\infty(t_{n_k})) \ge
d(x,\gamma_\infty(t_{n_k})) \ge d(x,x_{n_k}) + \epsilon$ (that last
inequality by assumption), so $d(y_k,\gamma_\infty(t_{n_k})) \ge
d(x,x_{n_k})  + \epsilon - L(c_k)$; thus, for $k$
sufficiently large, $d(y_k,\gamma_\infty(t_{n_k})) \ge
d(x,x_{n_k})  +  \epsilon/2$ (as $\{L(c_k)\}$ converges to 0).  On
the other hand, we can apply Lemma 6.4 to $y_k = (\rho_x,q_k)$,
$\gamma_\infty(t_{n_k}) = (\rho_{n_k},p_\infty)$, $x =
(\rho_x,p_x)$, and $x_{n_k} = (\rho_{n_k},p_{n_k})$, since
$d_K(q_k,p_\infty) \le d_K(p_x,p_{n_k})$; that yields 
$d(y_k,\gamma_\infty(t_{n_k}))\le d(x,x_{n_k})$ for all $k$.  Thus
the supposition leads to a contradiction.

Now suppose $\liminf \bigl(d(x,\gamma_\infty(t_n)) -
d(x,x_n)\bigl)\, < 0$, \ie, for some subsequence, for some
$\epsilon > 0$, for all $k$, $d(x,\gamma_\infty(t_{n_k})) \le
d(x,x_{n_k}) - \epsilon$.  Let $\Delta = d_K(p_x,p_\infty)$.  This
time choose the sequence $\{q_k\}$ convergent to
$p_x$ so that for all $k$, $q_k$ is no more than $K$-distance
$\Delta$ from $p_{n_k}$; since $\{p_{n_k}\}$
converges to $p_\infty$, this can be accomplished (if $p_{n_k}$ is
$\delta_k$ distant from $p_\infty$, it is no more than $\Delta +
\delta_k$ distant from $p_x$, so the $\Delta$-ball around $p_{n_k}$
contains points within $2\delta_k$ of $p_x$; let $q_k$ be one
such).   As before, let $y_k = (\rho_x,q_k)$ and $c_k$ be the
$\rho$-constant curve from $x$ to $y_k$.  But this time let
$\delta_k$ be a minimizing geodesic in $M$ from $y_k$ to $x_{n_k}$,
with $\beta_k$ the concatenation of
$c_k$ with $\delta_k$, running from $x$ to $x_{n_k}$.  Then we have,
by assumption, $d(x,\gamma_\infty(t_{n_k})) + \epsilon \le
d(x,x_{n_k}) \le L(\beta_k) = d(y_k,x_{n_k}) + L(c_k)$, so
$d(y_k,x_{n_k}) \ge d(x,\gamma_\infty(t_{n_k})) + \epsilon -
L(c_k)$; thus, for $k$ sufficiently large, $d(y_k,x_{n_k}) \ge
d(x,\gamma_\infty(t_{n_k})) + \epsilon/2$.  Again we employ Lemma
6.4:  We have the same coordinates in $\R$ and $K$ as before, but
this time $d_K(q_k,p_{n_k}) \le d_K(p_x,p_\infty)$, so we
conclude $d(y_k,x_{n_k}) \le d(x,\gamma_\infty(t_{n_k}))$.  Thus
this supposition also leads to contradiction.  (Actually, this part
can be proved more generally, \ie, without reference to Lemma 6.4;
but that achieves no simplicity of proof.)

The only remaining possibility is $\lim
\bigl(d(x,\gamma_\infty(t_n)) - d(x,x_n)\bigl)\, = 0$.  This
establishes the hypotheses in Proposition 6.1.  Thus, $\BB(M)$
is the same in the chronological topology as in the
function-space topology, and all its elements come from
geodesics---indeed, geodesics of the form $\gamma_\infty$, \ie,
radial geodesics of the form $\gamma_p^+ : [0,\omega-\rho_0) \to
M$ for $p \in K$, $\gamma_p^+(s) = (\rho_0+s,p)$, and $\gamma_p^-
: [0,\rho_0-\alpha) \to M$, $\gamma_p^-(s) = (\rho_0-s,p)$. 
Thus, $\BB(M)$ consists of $B_\omega = \{[b_{\gamma_p^+}] \st p
\in K \}$ and $B_\alpha  =  \{[b_{\gamma_p^-}] \st p
\in K \}$.  We need to see how integral conditions on $a$
determine these spaces to be either $K$ or a single point.  

\remark
{\bf Part II: Examining the boundary}\endremark

We will confine discussion to $B_\omega$, as the arguments are
identical for $B_\alpha$.  We will let $b_p = b_{\gamma_p^+}$.
Note that each $b_p$ is finite, as $\gamma^+_p$ is a ray.

We will need to make use of integral functions involved in the
distance formulas in $M$ and in $K$.  

Recall the  function
$$J_{\rho_1}^{\rho_2}(\theta) = 
\int_{\rho_1}^{\rho_2}
\frac{d\rho}{\sqrt{1-\left( \frac{a(\rho_1)\sin\theta}{a(\rho)}
\right)^2}}\;,$$
defined for $\rho_0 \le \rho_1 \le \rho_2$; for a geodesic
$\gamma$ in $M$ from $(\rho_1,p_1)$ to $(\rho_2,p_2)$ with
initial radial angle of $\theta$, if $\theta \le \pi/2$, then
$L(\gamma) = J_{\rho_1}^{\rho_2}(\theta)$.  Note that
$J_{\rho_1}^{\rho_2}$ is strictly increasing on $[0,\pi/2]$.

We know that the projection to $K$ of any
$M$-geodesic is a repara\-metrized $K$-geodesic.  Thus, we can
characterize all unit-speed geodesics in $M$ starting from a
point $x_1 = (\rho_1,p_1)$ as being of the form
$\gamma_\theta^\sigma$, where $\sigma$ is any unit-speed geodesic
in $K$ starting from $p_1$ and $\theta$ is any angle in $[0,\pi]$,
and $\gamma_\theta^\sigma$ is defined by
$\gamma_\theta^\sigma(s) = (r_\theta(s),c_\theta^\sigma(s))$ with
$c_\theta^\sigma = \sigma \circ t_\theta$, where $r_\theta$
and $t_\theta$ are functions characterized by
$$\alignat 2
(\dot r_\theta)^2 & =
1-\left(\frac{a(\rho_1)\sin\theta}{a(r_\theta)}
\right)^2, & \quad r_\theta(0) & = \rho_1 \\
\dot t_\theta & = \dfrac{a(\rho_1)\sin\theta}{a(r_\theta)^2}\,, 
& \quad t_\theta(0) & = 0.
\endalignat$$
(Note that $|\dot c_\theta^\sigma| = \dot t_\theta$, so we are
just using the same differential equations as before.)  
Assuming $\theta \le \pi/2$, let $\bar c$ be the reparametrization
of $c^\sigma_\theta$ by $\rho$. Then we have
$$\align
L(\sigma) & = L(c^\sigma_\theta) = L(\bar c) \\
& = \int_{\rho_1}^{\rho_2}  \left|\frac{d\bar c}{d\rho}\right|
\,d\rho \\
& = \int_{\rho_1}^{\rho_2}
\left|\frac{ds}{d\rho}\frac{dc^\sigma_\theta}{ds}\right|
\,d\rho\\   
& = \int_{\rho_1}^{\rho_2}
\frac{|\dot c^\sigma_\theta|}{\left|\frac{d\rho}{ds}\right|}
\,d\rho \\
& = \int_{\rho_1}^{\rho_2}
\frac{\dot t_\theta}{\dot r_\theta} \,d\rho \\
& = \int_{\rho_1}^{\rho_2}
\frac1{{\sqrt{1-\left( \frac{a(\rho_1)\sin\theta}{a(\rho)}
\right)^2}}}\frac{a(\rho_1)\sin\theta}{a(\rho)^2}
\,d\rho \\
& = \int_{\rho_1}^{\rho_2}\frac{d\rho}
{a(\rho)\sqrt{\left(
\frac{a(\rho)}{a(\rho_1)\sin\theta} \right)^2-1}} \; =
I_{\rho_1}^{\rho_2}(\theta).
\endalign$$
Note $I_{\rho_1}^{\rho_2}$, like $J_{\rho_1}^{\rho_2}$, is strictly
increasing on $[0,\pi/2]$. 

It is also convenient to express $I_{\rho_1}^{\rho_2}$ thus:
$$I_{\rho_1}^{\rho_2}(\theta) = a(\rho_1)\sin\theta 
\int_{\rho_1}^{\rho_2}\frac{d\rho}{a(\rho)^2\sqrt{1-\left(
\frac{a(\rho_1)\sin\theta}{a(\rho)}\right)^2}}.$$

For an arbitrary geodesic $\gamma$ in $M$, its projection to
$K$ is a reparametrization of a geodesic $\sigma_\gamma$ in
$K$.  In general, if $\gamma$ is minimizing, it does not
necessarily follow that $\sigma_\gamma$ is minimizing.  But it
does if the initial radial angle $\theta_\gamma$ is no more than
$\pi/2$.

\proclaim{Lemma 6.5} Let $\gamma$ be a geodesic in $M$ from
$x_1 = (\rho_1,p_1)$ to $x_2 = (\rho_2,p_2)$ with $\rho_0\le\rho_1
\le \rho_2$.  Assume the initial radial angle $\theta_\gamma \le
\pi/2$.  If $\gamma$ is minimizing, then so is the projected
geodesic $\sigma_\gamma$ in $K$. \endproclaim 

\demo{Proof of lemma} Suppose $\sigma_\gamma$ is not minimizing;
then there is a geodesic $\bar\sigma$ from $p_1$ to $p_2$ which is
minimizing.  We have $L(\bar\sigma) < L(\sigma_\gamma) =
I_{\rho_1}^{\rho_2}(\theta_\gamma)$.  Then there is a $\bar\theta <
\theta_\gamma$ with $I_{\rho_1}^{\rho_2}(\bar\theta) =
L(\bar\sigma)$.  Note that $\gamma_{\bar\theta}^{\bar\sigma}$ is a
geodesic from $x_1$ to $x_2$.  We have
$L(\gamma_{\bar\theta}^{\bar\sigma}) =
J_{\rho_1}^{\rho_2}(\bar\theta) <
J_{\rho_1}^{\rho_2}(\theta_\gamma) = L(\gamma)$, implying $\gamma$
is not minimizing.
\qed \enddemo

It follows that if there is a minimizing geodesic $\gamma$
from $(\rho_1,p_1)$ to $(\rho_2,p_2)$ which has initial radial
angle $\theta_\gamma \le \pi/2$, then the distance between $p_1$ and
$p_2$ is given by $d_K(p_1,p_2)
= I_{\rho_1}^{\rho_2}(\theta_\gamma)$, since Lemma 6.5 yields that
the projection $\sigma_\gamma$ is minimizing.

\remark{Note}  While $\sigma_\gamma$, for minimizing $\gamma$ but
$\theta_\gamma > \pi/2$, need not be minimizing, it never the less
follows from Lemma 6.5 that $\sigma_\gamma$ is the union of at most
two minimizing geodesics:  For consider that $\gamma$ is the union
of two segments, $\gamma_1$ and $\gamma_2$, with $\gamma_1$ going
from $(\rho_1,p_1)$ to some $(\rho_{\min},\bar p)$ and $\gamma_2$
from $(\rho_{\min},\bar p)$ to $(\rho_2,p_2)$.  Note that
$\theta_{\gamma_2} = \pi/2$, so $\sigma_{\gamma_2}$ is minimizing. 
Now consider that $\gamma_1$ can be considered in its reverse
parametrization, $-\gamma_1$, running from $(\rho_{\min},\bar
p)$ to $(\rho_1,p_1)$; this produces $\theta_{-\gamma_1} = \pi/2$, so
$\sigma_{-\gamma_1}$ is minimizing.  Thus, $\sigma_\gamma$ is
the concatenation of the two minimizing geodesics
$-\sigma_{-\gamma_1}$ and $\sigma_{\gamma_2}$.  
(This perspective will be useful in the consideration of
Reissner-Nordstr\"om.)
\endremark

\remark
{\bf IIa: Infinite integral}\endremark

Suppose first that $\int_{\rho_0}^\omega 1/a(\rho)^2 \,d\rho =
\infty$; we need to see that all $[b_p]$ are the same. 

First note that we must have $\omega = \infty$:  Since $a$ is
increasing on $[\rho_0,\omega)$, $1/a(\rho)^2$ is decreasing
for $\rho$ approaching $\omega$, so only integration over
an infinite interval can yield an infinite integral.  

Let us evaluate $b_p$ on an arbitrary point $(\bar\rho,q)$:
$$\align
b_p((\bar\rho,q)) & = \lim_{s\to\infty} (s - 
d((\rho_0+s,p), (\bar\rho,q))) \\ 
& = \lim_{s\to\infty} (-\rho_0 + s - d((s,p), (\bar\rho,q))) 
\\ & = \lim_{s\to\infty} \(-\rho_0 + s - \oint_{\bar\rho}^s
\frac{d\rho}{\sqrt{1-\(\frac
{\bar a\sin\theta_s}{a(\rho)}\)^2}}\),
\endalign$$
where $\bar a = a(\bar\rho)$, $\theta_s$ is the initial radial
angle for a minimizing geodesic from
$(\bar\rho,q)$ to $(s,p)$, and $\oint$ represents either one or
two integrals depending on whether $\theta_s$ is, respectively,
$\le \pi/2$ or $> \pi/2$.
 
Let us first assume that eventually $\theta_s \le \pi/2$.  Then
$$\align
b_p((\bar\rho,q)) & = -\rho_0 + \lim_{s\to\infty}
\(s - \int_{\bar\rho}^s
\frac{d\rho}{\sqrt{1-\(\frac{\bar a\sin\theta_s}{a(\rho)}\)^2}}\) \\
& = -\rho_0 + \bar\rho + \lim_{s\to\infty} \int_{\bar\rho}^s\(1-
\frac1{\sqrt{1-\(\frac{\bar a\sin\theta_s}{a(\rho)}\)^2}}\)d\rho.
\endalign$$
Let $A(\theta) = \sqrt{1-\(\frac{\bar a\sin\theta}{a(\rho)}\)^2}$;
then, using $1-\dfrac1{\sqrt{1-x}} =
\dfrac{-x}{\sqrt{1-x}\(\sqrt{1-x}+1\)}$, we get
$$b_p((\bar\rho,q)) = -\rho_0 + \bar\rho - 
\lim_{s\to\infty}\(
\bar a^2\sin^2\theta_s\int_{\bar\rho}^s\frac{d\rho}{a(\rho)^2
A(\theta_s)(A(\theta_s)+1)}\).$$
Let $J_s$ denote the integral above.  We have $A(\theta_s) \le 1$, 
so $J_s \ge (1/2)\int_{\bar\rho}^s\frac{d\rho}{a(\rho)^2}$;
therefore, our integral assumption on $a$ implies that
$\lim_{s\to\infty} J_s = \infty$.  Since we know that $b_p$ cannot
have $-\infty$ as a value, we conclude that $\sin\theta_s$ must
approach 0.  Since $\theta_s \le \pi/2$, this means
$\lim_{s\to\infty} \theta_s = 0$.

From Lemma 6.5, we know that for all $s$, $d_K(q,p) =
I_{\bar\rho}^s(\theta_s)$, i.e.,
$$d_K(q,p) = \bar a\sin\theta_s\int_{\bar\rho}^s
\frac{d\rho}{a(\rho)^2A(\theta_s)}.$$  
Denote this integral by $I_s$; we have $\lim_{s\to\infty} I_s
= \infty$, also. Solving this equation for
$\bar a\sin\theta_s$ and inserting in the equation above it yields
$$\align
b_p((\bar\rho,q)) & = -\rho_0 + \bar\rho - \lim_{s\to\infty}
\(\frac{d_K(q,p)}{I_s}\)^2J_s \\
& = -\rho_0 + \bar\rho - d_K(q,p)^2 \lim_{s\to\infty}
\(\frac{J_s}{I_s}\)\frac1{I_s}.
\endalign$$
The only difference between $J_s$ and $I_s$ is that $J_s$ has an
additional factor of $1/(A(\theta_s) + 1)$ in the integrand;
therefore, $J_s \le I_s$.  Since $I_s$ has $\infty$ as limit, we
conclude $$b_p((\bar\rho,q)) = -\rho_0 + \bar\rho$$ under the
assumption that eventually $\theta_s \le \pi/2$.

Now we look at the possibility that for some sequence of numbers
$\{s_n\}$ approaching $\infty$, $\theta_{s_n} > \pi/2$ for all
$n$.  We proceed much as before, utilizing the fact that
for $\theta > \pi/2$, $\oint_\alpha^\beta = \int_\alpha^\beta +
2\int_{\rho_{\text{min}}}^\alpha$, where $a(\rho_{\text{min}}) =
\bar a\sin\theta$. Using $\theta_n$ for $\theta_{s_n}$ and
$\rho_n$ for $a^{-1}(\bar a\sin\theta_n)$, we get
$$\align
b_p((\bar\rho,q)) & = \lim_{n\to\infty}\(-\rho_0 + s_n 
-\oint_{\bar\rho}^{s_n}\frac{d\rho}{A(\theta_n)}\) \\
& =  \lim_{n\to\infty}\(-\rho_0 + s_n 
-\int_{\bar\rho}^{s_n}\frac{d\rho}{A(\theta_n)}
-2\int_{\rho_n}^{\bar\rho}\frac{d\rho}{A(\theta_n)}\) \\
& = -\rho_0 + \bar\rho - 
\lim_{n\to\infty}\(
\bar a^2\sin^2\theta_n\int_{\bar\rho}^{s_n}\frac{d\rho}{a(\rho)^2
A(\theta_n)(A(\theta_n)+1)}\) \\
& \qquad\qquad\quad - 
\lim_{n\to\infty}2\int_{\rho_n}^{\bar\rho}\frac{d\rho}{A(\theta_n)}.
\endalign$$
The only difference from before is the last term.  That has no
effect on the observation that since $b_p$ cannot take $-\infty$ as
a value, we must have $\sin\theta_n$ approach 0 (which, under the
current assumption, means $\theta_n$ approaches $\pi$).  But 
that means $a(\rho_n) = \bar a \sin\theta_n$ approaches 0, which is
impossible, as $a(\rho) \ge a(\rho_0)$ for all $\rho$.  We conclude
that we cannot have the sequence $\{s_n\}$ after all.

It follows that $b_p((\bar\rho,q)) = -\rho_0 + \bar\rho$ for all
$p$, $\bar\rho$, and $q$, \ie, $b_p$, up to a constant, is just
projection onto the first coordinate---independent of $p$. 
Therefore, all $[b_p]$ are equal: $B_\omega$ is a point.

\remark
{\bf IIb: Finite integral}\endremark

Now suppose $\int_{\rho_0}^\omega 1/a(\rho)^2\,d\rho < \infty$; we
need to see that $B_\omega = \{[b_p] \st p \in K\}$ is homeomorphic
to $K$.  First note that each $c_p : s \mapsto (\rho_0 + s, p)$ is
a ray, so $b_p$ is a finite Busemann function; thus, we have the
map $i: K \to B_\omega$ given by $i: p \mapsto [b_p]$, and this is
onto in virtue of the applicability of Proposition 6.1.  Our task
is to show $i$ is continuous and injective and that $i^{-1}$ is
continuous.

\remark
{\it Showing $i$ is continuous}\endremark

We can express $i$ as $i = \pi \circ j$, where $j : K \to \Cal B(M)$
is the map into the space of Busemann functions, $j : p \mapsto
b_p$, and $\pi$ is the quotient by the
$\euc{}$-action.  We need to show that $j$ is continuous; as
$V_0^*$ carries the quotient topology from the space of Busemann
functions, $\pi$ is necessarily continuous, so that will suffice. 
Again using Proposition 6.1, we know that the
function-space topology on $\Cal B(M)$ is appropriate:  We need to
show that as $p'$ approaches $p$, $b_{p'}$ approaches $b_p$ in a
pointwise fashion.  We will accomplish this by showing, for any
point $x = (\rho_x,p_x) \in M$,
$$|b_p(x)-b_{p'}(x)| \le a(\rho_x)d_K(p,p').$$

We will not need the exact nature of the distance function in $M$,
except for making use of Lemma 6.4.  Since $\displaystyle b_p(x) =
\lim_{s\to\omega}(s-d(x,(\rho_0+s,p))) =
-\rho_0 + \lim_{\rho\to\omega-\rho_0}(\rho-d(x,(\rho,p)))$ and
similarly for $b_{p'}(x)$, what we need to show is for all $\rho$,
$$|d(x,(\rho,p)) - d(x,(\rho,p'))| \le
a(\rho_x)d_K(p,p').$$
We will show $d(x,(\rho,p)) \le a(\rho_x)d_K(p,p') +
d(x,(\rho,p'))$ by exhibiting a point $q \in K$ such that 
$$\align
d(x,(\rho_x,q)) & \le a(\rho_x)d_K(p,p') \quad \text{and} \\
d((\rho_x,q),(\rho,p)) & \le d(x,(\rho,p')).
\endalign$$
Reversing $p$ and $p'$ will then establish $d(x,(\rho,p')) \le 
a(\rho_x)d_K(p,p') + d(x,(\rho,p))$, and we will be done.

Clearly, $d(x,(\rho_x,q)) = d((\rho_x,p_x),(\rho_x,q)) \le
a(\rho_x)d_K(p_x,q)$ (just consider the constant-$\rho$ curve
between the points); we still need $d_K(p_x,q) \le d_K(p,p')$. In
order to arrive at
$d((\rho_x,q),(\rho,p))
\le d((\rho_x,p_x),(\rho,p'))$, Lemma 6.4 says that all we need to
do is to ensure $d_K(q,p) \le d_K(p_x,p')$.  Thus, the problem
reduces to finding $q \in K$ so that
$$\align
d_K(p_x,q) & \le d_K(p,p') \quad \text{and} \\
d_K(q,p) & \le d_K(p_x,p').
\endalign$$
Consider a minimizing geodesic $\sigma$ from $p_x$ to $p$, with
$\sigma(0) = p_x$ and $\sigma(1) = p$; we will locate $q$ along
$\sigma$. Let $l = d_K(p_x,p')$ and $\bar l = d_K(p,p')$.  For
each $t \in [0,1]$, let $l_t$ be the length of $\sigma$ from 0 to $t$
and $\bar l_t$ its length from $t$ to 1; then $l_t + \bar l_t =
d_K(p_x,p) \le l + \bar l$.  If we can find $t$ so that $l_t \le \bar
l$ and $\bar l_t \le l$, then $q = \sigma(t)$ will do the job. 

If $L(\sigma) \ge \bar l$, then there is some point $\sigma(t)$ along
$\sigma$ so that $l_t = \bar l$; we then have $\bar l_t \le l$,
and we are done.  If, on the other hand, $L(\sigma) < \bar l$, then
let $t = 1$: $l_1 = L(\sigma) < \bar l$ and $\bar l_1 = 0$, and again
we are done.

\remark
{\it Showing $i$ is injective}\endremark

We must show that for $p \neq p'$, $[b_p] \neq [b_{p'}]$ , \ie, that
$b_p$ and $b_{p'}$ do not differ merely by a constant.  We will
evaluate $b_p - b_{p'}$ on two points: $x_0 = (\rho_0,p)$ and
$x'_0 = (\rho_0,p')$.  

It's fairly easy to evaluate $b_p((\rho_0,q))$ for any $q$:
$$\align
b_p((\rho_0,q)) & = \lim_{s\to\omega-\rho_0}
(s - d((\rho_0,q),(\rho_0 + s,p))) \\
& = -\rho_0 + \lim_{s\to\omega} \(s - \int_{\rho_0}^s
\frac1{A(\theta^q_s)}\,d\rho\),
\endalign$$
where $A(\theta) = \displaystyle
\sqrt{1-\(\frac{a_0\sin\theta}{a(\rho)}\)^2}$ for $a_0 =
a(\rho_0)$, and $\theta^q_s$ is the initial radial angle for a
minimizing geodesic from $(\rho_0,q)$ to $(s,p)$.  Note that we are
justified in using $\int$ instead of the more general $\oint$,
since a geodesic starting from $\rho = \rho_0$ cannot have an
initial radial angle greater than $\pi/2$, if it is to reach any
points at all in $M^+$.  

Also worth noting: (1)  For any $\bar\rho \in (\rho_0,\omega)$,
$I_{\rho_0}^{\bar\rho}$ is increasing (and continuous) on
$[0,\pi/2)$, with $I_{\rho_0}^{\bar\rho}(0) = 0$ and
$I_{\rho_0}^{\bar\rho}(\pi/2) = \infty$. (2) For any $q \in K$,
$\theta_s^q$ is decreasing in $s$:  For any $s <\omega$, $d_K(q,p) =
I_{\rho_0}^s(\theta_s^q)$, so $s_1 < s_2$ implies
$I_{\rho_0}^{s_1}(\theta_{s_1}^q) = d_K(q,p) =
I_{\rho_0}^{s_2}(\theta_{s_2}^q) >
I_{\rho_0}^{s_1}(\theta_{s_2}^q)$, whence $\theta_{s_1}^q >
\theta_{s_2}^q$. (3)  For each $\bar\rho$, there is some $\bar\theta
< \pi/2$ such that $I_{\rho_0}^{\bar\rho}(\bar\theta) =
\text{diam}(K)$:  Just pick $\bar p$ and
$\bar q$ so that $d_K(\bar q,\bar p)$ realizes the diameter of $K$;
$\bar\theta =
\theta_{\bar\rho}^{\bar q}$ for that choice of $\bar p$. 
Then for any choice of $p$ and $q$, $\theta_{\bar\rho}^q \le
\bar\theta$ (since
$I_{\rho_0}^{\bar\rho}(\theta_{\bar\rho}^q) = d(q,p) \le
\text{diam}(K) = I_{\rho_0}^{\bar\rho}(\bar\theta)$).

We clearly have $\theta^p_s = 0$ for all
$s$, so $A(\theta^p_s) = 1$ (we could, of course, just note
$b_p((\rho_0,p)) = 0$, but it is convenient to treat it this way). 
Thus,
$$(b_p-b_{p'})(x_0) = \lim_{s\to\omega} 
\int_{\rho_0}^s\(\frac1{A(\theta^{p'}_s)}-1\)d\rho.$$
Using $\displaystyle \frac1{\sqrt{1-x}} - 1 =
\frac{x}{\sqrt{1-x}\(1 + \sqrt{1-x}\,\)}$, we get (writing $A_s$
for $A(\theta^{p'}_s)$),
$$(b_p-b_{p'})(x_0) = \lim_{s\to\omega}
\(a_0^2\sin^2\theta^{p'}_s\int_{\rho_0}^s 
\frac{d\rho}{a(\rho)^2A_s(1 + A_s)}\).$$
By Lemma 6.5, we can express the distance between $p$ and $p'$ in
terms of a minimizing geodesic from $x_0$ to $(s,p')$:
$$d_K(p,p') = a_0\sin\theta^{p'}_s\int_{\rho_0}^s
\frac{d\rho}{a(\rho)^2A_s}.$$ 
Solving for $a_0\sin\theta^{p'}_s$ and using that in the expression
above, we get
$$(b_p-b_{p'})(x_0) = d_K(p,p')^2 \lim_{s\to\omega}
\frac{\int_{\rho_0}^s\frac{d\rho}{a(\rho)^2A_s(1+A_s)}}
{\(\int_{\rho_0}^s\frac{d\rho}{a(\rho)^2A_s}\)^2}.$$
Since $a(\rho) \ge a(\rho_0)$, we have $1 \ge A_s \ge
\cos\theta^{p'}_s$, yielding  
$$\align 
(b_p-b_{p'})(x_0) & \ge 
d_K(p,p')^2 \lim_{s\to\omega}
\frac {\frac12 \int_{\rho_0}^s \frac{d\rho}{a(\rho)^2}}
{\(\int_{\rho_0}^s\frac{d\rho}{a(\rho)^2}\)^2\sec^2\theta_s^{p'}} \\
& = d_K(p,p')^2 \frac{\cos^2\theta^{p'}_\omega}
{2\int_{\rho_0}^\omega \frac{d\rho}{a(\rho)^2}}.  
\endalign$$

Pick some $\bar\rho$ with $\rho_0 < \bar\rho < \omega$.  As in
(3) above, there is some maximum value $\bar\theta$ to
$\theta^q_{\bar\rho}$ for all $q
\in K$, $\bar\theta < \pi/2$; note that $\theta^{p'}_\omega \le
\theta^{p'}_{\bar\rho}$ ($\theta^q_s$ is decreasing in $s$). With
$\int_{\rho_0}^\omega 1/a(\rho)^2\,d\rho = I < \infty$, this gives
us
$$(b_p-b_{p'})(x_0) \ge \frac{\cos^2\bar\theta}{2I}d_K(p',p)^2.$$
Reversing $p$ and $p'$ gives us
$$(b_{p'}-b_p)(x'_0) \ge \frac{\cos^2\bar\theta}{2I}d_K(p,p')^2$$ or
$$(b_p-b_{p'})(x'_0) \le -\frac{\cos^2\bar\theta}{2I}d_K(p,p')^2 .$$
It follows that for $p \neq p'$, $b_p-b_{p'}$ is non-constant.

\remark
{\it Showing $i^{-1}$ is continuous}\endremark

We must show that given any sequence $\{[b_{p_n}]\}$ converging to
$[b_p]$ in $B_\omega$, that $\{p_n\}$ converges to $p$.  Recall
that in the function-space topology on $\Cal L_1(M)$, we have a
continuous cross-section $\zeta: \Cal L_1(M)/\euc{} \to \Cal
L_1(M)$ given by $\zeta: [f] \mapsto f - f(\bar x)$, where $\bar x$
is any point we care to fix in $M$.  It follows that
$\{\zeta([b_{p_n}])\}$ converges to $\zeta([b_p])$, \ie, $\{b_{p_n}
- b_{p_n}(\bar x)\}$ converges pointwise to $b_p-b_p(\bar x)$. In
particular, we may choose to let
$\bar x = x_0 = (\rho_0,p)$.  This gives us
$\{b_{p_n}-b_{p_n}(x_0)\}$ converging pointwise to
$b_p - b_p(x_0)$.  Apply this to any point $y_0 =
(\rho_0,q)$: $\{b_{p_n}(y_0) - b_{p_n}(x_0) - (b_p(y_0) -
b_p(x_0))\}$ converges to 0.  (Actually, $b_p(x_0) = 0$, but
preserving the symmetry of expression is more clarifying than using
this fact.)

Suppose that $\{p_n\}$ fails to converge to $p$; then there is a
subsequence, which we will also call $\{p_n\}$, such that for all
$n$, $d_K(p_n,p) \ge \delta$ for some $\delta > 0$.  Furthermore,
we can find a further subsequence---still denoted
$\{p_n\}$---converging to some point $q \neq p$.  Applying the
inequalities at the end of the last section yields
$(b_p-b_{p_n})(x_0) \ge (\cos^2\bar\theta/(2I))d_K(p_n,p)^2$ and
$(b_{p_n} - b_p)(x_n) \ge (\cos^2\bar\theta/(2I))d_K(p_n,p)^2$,
where $x_n = (\rho_0,p_n)$.  Therefore, for some $\epsilon > 0$, we
have (for $n$ sufficiently large)
$$\align
b_p(x_0) - b_{p_n}(x_0) & \ge 2\epsilon \quad \text{and} \\
b_{p_n}(x_n) - b_p(x_n) & \ge 2\epsilon.
\endalign$$
Let $y_0 = (\rho_0,q)$, which is $\lim x_n$ for our
subsequence; then, as Busemann functions are Lipschitz-1, that last
inequality yields (for
$n$ sufficiently large that $d(x_n,y_0) < \epsilon/2$)
$$b_{p_n}(y_0) - b_p(y_0) \ge \epsilon.$$
That gives us $b_{p_n}(y_0) - b_{p_n}(x_0) -
(b_p(y_0) - b_p(x_0)) \ge 3\epsilon$, contradicting
$\{b_{p_n}(y_0) - b_{p_n}(x_0) - (b_p(y_0) - b_p(x_0))\}$
converging to 0.  It follows that $\{p_n\}$ must converge to $p$.

Theorem 6 of \cite{H3} establishes the causal/chronological nature
of the lines forming the cones:  A line-element of the cone is the
$\euc{}$-orbit of a Busemann function $b_c$; if $c$ is a curve of
finite length, the line is timelike ($b_c + t \ll b_c + t'$ for $t
< t'$), while if $c$ is of infinite length, the line is null ($b_c
+ t \prec b_c + t'$ for $t < t'$, but no $\ll$ relations).
\qed\enddemo

\proclaim{Corollary 6.6} Let $V$ be a spacetime with topology
$\euc1\times(r_{\text{min}},r_{\text{max}})\times\sph2$
($r_{\text{min}} > 0$) and metric $g = -f(r)dt^2 + (1/f(r))dr^2 +
r^2d\,\sph2$, where $f$ is some positive function on
$(r_{\text{min}},r_{\text{max}})$ and
$d\,\sph2$ is the round metric on the sphere.  Let $\phi(r) =
r^2/f(r)$.  If $\phi$ has minimum value at $r_0$ and is
decreasing on $(r_{\text{min}},r_0)$ and increasing on
$(r_0,r_{\text{max}})$, then the future-completion of
$V$ (in the future chronological topology) is Hausdorff, and the
future causal boundary consists of two cones on
$\sph2$---corresponding to going out to infinity in the $r =
r_{\text{min}}$ and $r = r_{\text{max}}$ directions,
respectively---conjoined at their common vertex.  
The $r_{\max}$ cone is null if $\int_{r_0}^{r_{\max}} 1/f(r) dr =
\infty$, otherwise timelike; similarly for the $r_{\min}$ cone. 

If $V = \euc1\times(0,r_{\max})\times\sph2$ with metric, $f$,
$\phi$, and $r_0$ all as above, then the result changes
only in that the $r_{\min}$ cone is replaced by a single line---null
if $\int_0^{r_0} 1/f(r) dr = \infty$, otherwise timelike.
\endproclaim

\demo{Proof} The metric on $V$ is conformal to $\bar g = -dt^2 +
(1/f(r))^2dr^2 + (r^2/f(r))d\,\sph2$.  If we define $\rho$ by
$d\rho = dr/f(r)$, then we have the situation prescribed by the
hypotheses of Theorem 6.2, with $\alpha = 
\int^{r_0}_{r_{\text{min}}} 1/f(r)\,dr$, $\omega = 
\int_{r_0}^{r_{\text{max}}} 1/f(r)\,dr$, and $a(\rho) =
\sqrt{\phi(r)}$. We have $\int_{\rho_0}^\omega \dfrac1{a(\rho)^2}
d\rho = \int_{r_0}^{r_{\text{max}}} \dfrac1{\phi(r)} \dfrac{dr}{f(r)}
= \int_{r_0}^{r_{\text{max}}} \dfrac1{r^2}dr = 1/r_0 -
1/r_{\text{max}}$, which is necessarily finite; similarly for the
other endpoint if
$r_{\min} > 0$.  If $r_{\min} = 0$, then $\int_\alpha^{\rho_0}
1/a(\rho)^2 d\rho = \int_0^{r_0} 1/r^2 dr = \infty$.
\qed \enddemo

We will explore applications of Corollary 6.6 in a discussion
subsection.  Before that, we present an extension of Theorem 6.2 to
more general warping functions; though it appears no classical
spacetimes are examples of this extension, it lends itself to a
useful perspective.

\proclaim{Theorem 6.7} Let $M$ be a Riemannian manifold expressible
as a warped product with a compact manifold, $(\alpha,\omega)
\times_a K$, metric $d\rho^2 + a(\rho)^2j_k$.  Suppose for some
$\rho_- < \rho_+$ in $(\alpha,\omega)$, $a$ is decreasing on
$(\alpha,\rho_-)$ and increasing on $(\rho_+,\omega)$.  Then the
same conclusions from Theorem 6.2 apply:  The Busemann boundary of
$M$ is Hausdorff and consists of spaces $B_\iota$ attaching at
$\{\iota\} \times K$, $\iota = \alpha,\omega$, with $B_\iota$ a
point if $|\int_{\rho_+}^\iota 1/a(\rho)^2 \,d\rho| = \infty$ and
$B_\iota \cong K$ otherwise.  The future causal boundary of
$V = \mk1\times (\alpha,\omega) \times_a K$ is a pair of conjoined
cones over $B_\alpha$ and $B_\omega$, the respective cone being null
if the endpoint ($\alpha$ or $\omega$) is infinite, otherwise
timelike.
\endproclaim

\demo{Proof}  All that was done in Theorem 6.2 with respect to
$(\rho_0,\omega)$ can be done here with $(\rho_+,\omega)$. \qed
\enddemo

An alternative view:  $[\rho_-,\rho_+]$ can be decomposed into a
finite number of subintervals $\bigcup_{1=1}^{2m}
[\rho_{i-1},\rho_i]$ ($\rho_0 = \rho_-$, $\rho_{2m} = \rho_+$) on
each of which $a$ is monotonic. We can apply Thereom 6.2 separately
to $(\alpha,\rho_1) \times_a K$, to $(\rho_1, \rho_3) \times_a K$,
and so on up through $(\rho_{2m-1},\omega) \times_a K$.  The
spacetimes $V_i = \mk1 \times (\rho_{2i-1},\rho_{2i+1}) \times_a K$
each have, as future causal boundary, timelike cones over two
copies of $K$, one each at $\{\rho_{2i-1}\} \times K$ and
$\{\rho_{2i+1}\} \times K$ (and similarly, {\it mutatis
mutandis\/}, for $\mk1 \times (\alpha, \rho_1) \times K$ and $\mk1
\times (\rho_{2m-1},\omega) \times_a K$). Thus, the
future-completions $V^+_i$ fit together quite neatly, boundaries
all matching in topology and causality.  The union of all those
is $V^+$. 

\proclaim{Corollary 6.8} Let $V$ be a spacetime with toplogy $\euc1
\times (r_{\text{min}},r_{\text{max}}) \times \sph2$
($r_{\text{min}} > 0$) and metric $-f(r)dt^2 + (1/f(r))dr^2 +
r^2d\sph2$ for $f$ a positive function on
$(r_{\text{min}},r_{\text{max}})$. If $r^2/f(r)$ is decreasing on
some interval $(r_{\text{min}},r_-)$ and increasing on some
interval $(r_+,r_{\text{max}})$, then the future-completion of $V$
is Hausdorff, and the future causal boundary consists of two
conjoined cones on $\sph2$; the cone at $r_{\text{max}}$ is null if
$\int_{r_0}^{r_{\text{max}}} 1/f(r)\,dr = \infty$, otherwise
timelike (and similarly for the cone at $r_{\text{min}}$).

If $V = \euc{} \times (0,r_{\max}) \times \sph2$ with metric as
above, then the result changes only in that the $r_{\min}$ cone is
replaced by a single line---null if $\int_0^{r_-} 1/f(r) dr =
\infty$, otherwise timelike.
\endproclaim \qed

Further generalizations are possible. For instance, the
same results as in Theorems 6.2 or 6.7 can be proved with $a$ having
the property that for any $\rho$ in its domain, there exist $\rho^-$
and $\rho^+$ in the domain with $\rho^- < \rho < \rho^+$ such that
for any $s < \rho^-$, $a(s) > a(\rho^-)$ and for any $s > \rho^+$,
$a(s) > a(\rho^+)$ (one proves that lengths are realized by
geodesics in the same manner as in Theorem 6.2, and sections I and
IIa follow nearly the same as before; IIb is handled by observing
that a geodesic which starts at a point where $a$ has its global
minimum can be parametrized by $\rho$ on its entire length).  Of far
more practical use are versions where $a$ is monotonic.  This is
much more difficult in general; a version for monotonic $a$ in which
$a$ approaches 0 is discussed in some detail in the discussion
section, at least for the case of $\int 1/a(\rho)^2 d\rho = \infty$,
as this is important for Reissner-Nordstr\"om.

\subhead
Discussion: Three classical spacetimes
\endsubhead

A number of classical spacetimes have sections that have the form
specified in Corollary 6.6.  We will briefly examine three classes
of such spacetimes and also look at what is known by \cite{H4}
about the other sections of those spacetimes and how the sections
fit together.  The object in all of these examinations is not to
discover previously unknown properties, as most of these boundaries
are already widely known or at least suspected; rather, the point
is to show how the methods of this paper and of \cite{H4} can be
easily applied.   But it should be noted that what is established
here are universal results---in the topological sense of
\cite{H2} for spacelike boundaries and at least in the
chronological sense of \cite{H1} for timelike and null boundaries. 
This tells us that any other completions must be derived from these
natural ones, such as by quotients. 

Some of the boundaries we will exhibit are
interior boundaries, in the sense of being boundaries of sections of
the spacetime that, in the maximally extended model, have
another section on the other side, and the boundaries exemplified
here are illustrative for showing how the different sections fit
together.  These are not intended as proofs of the global structure
of these spacetimes, and we are relying on the reader's prior
understanding of the global nature; the point is to show how these
boundary results fit into the general picture.

\subsubhead
Schwarzschild
\endsubsubhead

As mentioned above, the Schwarzschild metric, representing a
spherically symmetric, uncharged black hole of mass $m > 0$ in a
vacuum spacetime which is asymptotically flat, is
$$ds^2 = -\(1-\dfrac{2m}r\)dt^2 + \(1-\dfrac{2m}r\)^{-1}dr^2
+ r^2 d\sph 2$$ 
where $d\sph 2$ denotes the round metric on the sphere.  This has
the form of Corollary 6.6 for $f(r) = 1 - \dfrac{2m}r$, which is
positive for $r$ in $(2m,\infty)$.  Thus, we are looking here at
external Schwarzschild (that is to say, external to the event
horizon), $r > 2m$; the manifold is $\sch_\ext =
\euc{} \times (2m,\infty) \times \sph2$.  For
$\phi(r) = r^2/f(r)$ we easily note that $\phi$ has a minimum at $r
= 3m$, with $\phi$ monotone on either side of that minimum.  This
puts Corollary 6.6 into play.  We have $\int_{3m}^\infty 1/f(r) dr =
\infty$ (just note that $f(r) < 1$) and $\int_{2m}^{3m} 1/f(r) dr =
\infty$ (for $r$ close to $2m$, $1/f(r) = r/(r-2m)$ behaves like
$2m/(r-2m)$).  Therefore $\FB(\sch_\ext)$ is a pair
of conjoined null cones (cones on $\sph2$), one at infinity and one
at the event horizon; the conjunction, a single point, may
reasonably be labled $i^+$, future timelike infinity.

Of course, we fully expect a Minkowski-like future causal boundary
for Schwarzschild, and that's the null cone at infinity---but what's
this other boundary at $r = 2m$?  That is an artifact of cutting
off the full Schwarzschild spacetime at the event horizon; all it
does is tell us that we cut along a null cone to obtain the
exterior sector.  What about looking at the full spacetime---or, at
least, other sectors?  (See \cite{HE} for a thorough discussion of
maximally extended Schwarzschild.)

Let us first examine interior Schwarzschild:  With $0 < r < 2m$,
the metric is perhaps more readily understood as
$$ds^2 = -\(\dfrac{2m}r-1\)^{-1}dr^2 + \(\dfrac{2m}r-1\)dt^2
+ r^2 d\sph 2$$ 
on a manifold $\sch_\intr = (0,2m)
\times \euc{} \times \sph2$, with the time orientation having $0$
as the future endpoint of the first factor (particles fall
inward).  This is analyzed in section 5.2 of \cite{H2}: 
$\sch_\intr$ is conformal to a warped product, metric
$-dr^2 + (\frac{2m}r-1)^2 dt^2 + r^2(\frac{2m}r-1) d\sph2$.  We
apply Proposition 5.2 in \cite{H2}: The spacelike factors $\euc{}$
and $\sph2$ are both complete and the integrals of the warping
functions, near the future endpoint of $(0,2m)$, yield
$\int_0^\epsilon (\frac{2m}r -1)^{-1} dr < \infty$ and
$\int_0^\epsilon r^{-1} (\frac{2m}r -1)^{-\frac12} dr < \infty$. 
This implies $\FB(\sch_\intr)$ is spacelike and is the
product of the spacelike factors of the spacetime, $\euc{} \times
\sph2$.

But this analysis tells us nothing of how $\sch_\intr$
and $\sch_\ext$ fit together.  Since the exterior
portion joins the interior portion at $r= 2m$, which is at the past
endpoint of the timelike factor in $\sch_\intr$, it is
$\PB(\sch_\intr)$ we need to look at.  When we look at
the integrals of the warping functions near the other endpoint, we
discover $\int_{2m-\epsilon}^{2m} (\frac{2m}r-1)^{-1} dr = \infty$
and $\int_{2m-\epsilon}^{2m} r^{-1} (\frac{2m}r-1)^{-\frac12} dr <
\infty$.  Proposition 5.2 of \cite{H2} doesn't apply in this case;
but its extension in Proposition 3.5(b) of \cite{H4} does, telling us
that, with a non-compact spacelike factor having an infinite
integral for its warping function,
$\PB(\sch_\intr)$ has null-related elements.

That proposition concerns the causal boundary of a spacetime of
the form $V = (a,b)\times_{f_1} K_1 \times \dots \times_{f_k} K_k
\times_{f_{k+1}} K_{k+1} \times \dots \times_{f_m} K_m$, where each
$f_i: (a,b) \to \euc{}$ is a positive function with $\int_{b^-}^b
f_i^{-1/2} < \infty$ for $i \le k$ and $\int_{b^-}^b f_i^{-1/2}  =
\infty$ for $i > k$, each $K_i$ is a manifold with a complete
Riemannian metric $h_i$, and the spacetime metric is
$-dt^2 + \sum_i f_i(t) h_i$.  If $K_i$ is non-compact for some $i >
k$, the proof there shows how to construct a null line in $\FB(V)$
out of any choice of point $x^0 \in K^0 =  K_1 \times \dots \times
K_k$ and the Busemann function for any infinite-length ray $c_i \in
K_i$.  This can be amplified a bit in the case
that $m = k+1$:  $\FB(V) = (K^0 \times \Cal B(K_{k+1}) \times
\euc{}) \cup K^0$, where $\Cal B(K_{k+1})$ is the set of equivalence
classes of finite Busemann functions on $K_{k+1}$, the $\euc{}$
factor is a set of null lines on $K^0 \times \Cal B(K_{k+1})$, and
those lines all converge to a spacelike image of
$K^0$; in effect, this is a null cylinder on $K^0 \times \Cal
B(K_{k+1})$, capped off by squeezing the $\Cal B(K_{k+1})$ factor to
a point.  (The picture is more complex when $m > k+1$).

We can apply this for $\PB(\sch_\intr)$ (time-reversing the result
above):  The warping function for the $\sph2$ factor yields a finite
integral, that for the $\euc{}$ factor an infinite integral.  There
are exactly two classes of finite Busemann functions on $\euc{}$,
associated with the two rays, one for each end
of $\euc{}$.  We conclude that $\PB(\sch_\intr)$ consists of a pair
of null cylinders on $\sph2$, one for $t = \infty$ and one for $t =
-\infty$, conjoined in a spacelike $\sph2$, much like an extended
$i^-$.  The $t = \infty$ null cylinder in $\PB(\sch_\intr)$ is
identified with the $r = 2m$ null cone in $\FB(\sch_\ext)$
(external time coordinate is infinite for passage through the event
horizon); as it is only the null lines in either cylinder or cone
which are identified, there is no problem with the vertex of the
cone or the cap on the cylinder.  Thus do
$\sch_\ext$ and $\sch_\intr$ fit together across the evident
topological and causal match of their boundary components (though it
takes other considerations to guarantee that we have a geometric
match).  

The $t = -\infty$ null cylinder in $\PB(\sch_\intr)$ fits
with a corresponding null cone in the future causal boundary of
$\sch_\alt$, the ``alternate universe" sector of
maximally extended Schwarzschild (isometric with
$\sch_\ext$, though $t$-reversed).  The $r=2m$ null cones in 
$\PB(\sch_\ext)$ and $\PB(\sch_\alt)$
fit with the two null cylinders in the future causal boundary of
$\sch_\wth$, the ``white hole" sector of maximal
Schwarzschild (isometric with a time-reversal of
$\sch_\intr$).  The two spacelike $\sph2$
caps---one from $\PB(\sch_\intr)$ and one from
$\FB(\sch_\wth)$---become identified as a sphere in the spacetime at
the juncture of the four sectors. Thus, the entire causal boundary of
$\sch = \sch_\ext \cup \sch_\intr \cup \sch_\alt \cup \sch_\wth$
consists of two null cones in the future boundary joining a future
singularity of $\euc{} \times \sph2$ (with the joins being points,
$i^+$ and $i^+_\alt$, closing off either end of that
cylinder) and the past mirror of all that.

\subsubhead
Schwarzschild--de Sitter
\endsubsubhead

Schwarzschild--de Sitter models a spherically symmetric, uncharged,
black hole of mass $m > 0$ in a vacuum spacetime with cosmological
constant $\Lambda > 0$, asymptotically like de Sitter space
(a universe with accelerating expansion); see, for instance
\cite{R}.  Its metric is
$$ds^2 = -\(1-\dfrac{2m}r - \dfrac{\Lambda}3r^2\)dt^2 +
\(1-\dfrac{2m}r - \dfrac{\Lambda}3r^2\)^{-1}dr^2 + r^2 d\sph2;$$
this gives us $f(r) = 1-\dfrac{2m}r - \dfrac{\Lambda}3r^2$, which
has an interval for which it is positive if and only if $\Lambda <
1/(9m^2)$.  Assuming, then, that $0 < \Lambda < 1/(9m^2)$, we have
$f > 0$ on an interval $(r_-,r_+)$, where $0 < r_- < 3m < r_+$.  This
yields $\phi(r) = r^2/f(r)$ having a minimum at $r = 3m$ and going
to infinity monotonically at both $r_-$ and $r_+$.  Thus, we can
apply Corollary 6.6 to what we may perhaps call medial
Schwarzschild--de Sitter, $\sds_\med = \euc{} \times
(r_-,r_+) \times \sph2$:  Since $\int_{r_-}^{3m} 1/f(r) dr =
\int_{3m}^{r_+} 1/f(r) dr = \infty$ (just note that for some constant
$A$, $1/f(r) = r(r-2m-(\Lambda/3)r^3)^{-1}$ behaves like
$A(r-r_-)^{-1}$  for $r$ close to $r_-$, and so on), we get
$\FB(\sds_\med)$ is a pair of conjoined null cones
on $\sph2$, one each at $r = r_-$ and $r = r_+$ and meeting in
a point which we'll call $i^+$ by analogy with Schwarzschild.

As with Schwarzschild, we can examine the causal boundary of other
sectors, where $f < 0$, by using Proposition 3.5(b) of \cite{H4}. 
For interior Schwarzschild--de Sitter, $\sds_\intr = (0,r_-) \times
\euc{} \times \sph2$, we get the same as for interior Schwarzschild:
$\FB(\sds_\intr)$ is a spacelike cylinder $\euc{} \times \sph2$, the
singularity at $r = 0$, while $\PB(\sds_\intr)$
is a pair of null cylinders on $\sph2$, one each at $t =
\infty$ and $t = -\infty$, sharing a mutual cap of a spacelike
$\sph2$; and these all match up with the boundaries of $\sds_\med$
just as the boundaries of interior and exterior Schwarzschild match
up.  We also have exterior Schwarzschild--de Sitter, $\sds_\ext =
(r_+,\infty) \times \euc{} \times \sph2$, which behaves very
similarly: $\FB(\sds_\ext)$ is the de Sitter--like boundary at
infinity, a spacelike cylinder $\euc{} \times \sph2$ at $r = \infty$,
while $\PB(\sds_\ext)$ is a pair of null cylinders on $\sph2$ at $t
= \infty$ and $t = -\infty$, meeting in a spacelike $\sph2$; these
also join up with $\sds_\med$ in a similar fashion, with the $t
=\infty$ end of the cylinder in
$\FB(\sds_\ext)$ joining to the point $i^+$ in
$\FB(\sds_\med)$, and the null cylinder in
$\PB(\sds_\ext)$ at $t = \infty$ matching the null cone in
$\FB(\sds_\med)$ at $r = r_+$.  

This still leaves the $t = -\infty$ null cylinder in
$\PB(\sds_\ext)$, which analysis shows not to be at geometric
infinity (for instance, radial null geodesics are affinely
parametrized by $r$, so we should expect continuance beyond $r =
r_+$):  In maximally extended Schwarzschild--de Sitter (see
\cite{GiH}), another ($t$-reversed) copy of
$\sds_\med$ attaches to that cylinder, via that copy's $r = r_+$ null
cone in $\FB(\sds_\med)$; likewise, the $r=r_-$ cone attaches to
another copy of $\sds_\intr$ via one cylinder in $\PB(\sds_\intr)$,
and the whole process repeats at the other cone in that boundary,
yielding further copies of $\sds_\med$, then $\sds_\ext$, then
$\sds_\med$, $\sds_\intr$, and so forth.  Looking back at the
original $\sds_\intr$, we find yet another copy of $\sds_\med$
attaching (just as $\sch_\alt$ attaches to
$\sch_\intr$), along with a corresponding string of further copies
of $\sds_\ext$, $\sds_\med$, $\sds_\intr$, so on.   
 
And just as $\sch_\wth$ attaches to $\sch_\ext$ and $\sch_\alt$, the
maximal extension also has a time-reversed copy of $\sds_\intr$,
which we may call $\sds_\wth$, attaching to neighboring copies of
$\sds_\med$: each of the two null cylinders in $\FB(\sds_\wth)$
attaching to one of the null cones in each of the copies of
$\PB(\sds_\med)$. Thus, the entire future causal boundary of maximal
$\sds$ consists of a string of spacelike cylinders
$\euc{}\times\sph2$ (alternating as spatially extended sigularities
and spatially extended infinities) joined together with points $i^+$
which close off the cylinders; the past causal boundary is the
mirror of that with no connection to the future causal boundary.  

One may also consider Schwarzschild--Anti-de Sitter space, which is
the same metric but with $\Lambda < 0$, \ie, a negative cosmological
constant and a spacetime which is asymptotically like anti-de Sitter
space.  We have $f > 0$ on $(r_0,\infty)$ for some $r_0
< 2m$, and $\phi(r) = r^2/f(r)$ has a minimum at $r = 3m$, going
monotonically to inifinty at $r = r_0$ and monotonically to
$3/(-\Lambda)$ at $r  = \infty$.  Thus, $\sads_\ext = \euc{} \times
(r_0,\infty) \times \sph2$ comes under the conditions of Corollary
6.6.  We have $\int_{r_0}^{3m} 1/f(r) dr = \infty$, but
$\int_{3m}^\infty 1/f(r) dr < \infty$ (as $1/f(r)$ behaves like
$r^{-2}$ for $r$ near $\infty$).  This tells us $\FB(\sads_\ext)$ has
a null cone on $\sph2$ at $r = r_0$ but a timelike cone on $\sph2$
at $r = \infty$ (the two cones conjoined at the vertex $i^+$).  The
same is true for $\PB(\sads_\ext)$, of course (with vertex $i^-$). 
It is common wisdom to combine timelike elements of the future and
past causal boundaries, yielding a timelike suspension (cylinder
closed at both ends) on $\sph2$ at $r = \infty$ serving duty (along
with the respective null cones at $r = r_0$) for both
$\FB(\sads_\ext)$ and $\PB(\sads_\ext)$.  This timelike suspension
on $\sph2$ lies at geometric infinity; there is no extension
possible beyond it (null geodesics cannot extend beyond $r =
\infty$; it is essentially the anti-de Sitter horizon at spatial
infinity).  But the $r = r_0$ null cone attaches, just as before, to
a  black-hole--like $\sads_\intr$ and white-hole--like $\sads_\wth$,
with a ($t$-reversed) $\sads_\alt$ attaching on the other boundaries
of
$\sads_\intr$ and $\sads_\wth$.

This yields a total causal boundary for maximal $\sads$ (assuming
the standard identification) as two timelike cyinders $\euc{} \times
\sph2$ (at infinity) joined to two spacelike cylinders
$\euc{}\times\sph2$ (singularities) at points $i^\pm$ and
$i^\pm_\alt$, closing off the cylinders.

It should be noted that issues of identification of various parts of
the future and past causal boundaries typically become important, and
possibly controversial, when portions of the boundaries have timelike
character.  Some discussion of appropriate conditions for making
identifications within the causal boundary construction can be found
in \cite{S}.  A somewhat different (but closely related) approach
can be found in \cite{MR} and \cite{F}.

\subsubhead
Reissner-Nordstr\"om
\endsubsubhead

Reissner-Nordstr\"om models a spherically symmetric black hole of
mass $m > 0$ and electric charge $q \neq 0$ in an otherwise empty,
asymptotically flat universe (extensively described in \cite{HE}). 
The metric is
$$ds^2 = -\(1-\dfrac{2m}r + \dfrac{q^2}{r^2}\)dt^2 +
\(1-\dfrac{2m}r + \dfrac{q^2}{r^2}\)^{-1}dr^2 + r^2 d\sph2,$$
\ie, $f(r) = 1-\dfrac{2m}r + \dfrac{q^2}{r^2}$.  The analysis of $f$
depends on the relative sizes of $q$ and $m$.

First assume the undercharged case, $|q| < m$; then $f > 0$ on two
intervals $(0,r_-)$ and $(r_+,\infty)$, where $r_\pm = m\(1 \pm
\sqrt{1-(\frac{q}{m})^2}\,\)$.  We find $\phi(r) = r^2/f(r)$ has a
local minimum at $r_+^0 = \frac32m\(1 +
\sqrt{1-(\frac23\frac{q}{m})^2}\,\) > r_+$ and is monotonic on either
side of $r_+^0$ within $(r^+,\infty)$.  Thus, Corollary 6.6 applies
to exterior undercharged Reissner-Nordstr\"om $\rn^\und_\ext =
\euc{} \times (r_+,\infty) \times \sph2$.  We have both
$\int_{r_+^0}^\infty 1/f(r) dr = \infty$  and $\int_{r_+}^{r_+^0}
1/f(r) dr = \infty$, so $\FB(\rn^\und_\ext)$ consists of two null
cones on $\sph2$ conjoined at the vertex $i^+$, with
$\PB(\rn^\und_\ext)$ analogous.  

The medial portion $\rn^\und_\med = (r_-,r_+)\times\euc{}\times\sph2$
can be analyzed with the means in \cite{H4}:  The metric is
conformal to the warped product metric $-dr^2 + (-(\frac{q}r)^2 +
\frac{2m}r -1)^2dt^2 + r^2(-(\frac{q}r)^2 + \frac{2m}r -1)d\sph2$;
but in this case, the appropriate integrals of the warping function
for the $\euc{}$ factor are finite at both the future ($r_-$) and
past ($r_+$) endpoints of the interval, so both $\FB(\rn^\und_\med)$
and $\PB(\rn^\und_\med)$ consist of two null cylinders on
$\sph2$ (one each for rays going out to $t =\infty$ and $t =
-\infty$), joined at a spacelike $\sph2$.  The $t = \infty$ cylinder
in $\PB(\rn^\und_\med)$ attaches to the $r = r_+$ cone in
$\FB(\rn^\und_\ext)$.  

The inner sector $\rn^\und_\intr = \euc{}\times (0,r_-)\times\sph2$
has $\phi(r)$ behaving in a way that is not covered by Corollary
6.6:  $\phi$ is increasing on the entire range, approaching 0 at $r
= 0$ and infinity at $r = r_-$.  A full treatment here would require
a theorem analogous to Theorem 6.2 covering the case of $a$ being
monotonic on $(\alpha,\omega)$ with $a(\alpha) = 0$.  This is a bit
more complicated in that the Riemannian manifold $(\alpha,\omega)
\times_a K$ does not have all distances realized by geodesics; a
typical example is with $(\alpha,\omega) = (0,\infty)$, $a(r) = r$,
and $K = \sph{n-2}$, yielding Euclidean space $\euc{n-1}$ minus the
origin.  But this is not unmanageable, and there appears to be no
major obstacle to achieving very similar results (though the case of
$a(\alpha) > 0$ is considerably messier and the results less clear).

Other than that missing point, how do the proofs change for a
monotonic analogue of Theorem 6.2?  The first thing to note is that
for geodesics heading off to $\alpha$, the limiting radial geodesic
$\gamma$ necessarily has $\theta_\gamma = \pi$; the analogue of Lemma
6.3 must include inward-radial as a possible (indeed,
necessary) conclusion.   Lemma 6.4 works the same as before, but use
of Lemma 6.5 becomes more complicated:  We must instead make use of
the note following Lemma 6.5 in an analogue of Part IIa of the
Theorem 6.2 proof.  Here is how that runs:

We get the same equation for $b_p((\bar\rho,q))$, in terms of a
sequence of numbers $\{s_n\}$ approaching infinity with $\theta_n >
\pi/2$, that appears at the end of Part IIa, including the two limit
terms involving integrals.  Unlike in the original version of
Theorem 6.2, in which it was shown that a contradiction arises from
the existence of the sequence $\{s_n\}$, now we must show why each of
these limits is independent of $p$ (the apparent $p$-dependence comes
in $\theta_n$, the intial radial angle of the minimizing geodesics
$\gamma_n$ running from $(\bar\rho,p)$ to $(s_n,q)$).  This is easy
for the second limit:  We are concerned with
$\lim_{n\to\infty}\int_{\rho_n}^{\bar\rho} \(1/\sqrt{1-(\bar
a\sin\theta_n / a(\rho))^2}\)\,d\rho$, where $\rho_n = a^{-1}(\bar
a\sin\theta_n)$.  We just rewrite this as $\lim_{n\to\infty}
\int_{\rho_n}^{\bar\rho} \(a(\rho) / \sqrt{a(\rho)^2 -
a(\rho_n)^2}\) d\rho$, with $\{\rho_n\}$ being some sequence
approaching $\alpha$, and we see it has no dependence on $p$.  (If
$\int_x^{\bar\rho} \(a(\rho) / \sqrt{a(\rho)^2 - a(x)^2}\) d\rho$ is
continuous in $x$---not obvious, as this is an improroper
integral---then the limit is $\bar\rho - \alpha$; but we don't
need this.)

For the first limit, $\lim_{n\to\infty}\(\bar a^2 \sin^2\theta_n
\int_{\bar\rho}^{s_n}
\dfrac{d\rho}{a(\rho)^2A(\theta_n)(A(\theta_n) + 1)}\)$, let us
denote the integral by $J_n$.  As $A(\theta_n) \le 1$, we must have
$\lim_n J_n = \infty$, as we are assuming $\int_{\bar\rho}^\infty
1/a(\rho)^2 d\rho = \infty$.  This forces
$\{\theta_n\}$ to approach $\pi$ (given that all $\theta_n > \pi/2$),
since otherwise we would have $b_p((\bar\rho,q)) = -\infty$, and
that is impossible.  Thus, all we need to do is show that the
sequence $\{\bar a \sin\theta_n J_n\}$ is bounded above, and we will
have $\lim_n (\bar a^2 \sin^2\theta_n J_n) = 0$, independent of
$p$.  It will suffice to have $\bar a \sin\theta_n
\int_{\bar\rho}^{s_n} \dfrac{d\rho}{a(\rho)^2A(\theta_n)}$ bounded
above; note that this is exactly the same as $L(\sigma_n)$, where
$\sigma_n$ is the projection of $\gamma_n$ to $K$, a pregeodesic
from $p$ to $q$.  While we don't know that $\sigma_n$ is minimizing,
we do know, from the note following Lemma 6.5, that it is the union
of at most two minimizing geodesics.  It follows that $L(\sigma_n)
\le 2\,\text{diam}(K)$, and we are done.

So a monotonic analogue of Corollary 6.6 necessitates we need
consider only $\int_{r_--\epsilon}^{r_-} 1/f(r) dr$ and
$\int_0^\epsilon 1/f(r) dr$; the first is infinite, the second
finite.  This gives us, for both $\FB(\rn^\und_\intr)$ and
$\PB(\rn^\und_\intr)$, a null cone on $\sph2$ at $r = r_-$
conjoined at the vertex $j^+$ or $j^-$ to a timelike line at
$r = 0$ (the central singularity---or one instance of it).  The $r =
r_-$ cone in $\PB(\rn^\und_\intr)$ attaches to the $t = \infty$
cylinder in $\FB(\rn^\und_\med)$.  Conventional wisdom is to identify
the $r=0$ lines in $\FB(\rn^\und_\intr)$ and $\PB(\rn^\und_\intr)$,
yielding a timelike line with two endpoints, $j^+$ and $j^-$, for the
singularity. 

In the maximal extension of $\rn^\und$, the $t= -\infty$ cylinder in
$\PB(\rn^\und_\med)$ attaches to the $r = r_+$ cone in the future
boundary of a $t$-reversed copy $\rn^\und_\alt$ of $\rn^\und_\ext$. 
We also have the $t = \infty$ cylinder in $\FB(\rn^\und_\med)$
attaching to the $r=r_-$ cone in the past boundary of a $t$-reversed
copy $\rn^\und_{\altint}$ of $\rn^\und_\intr$. A typical Penrose 
diagram also suggests identifying $i^+$ in $\rn^\und_\ext$ with
$j^-_\alt$, the past endpoint (at $t = \infty)$ of the singularity
in $\rn^\und_\altint$; this is perhaps justified by the observation
that the elements of the alternate singularity, parametrized by $t$,
do approach $i^+$ as $t$ goes to $\infty$.  In similar fashion,
$i^+_\alt$ (future timelike infinity, at $t = -\infty$) in
$\rn^\und_\alt$ may be identified with $j^-$ in $\rn^\und_\intr$.

Continuing in the maximal extension:  A $t$-reversed copy of
$\rn^\und_\med$ attaches via the two cylinders of its past boundary
to the $r = r_-$ cones in $\FB(\rn^\und_\intr)$ and
$\FB(\rn^\und_{\altint})$.  The future boundary of that copy of
$\rn^\und_\med$ attaches to the past boundaries of copies of
$\rn^\und_\ext$ and $\rn^\und_\alt$, another copy of
$\rn^\und_\med$ attaches via its past boundary to the future
boundaries of those, and so on, futurewards.  Similarly, the $r =
r_+$ cones in $\PB(\rn^\und_\ext)$ and $\PB(\rn^\und_\alt)$ attach to
the future boundary of another $t$-reversed copy of $\rn^\und_\med$,
whose past boundary attaches to a continued pastwards extension.

Thus, the entire causal boundary of $\rn^\und$ consists of the
following: a sequence of connected components, each component
consisting of a null cone (on $\sph2$) with vertex $i^+/j^-$ to the
future, timelike line leading futurewards from that $i^+/j^-$ to a
point $i^-/j^+$, and a null cone with that $i^-/j^+$ as vertex to the
past; and a duplicate of this sequence on the ``other side". 

Critically charged Reissner-Nordstr\"om has $|q| = m$, yielding
$f(r) = (1-\frac{m}r)^2$.  On the exterior portion, $\rn^\crt_\ext
= \euc{} \times (m,\infty) \times \sph2$, we have $\phi(r) =
(r^2/(r-m))^2$, and we apply Corollary 6.6 as before, with the same
result: $\FB(\rn^\crt_\ext)$ and $\PB(\rn^\crt_\ext)$ each consist
of two null cones on $\sph2$ (one for $r = \infty$ and one for $r = 
m$) conjoined at what might respectively be called $i^+$ and $i^-$. 
The interior portion, $\rn^\crt_\intr =
\euc{}\times(0,m)\times\sph2$, has the same issue as with 
$\rn^\und_\intr$\,:  $\phi(r)$ decreases to 0 at $r = 0$.  Using the
same monontonic analogue of Corollary 6.6, we obtain, similarly, a
null cone on $\sph2$ at $r = m$ conjoined at its vertex ($j^+$ or
$j^-$) to a timelike line at $r= 0$, for both $\FB(\rn^\crt_\intr)$
and $\PB(\rn^\crt_\intr)$.  The null cone in $\PB(\rn^\crt_\intr)$
attaches to the $r=m$ cone in
$\FB(\rn^\crt_\ext)$.  Convention dictates identifying the timelike
lines in $\FB(\rn^\crt_\intr)$ and $\PB(\rn^\crt_\intr)$.  In a
maximal extension, the $r = m$ cone in
$\FB(\rn^\crt_\intr)$ attaches to the $r=m$ cone in the past
boundary of another copy of $\rn^\crt_\ext$, which attaches to
another copy of $\rn^\crt_\intr$, and  so on furturewards; and
similarly pastwards from the $r=m$ cone $\PB(\rn^\crt_\ext)$.  A
typical Penrose diagram suggests identifying the $i^-$ of one copy of
$\rn^\crt_\ext$ with the $i^+$ of the copy to its immediate past,
but justification for this is questionable.  However, topology
justifies identifying the past endpoint $j^-$ of the singularity in
each copy of $\rn^\crt_\intr$ with the future endoint $j^+$ of the
singularity in the copy to its immediate past, much as in
$\rn^\und$.  We obtain, then, for the full causal boundary, the
following: in the exterior, a sequence of null cones on $\sph2$,
alternating with vertex at future or at past (unlcear whether to
identify pairs of verticies $i^-$ and $i^+$) ; and in the interior, a
single timelike line (as paired $j^-$ and $j^+$ are identified).

Overcharged Reissner Nordstr\"om has $|q| > m$, yielding $f(r) > 0$
for all $r$: $\rn^\ovr = \euc{}\times(0,\infty)\times\sph2$.  We
have $\phi(r)$ decreasing to 0 at $r = 0$ and eventually increasing
to infinity at $r = \infty$.  We have $\int_0^m 1/f(r) dr < \infty$
 and $\int_m^\infty 1/f(r) dr = \infty$.  A monotonic analogue of
Corollary 6.8 tells us $\FB(\rn^\ovr)$ and $\PB(\rn^\ovr)$ are each
a null cone on $\sph2$ at $r = \infty$ joined, at the vertex $i^+$
or $i^-$, to a timelike line at $r = 0$; conventionally, the timelike
lines are identified yielding a single timelike line from $i^-$ to
$i^+$ for the singularity.  Thus, the entire entire causal boundary
consists of a null cone (on $\sph2$) with vertex $i^-$ to the past, a
timelike line leading futwards from $i^-$ to a point $i^+$, and a
null cone with that $i^+$ as vertex to the future.

What is perhaps most noteworthy here is that the singularity in 
$\rn$ is one-dimensional, as opposed to the $\euc{}\times\sph2$
structure of the singuarity in $\sch$ (or $\sds$ or $\sads$).  For
the spacelike singularities (of Schwarzschild character), this is a
topologically universal result, as detailed in \cite{H2}; for the
timelike Reissner-Nordstr\"om singularity, it is at least causally
universal, as detailed in \cite{H1}.

\enddocument